\documentstyle[prd,aps,preprint,psfig,floats]{revtex}
\begin{document}
\newcommand{\Rz}{\left(\frac{\rho_\phi}{s}\right)_{0}}
\newcommand{\RBB}{\left(\frac{\rho_\phi}{s}\right)_{BB}}
\newcommand{\RBBm}{\left(\frac{\rho_\phi}{s}\right)_{BBm}}
\newcommand{\RTI}{\left(\frac{\rho_\phi}{s}\right)_{TI}}
\newcommand{\TC}{\left(\frac{T_c}{m_0}\right)}
\newcommand{\mgra}{\left(\frac{m_{3/2}}{m_\phi}\right)}
\newcommand{\phiz}{\left(\frac{\phi_0}{M_G}\right)}
\newcommand{\eV}{~\mbox{eV}}
\newcommand{\keV}{~\mbox{keV}}
\newcommand{\MeV}{~\mbox{MeV}}
\newcommand{\GeV}{~\mbox{GeV}}
\newcommand{\TeV}{~\mbox{TeV}}
\newcommand{\for}{~~\mbox{for}~}
\newcommand{\fn}{~~\mbox{for}~n=1}
\tighten
\preprint{\begin{tabular}{l}
\hbox to\hsize{\hfill UT-851}\\
\hbox to\hsize{\hfill RESCEU-13/99}\\
\hbox to\hsize{\hfill}\\
\hbox to\hsize{\hfill}\\
\hbox to\hsize{\hfill}
\end{tabular}}
\title{Cosmological Moduli Problem and 
       Thermal Inflation Models}
\author{T. Asaka$^1$ and M. Kawasaki$^2$}
\address{$^1$Department of Physics,  University of
  Tokyo, Tokyo 113-0033, Japan}
\address{$^2$Research Center for the Early Universe (RESCEU),
  University of Tokyo, Tokyo 113-0033, Japan}
\date{May 25, 1999}

\maketitle
\begin{abstract}
    In superstring theories, there exist various dilaton and modulus
    fields which masses are expected to be of the order of the
    gravitino mass $m_{3/2}$.  These fields lead to serious
    cosmological difficulties, so called ``cosmological moduli
    problem'', because a large number of moduli particles are produced
    as the coherent oscillations after the primordial inflation.  We
    make a comprehensive study whether the thermal inflation can solve
    the cosmological moduli problem in the whole modulus
    mass region $m_\phi \sim 10$~eV--$10^4$~GeV predicted by both
    hidden sector supersymmetry (SUSY) breaking and gauge-mediated
    SUSY breaking models. In particular, we take into account the
    primordial inflation model whose reheating temperature is so low
    that its reheating process finishes after the thermal inflation
    ends.  We find that the above mass region $m_\phi ( \simeq
    m_{3/2}) \sim 10$~eV--$10^4$ GeV survives from various
    cosmological constraints in the presence of the thermal inflation.
\end{abstract}
\clearpage
\section{Introduction}
\label{sec-intro}
%
Supersymmetry (SUSY) is one of the most attractive extensions of the
standard model.  In virtue of the SUSY the electroweak scale can be
stabilized against the radiative corrections.  Furthermore, in the
SUSY grand unified theories the unification of the standard gauge
couplings can be realized.

However, overviewing the cosmology of the SUSY model, we are faced
with various difficulties.  One of the cosmological problems is the
gravitino problem \cite{Pagels-Primack,Weinberg,Krauss}.  This problem
still exists even if the universe experienced a primordial inflation,
since the gravitino is reproduced by scatterings with particles in the
thermal bath at the reheating epoch.  In the hidden sector SUSY
breaking (HSSB) models \cite{HSSB} the gravitino has a mass of the
order of the electroweak scale: $m_{3/2} \sim 10^{2}$--$10^3$ GeV, and
it decays soon after the big bang nucleosynthesis (BBN).  Since high
energy photons produced by the gravitino decay might destroy the light
elements (D, $^3$He, $^4$He) synthesized by the BBN, the reheating
temperature of the primordial inflation should be low enough not to
conflict with the observations (e.g., see a recent analysis in Ref.
\cite{Holtmann-Kawasaki-Kohri-Moroi}).  On the other hand, the gauge
mediated SUSY breaking (GMSB) models \cite{GMSB} predict a light
gravitino of mass $m_{3/2} \sim 10$~eV--$1$ GeV. If the gravitino mass
is $m_{3/2} \gtrsim 1$keV \cite{Pagels-Primack}, a low enough
reheating temperature of the inflation is also required to avoid the
overclosure by the stable gravitino \cite{Moroi-Murayama-Yamaguchi}.

Furthermore, when one considers the SUSY models in the framework of
the superstring theories \cite{Green-Schwartz-Witten}, they suffer
from a more serious problem, i.e., ``cosmological moduli problem''
\cite{Coughlan-Fischler-Kolb-Raby-Ross,%
Banks-Kaplan-Nelson,Carlos-Casas-Quevedo-Roulet}.  As a general
consequence of the superstring theories, various dilaton and modulus
fields appear.  (We call them as ``moduli'' throughout this paper.)
These moduli fields $\phi$ are expected to acquire masses of the order
of the gravitino mass ($m_{\phi} \simeq m_{3/2}$) from nonperturbative
effects of the SUSY breaking \cite{Carlos-Casas-Quevedo-Roulet}.
Their lifetimes, since they have only the gravitationally suppressed
interaction, are roughly estimated as
\begin{eqnarray}
    \label{tp_naive}
    \tau_\phi 
    ~\sim~ 
    N^{-1} \frac{ M_{pl}^2 }{ m_\phi^3 }
    ~\simeq~ 10^{17} ~\mbox{sec}~ N^{-1} 
    \left( \frac{ m_\phi }{ 100\MeV} \right)^{-3},
\end{eqnarray}
where $N$ denotes the number of the decay channels and $M_{pl}$ is the
Planck scale $M_{pl} = 1.2 \times 10^{19}$ GeV.  The existence of
moduli $\phi$ with such long lifetimes leads to various cosmological
difficulties, because a large number of moduli particles are produced
as the coherent oscillations after the primordial inflation.  In the
HSSB models $\phi$ decay soon after the BBN as the
gravitino, and hence the light elements might also be destroyed
\cite{Coughlan-Fischler-Kolb-Raby-Ross,%
Banks-Kaplan-Nelson,Carlos-Casas-Quevedo-Roulet}.  On the other hand,
for the lighter stable moduli predicted by the GMSB models, say
$m_\phi \lesssim 100$ MeV, the energy of the oscillation lasting until
the present overcloses the universe.  Moreover, moduli with masses
$m_\phi \sim 0.1$~MeV--$1$ GeV give too much contributions to
the x($\gamma$)-ray background spectrum
\cite{Kawasaki-Yanagida}.  Therefore, moduli particles typically
predicted by both HSSB and GMSB models bring a cosmological disaster.

One way to evade these difficulties due to the string moduli is to
give heavy masses ($\phi \gtrsim 100$ TeV) to all moduli, 
so that the decays of moduli take place before the BBN
\cite{Ellis-Nanopoulos-Quiros} 
and the lightest superparticles produced by the moduli decays
do not overclose the universe \cite{Kawasaki-Moroi-Yanagida}.
Although a large entropy
is produced by the moduli decay, the present observed baryon asymmetry
can be naturally explained by the Affleck-Dine mechanism
\cite{Affleck-Dine} as shown in Ref. \cite{Moroi-Yamaguchi-Yanagida}.
However, such heavy moduli masses (i.e., a heavy gravitino mass) could
be only achieved by some specific models.

Here we would like to emphasize that this cosmological moduli problem
could not be solved by the primordial inflation, even if one assumed
an extremely low reheating temperature about 10 MeV which is limited
from below by the BBN observation.  This is a crucial difference from
the gravitino problem.  Therefore, we require some extra mechanism
other than the primordial inflation to dilute the moduli mass density
sufficiently.\footnote{
Some other approach to the solution has been proposed
\cite{Dine-Randall-Thomas}, but it is difficult to construct a
realistic model.  }

One of such dilution mechanisms is the thermal inflation model
proposed by Lyth and Stewart \cite{Lyth-Stewart}.\footnote{
The other possible mechanism is the oscillating inflation by Moroi
\cite{Moroi} which works in the GMSB models.  A mini-inflation takes
place while a scalar field corresponding to the flat direction
oscillates along the logarithmic potential induced by the GMSB
mechanism and it significantly dilutes the relic density of the
moduli.  See also Ref. {\protect \cite{Asaka-Kawasaki-Yamaguchi}}.}
The thermal inflation occurs before the electroweak phase transition
and produces tremendous entropy (just) before the BBN epoch, which
leads to sufficient dilution of the moduli density.  
In Ref. \cite{Lyth-Stewart} it was shown
that the moduli problem in the HSSB
models can be solved by the thermal inflation.\footnote{
More precisely, the thermal inflation can solve the moduli problem if
one assumes the primordial inflation model which reheating process
completes after the moduli begin the coherent oscillations.  }

On the other hand, the attempts to solve the problem in the GMSB
models by the thermal inflation were done in Refs.
\cite{Gouvea-Moroi-Murayama,Kawasaki-Yanagida,%
Hashiba-Kawasaki-Yanagida,Asaka-Hashiba-Kawasaki-Yanagida1}.  
It was shown by Ref.
\cite{Gouvea-Moroi-Murayama} that it can solve the overclosure problem
for the stable moduli whose masses are $m_\phi \sim 10$~eV--$10^{-1}$
GeV.  However, in Refs.
\cite{Kawasaki-Yanagida,Hashiba-Kawasaki-Yanagida,%
Asaka-Hashiba-Kawasaki-Yanagida1}, the stringent x($\gamma$)-ray
background constraint was found to exclude the modulus mass region
$m_\phi \sim 10^{-4}$--$1$ GeV.  Thus the thermal inflation can cure
a part of the cosmological difficulties of the string moduli
predicted by both HSSB and GMSB models.

However, most of previous works (except Ref. \cite{Lyth-Stewart})
assumed that the reheating process of the primordial inflation be
completed before the moduli oscillations start, i.e., only the
inflation models with relatively high reheating temperatures were
considered.  Even in Ref. \cite{Lyth-Stewart}, the reheating process
is assumed to end before the thermal inflation starts.  Without this
restriction one might obtain wider allowed region for the modulus
mass.  In this paper, therefore, we make a comprehensive study whether
the thermal inflation can  solve cosmological problem of the moduli
particles with masses $m_\phi \sim 10$~eV--$10^4$~GeV (i.e., the whole
modulus mass region predicted by both HSSB and GMSB models),
especially taking into account the inflation model whose
reheating temperature is low enough so that its reheating process
finishes after the thermal inflation ends.  We find that the above mass
region $m_\phi ( \simeq m_{3/2}) \sim 10$~eV--$10^4$ GeV
survives from various cosmological constraints\footnote{
Here we do not include a constraint from the present baryon
asymmetry which excludes some modulus mass regions.  See the
discussion below.  }
if we consider the primordial inflation with sufficiently low
reheating temperature in addition to the thermal inflation.

The organization of this paper is as follows. In Sec. \ref{sec-mp} we first
review about the cosmological difficulties of the string moduli whose
masses are $m_\phi \sim 10$ eV--$10^{4}$ GeV.  The (original)
thermal inflation model proposed by Lyth and Stewart is explained in
Sec. \ref{sec-oti}.  Then, in Sec. \ref{sec-mpoti}, 
we study whether this original thermal
inflation could solve the moduli problem, considering various
models of the primordial inflation.  
In the original thermal inflation model, however,
there appears an $R$-axion which is the NG boson 
from the spontaneous breaking of the $R$-symmetry.
In the GMSB models the $R$-axion is light enough so that 
the flaton, which causes the thermal inflation, 
almost decays into them. Then, as shown in Ref. 
\cite{Asaka-Hashiba-Kawasaki-Yanagida1},
the original model could not produce a sufficient entropy 
to dilute the light moduli predicted by the GMSB.  
To avoid this difficulty we introduce in Sec. \ref{sec-mti}
the ``modified'' thermal inflation by Ref.
\cite{Asaka-Hashiba-Kawasaki-Yanagida1}, 
which forbids the flaton decay into $R$-axions.
Then, we investigate the solution of
the moduli problem by this modified model in Sec. \ref{sec-mpmti}.  
Finally, in Sec. \ref{sec-dis}, 
we discuss
two extra problems when one assumes the tremendous entropy production
enough to dilute the string moduli, i.e., problems of baryogenesis and
dark matter.  We find that the problem of the baryon asymmetry is so
serious that we have only two possible solution so far; 
(i) sufficient baryons can be
generated through the Affleck-Dine mechanism \cite{Affleck-Dine}
for the modulus mass
region $m_\phi (\simeq m_{3/2}) \lesssim 1$ MeV and (ii) for $m_\phi
(\simeq m_{3/2}) \sim 1$--$10^4$ GeV the electroweak baryogenesis can
produce the present observed baryon asymmetry.  
\section{Cosmological Difficulties of Moduli Particles}
\label{sec-mp}

First of all, we explain the cosmological difficulties due to the
string moduli particles.  In the following analysis we assume only one
modulus field, $\phi$, with the mass $m_\phi \simeq m_{3/2}$ to derive
the conservative constraints, although we can easily extend the
analysis for more general cases.

After the primordial inflation ends, the modulus field $\phi$ is
considered to be displaced from the true minimum 
and the displacement is of the order of $M_G$ 
($M_G$ is the reduced Planck scale $M_G = 2.4 \times
10^{18}$ GeV).  This is because of the additional SUSY breaking effect
due to large vacuum energy of the inflaton.  Then, when the expansion
rate of the universe (i.e., the Hubble parameter $H$) becomes
comparable to the modulus mass, the modulus starts to oscillate around
its true minimum with the initial amplitude $\phi_0 \sim M_G$.  At
this time, the energy density of this oscillation is given by
$\rho_\phi = m_\phi^2 \phi_0^2/2$ and the cosmic temperature of the
universe is estimated as
\begin{eqnarray}
    \label{T-phi}
    T_\phi ~\simeq ~
    \left( \frac{ 90 }{ \pi^2 g_\ast } \right)^{1/4}
    \sqrt{ m_\phi M_G }
    ~\simeq ~
    7.2 \times 10^{8}~\mbox{GeV} ~
    \left( \frac{ m_\phi }{ 1 ~\mbox{GeV} } \right)^{1/2},
\end{eqnarray}
where $g_\ast (\simeq 200)$ counts the effective degrees of freedom of
the radiation.  Thus the ratio between $\rho_\phi$ to the entropy
density $s$ is given by
\begin{eqnarray}
    \label{Rp1}
    \frac{ \rho_\phi }{ s } 
    ~\simeq~ \frac{ \frac{ 1 }{ 2 } m_\phi^2 \phi_0^2 }
           { \frac{ 2 \pi^2 }{ 45 } g_\ast T_\phi^3}
    ~\simeq~ 0.89 \times 10^8 ~\mbox{GeV}~
    \left( \frac{ m_\phi }{ 1~\mbox{GeV} } \right)^{1/2}
    \left( \frac{ \phi_0 }{ M_G } \right)^2.
\end{eqnarray}
Here it should be noted that $\rho_\phi$ and $s$ are both diluted 
like $R^{-3}$ ($R$ is the scale factor of the universe) as
the universe expands. Thus this ratio takes a constant value until
today (or until the modulus decays) if ``no'' extra entropy is
produced.  Since, at $T=T_{\phi}$, the energy density of the
radiation, $\rho_R$, is comparable to that of the modulus 
($\rho_R \sim T_\phi^4 \sim \rho_\phi$) and is diluted faster as $R^{-4}$,
the modulus oscillation soon dominates the whole energy of the
universe.

In deriving Eq.~(\ref{Rp1}) we have assumed that the reheating process
of the primordial inflation be completed before the beginning of the
modulus oscillation, i.e., the decay rate $\Gamma_{\varphi_I}$ of the
inflaton be larger than $m_\phi$.\footnote{
In other words the reheating temperature of the primordial inflation
is higher than $T_{\phi}$ given by Eq.
({\protect \ref{T-phi}}).  }
On the other hand, when $\Gamma_{\varphi_I} < m_\phi$, the energy of
the modulus oscillation is expected to be diluted
by the entropy production of the primordial inflation.\footnote{
Note that the modulus oscillation is considered to start 
at least after the primordial inflation ends, since $m_{\phi} \ll
H_{I}$ where $H_{I}$ denotes the Hubble parameter during the inflation
and $H_{I} = V_{I}/( \sqrt{3} M_{G} )$ with the inflaton's vacuum
energy $V_{I}$.  }
In this case,
when the reheating process completes at $H \simeq \Gamma_{\varphi_I}$,
the ratio $\rho_\phi / s$ is estimated as
\begin{eqnarray}
    \label{Rp2}
    \frac{ \rho_\phi }{ s } 
    &\simeq& \frac{1}{8} T_{RI} 
    \left( \frac{ \phi_0 }{ M_G } \right)^2.
\end{eqnarray}
Here $T_{RI}$ denotes the reheating temperature of the primordial
inflation which is given by
\begin{eqnarray}
    \label{TRI}
    T_{RI} ~\simeq~ 
    \left( \frac{ 90 }{ \pi^2 g_\ast} \right)^{1/4} 
    \sqrt{ \Gamma_{\varphi_I} M_G }.
\end{eqnarray}
In order to keep the success of the BBN it
should be higher than about 10 MeV and we obtain
\begin{eqnarray}
    \frac{ \rho_\phi }{ s } 
    \gtrsim 1.25 \times 10^{-3} ~\mbox{GeV} ~
    \left( \frac{ \phi_0 }{ M_G } \right)^2.
\end{eqnarray}
Comparing with Eq.(\ref{Rp1}), it is seen that the primordial
inflation with a low $T_{RI}$ does dilute the modulus energy
significantly.  However, even if we assume the primordial inflation
with extremely low reheating temperature, 
the energy of the modulus oscillation is still large.  
It leads to disastrous effect on the thermal history of
the universe after the BBN if the modulus lifetime is long enough.  

The modulus, in fact, has a very long lifetime since it has only
gravitationally suppressed interaction.  
From the naive dimensional analysis 
the modulus lifetime is estimated as shown in Eq. (\ref{tp_naive}).
Then we can roughly say that the modulus with a mass $m_\phi \lesssim
100$ MeV has a longer lifetime than the present age of the universe
$\sim 10^{17}$ sec.

We can say more precisely about the modulus lifetime for the dilaton
particle which is the most plausible candidate among various moduli.
The dilaton has the following couplings to the kinetic terms of the
gauge fields:
\begin{eqnarray}
    \label{Lint-pgg}
    {\cal L} = \frac{b}{4} \frac{\phi}{M_G}  F_{\mu \nu}^2,
\end{eqnarray}
where we introduce an order one parameter $b$ which is determined by
the string model and its compactification.  For example, the
dilaton has a coupling $b = \sqrt{2}$ \cite{Dilaton} in some
compactification of the M-theory \cite{M-theory}.  If the dilaton mass
is light enough, say $m_\phi \lesssim$ 1 GeV, it dominately decays
into two photons \cite{Kawasaki-Yanagida} and the lifetime of the
dilaton is estimated as
\begin{eqnarray}
    \label{tau-phi}
    \tau_\phi ~=~
    7.6 \times 10^{17}~\mbox{sec}~
    \frac{ 1 }{ b^2 }
    \left( \frac{m_\phi}{ 100~\mbox{MeV} } \right)^{-3}.
\end{eqnarray}
Moreover, the dilaton could decay into two gluons similarly if
kinematically allowed.  Then the dilaton lifetime becomes shorter by a
factor of 1/9 counting only the number of the final states. In
addition, if $m_\phi$ is heavier than the electroweak scale, various
decay modes may arise through the gravitational interaction with the
SUSY standard model particles, and its decay lifetime becomes shorter.
Comparing with the naive estimation Eq.  (\ref{tp_naive}), the both
results are not different very much.  
Thus, in the present analysis
we assume that the modulus lifetime be the same as the
dilaton one and it decay into only two photons and two gluons if
possible through the interaction (\ref{Lint-pgg}) with $b=1$ 
for simplicity.

We turn to see constraints on the modulus energy
from various cosmological observations.
First of all, the stable modulus of mass $m_\phi \lesssim 100$
MeV is constrained from the overclosure limit.
The present energy density of the modulus oscillation 
should be smaller than the critical density of the universe
$\rho_{cr}$.
This requirement leads to to
\begin{eqnarray}
    \label{overclose}
    \frac{ \rho_\phi }{ s } 
    ~\lesssim ~
    \frac{\rho_{cr} }{ s_0 } = 3.6 \times 10^{-9}~h^2~\mbox{GeV},
\end{eqnarray}
where $s_0$ is the present entropy density and $h$ denotes the present
Hubble parameter in units of 100 km/sec$\cdot$Mpc$^{-1}$.  
Note that even if the
modulus is unstable, we obtain similar upper bound on the modulus
abundance requiring that the decay products (i.e., photons or gluons)
should not overclose the universe.  Since the decay products are
relativistic and its energy density decreases faster than 
the modulus one,
the constraint on the abundance becomes weaker than
Eq.~(\ref{overclose}).

If the lifetime of the modulus is longer than the time of the
recombination $\tau_\phi \gtrsim 10^{12}$ sec, i.e., $m_{\phi}
\lesssim 1$ GeV, a stringent constraint comes from the the
observed cosmic x($\gamma$)-ray background spectrum
\cite{Kawasaki-Yanagida}.  The photon flux produced by the modulus
decay directly contributes to the x($\gamma$)-ray backgrounds and we
obtain the upper bound on the modulus abundance by requiring that they
should not exceed the present observed spectrum.  (One may find the
detail analysis in
Refs.~\cite{Kawasaki-Yanagida,Asaka-Hashiba-Kawasaki-Yanagida1}.)  In
fact this gives us a more stringent constraint than the above
overclosure limit (\ref{overclose}) for 1 GeV$ \gtrsim m_\phi \gtrsim
100$ keV.  Therefore, this is very stringent constraint on the models
of the GMSB where $m_\phi \simeq m_{3/2}
\lesssim 1$ GeV as first pointed out by Ref. \cite{Kawasaki-Yanagida}.

The modulus abundance is also constrained from the spectrum of the
cosmic microwave background radiation (CMBR).  If the modulus lifetime
is about $10^6$ sec $\lesssim \tau_{\phi} \lesssim 10^{12}$ sec, extra
radiation energy produced by the modulus decay may cause the spectral
distortion of the cosmic microwave background from the blackbody one.
The observation by the COBE satellite \cite{CMBR} gives the following
upper bounds on the modulus abundances:
\begin{eqnarray}
    \frac{ \rho_\phi }{ s }
    &\lesssim&
    2.8 \times 10^{-13}~\mbox{GeV}~
    \left( \frac{10^{10}~\mbox{sec}}{\tau_\phi} \right)^{1/2}\\
    & & ~~~~~~~~\mbox{for}~ 
    10^{6}\left(\frac{\Omega_Bh^2}{0.0125}\right)~\mbox{sec} 
    ~\lesssim \tau_\phi \lesssim
    1.4\times
    10^{9}\left(\frac{\Omega_Bh^2}{0.0125}\right)~\mbox{sec}.
    \nonumber \\
    \frac{ \rho_\phi }{ s }
    &\lesssim&
    2.6 \times 10^{-13}~\mbox{GeV}
    \left( \frac{10^{10}~\mbox{sec}}{\tau_\phi} \right)^{1/2}\\
    & & ~~~~~~~~\mbox{for}~ 
    1.4\times 10^{9}\left(\frac{\Omega_Bh^2}{0.0125}\right)\mbox{sec}
    ~\lesssim \tau_\phi \lesssim
    10^{12}~\mbox{sec}.\nonumber
\end{eqnarray}
Here $\Omega_B$ is the density parameter for baryons.

Furthermore, the BBN plays a significant role to set limits to the
moduli abundance.  First, if the modulus oscillation exits at the
cosmic temperature about 1 MeV, its extra energy accelerates the
expansion of the universe and the weak interaction freezes out
earlier.  This leads to higher neutron-proton number ratio 
and overproduction
of ${}^{4}\mbox{He}$.  To keep the success of the BBN, the modulus
energy density should be smaller than that of one neutrino species
[e.g., see Ref. \cite{Holtmann-Kawasaki-Kohri-Moroi}],
i.e.,
\begin{eqnarray}
    \frac{ \rho_\phi }{ s } 
    \lesssim 1.2 \times 10^{-4} ~\mbox{GeV},
\end{eqnarray}
for $\tau_{\phi} \gtrsim 1$~sec.

Furthermore, when the modulus mass becomes larger than about 10 GeV,
the modulus decays soon after the BBN.  If the lifetime of the
modulus is longer than about $10^4$~sec, the high energy photons
emitted from the modulus may destroy or overproduce the light elements
of the universe synthesized by the BBN.  Not to spoil the success of
the BBN, the modulus abundance is stringently constrained
[e.g., see the resent work \cite{Holtmann-Kawasaki-Kohri-Moroi}].
Moreover, when the modulus lifetime becomes shorter than about
$10^4$~sec, the hadronic cascade processes associated with the modulus
decay may modify the primordial abundances of light elements.  This
also puts the upper bound \cite{Reno-Seckel}.

Although we have neglected the modulus decay into the SUSY standard
model particles, the stable lightest SUSY particle (LSP) produced by
the modulus decay is potentially dangerous, since its energy density
might overclose the universe.  However, from the present mass limit 
of the LSP, we find that this constraint, if exists, is weaker than those
from the distortion of the CMBR and the photo (hadronic) dissociation
of the light elements.

It should be noted that the modulus with $m_\phi \gtrsim 10$ TeV is
cosmologically viable \cite{Ellis-Nanopoulos-Quiros}.
From Eq. (\ref{tp_naive}) the modulus with
such a heavy mass decays before the BBN and reheats the universe as $T
\gtrsim 10$ MeV which gives the initial condition of the BBN.
Then one can evade the previous constraints.\footnote{
However, in order to avoid the overclosure by the LSP produced
by the modulus decay $m_\phi \gtrsim 100$ TeV is required
\cite{Kawasaki-Moroi-Yanagida}.
Furthermore, if the modulus decays into
lighter gravitinos of mass $m_{3/2} \sim 100$ GeV--10 TeV, we are
faced with the gravitino problem {\protect
\cite{Hashimoto-Izawa-Yamaguchi-Yanagida}}.  }
\begin{figure}[t]
    \centerline{\psfig{figure=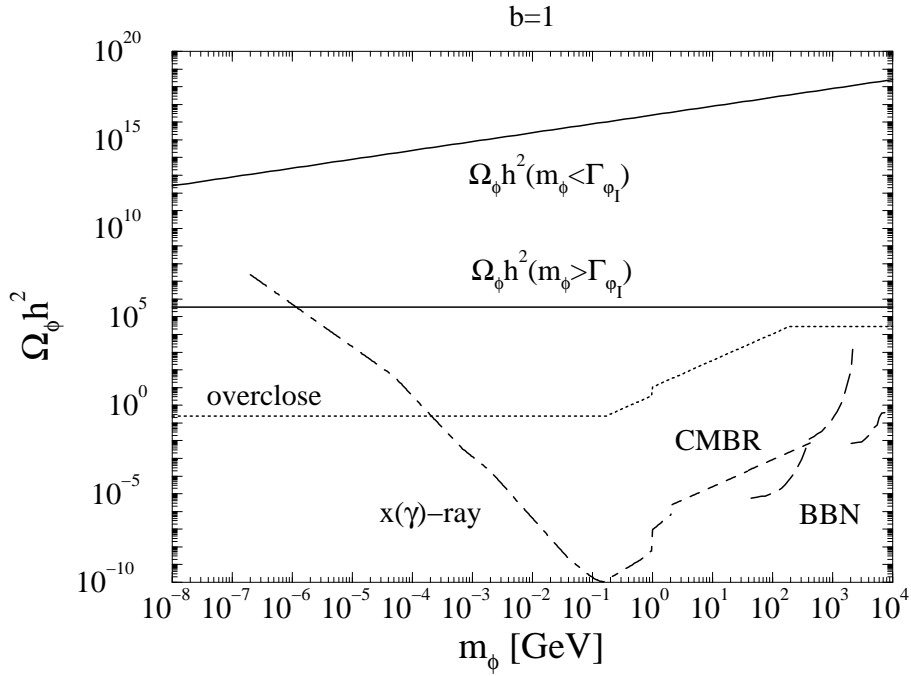,width=12cm}}
    \caption{
    Various cosmological upper bounds on the modulus abundance for the
    case that the modulus coupling is $b=1$.  The dotted line
    represents the upper bound from the overclosure limit on the
    abundance of the modulus (or its decay products) for $m_{\phi}
    < 200$ GeV, and the upper bound from the BBN speed up
    effects for $m_{\phi} > 200$ GeV.  The dot-dashed line
    represents the upper bound from the x($\gamma$)-ray backgrounds.
    The short dashed line represents the upper bound from the CMBR
    spectrum.  The long dashed line represents the upper bound from
    the dissociation of the BBN light elements.  Note that there exist
    two typical modulus masses; $m_{\phi} \simeq 100$ MeV (the modulus
    lifetime is equal to the age of the universe) and $m_{\phi} \simeq
    1$ GeV (the modulus decay into two gluons starts to open.)  We
    also show the predicted modulus abundances of $\phi_0 = M_G$ by
    the solid lines for the case $m_\phi < \Gamma_{\varphi_I}$
    [Eq. ({\protect \ref{Rp1}})] and for the case $m_\phi >
    \Gamma_{\varphi_I}$ with the reheating temperature $T_{RI}$ = 10
    MeV [Eq. ({\protect \ref{Rp2}})]. }
    \label{fig:ompb1}
\end{figure}
\begin{figure}[t]
    \centerline{\psfig{figure=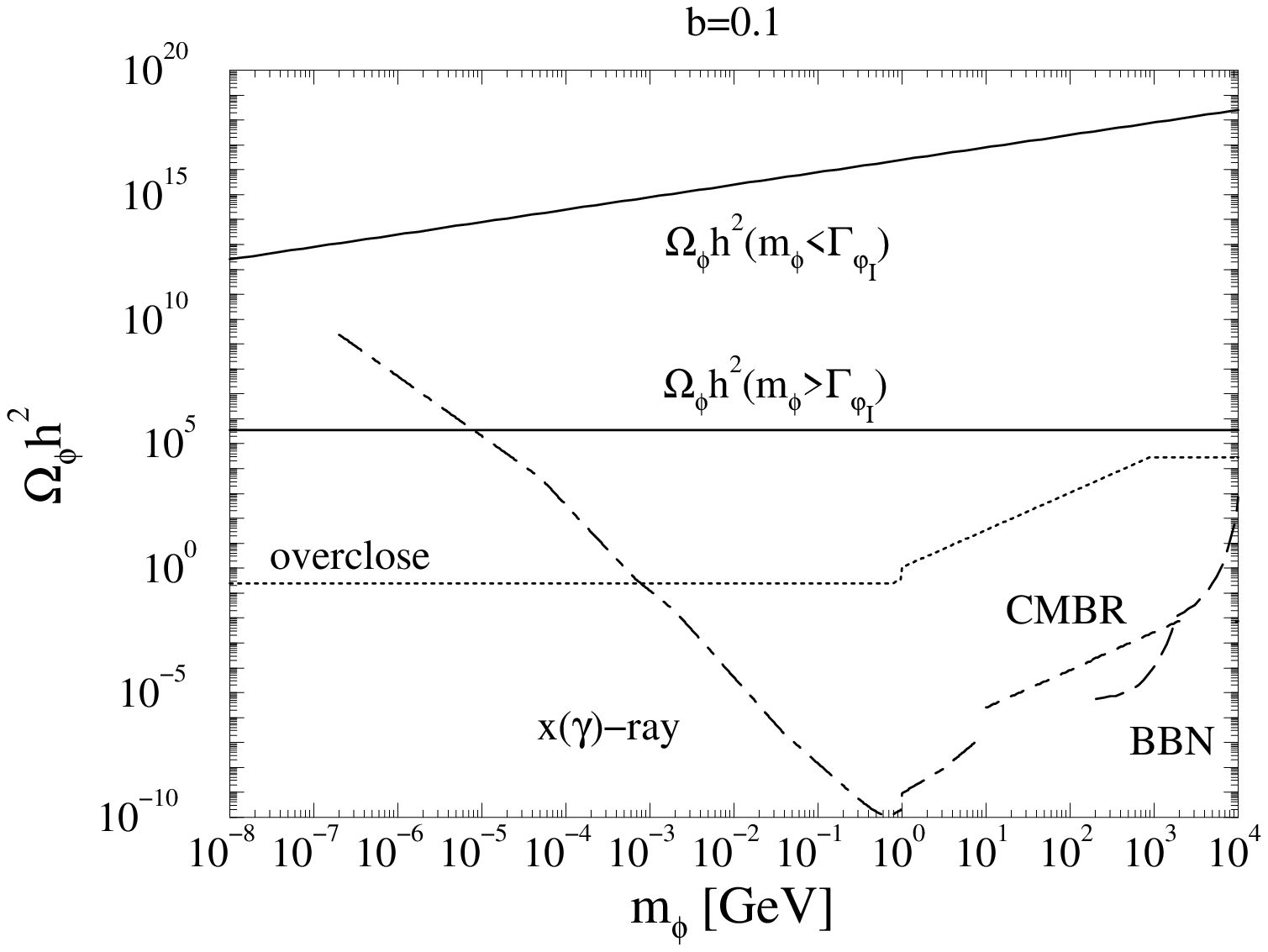,width=12cm}}
    \caption{
    Same figure as Fig. \ref{fig:ompb1} except for the value of the
    modulus coupling $b=0.1$.}
    \label{fig:ompb01}
\end{figure}
\begin{figure}[ht]
    \centerline{\psfig{figure=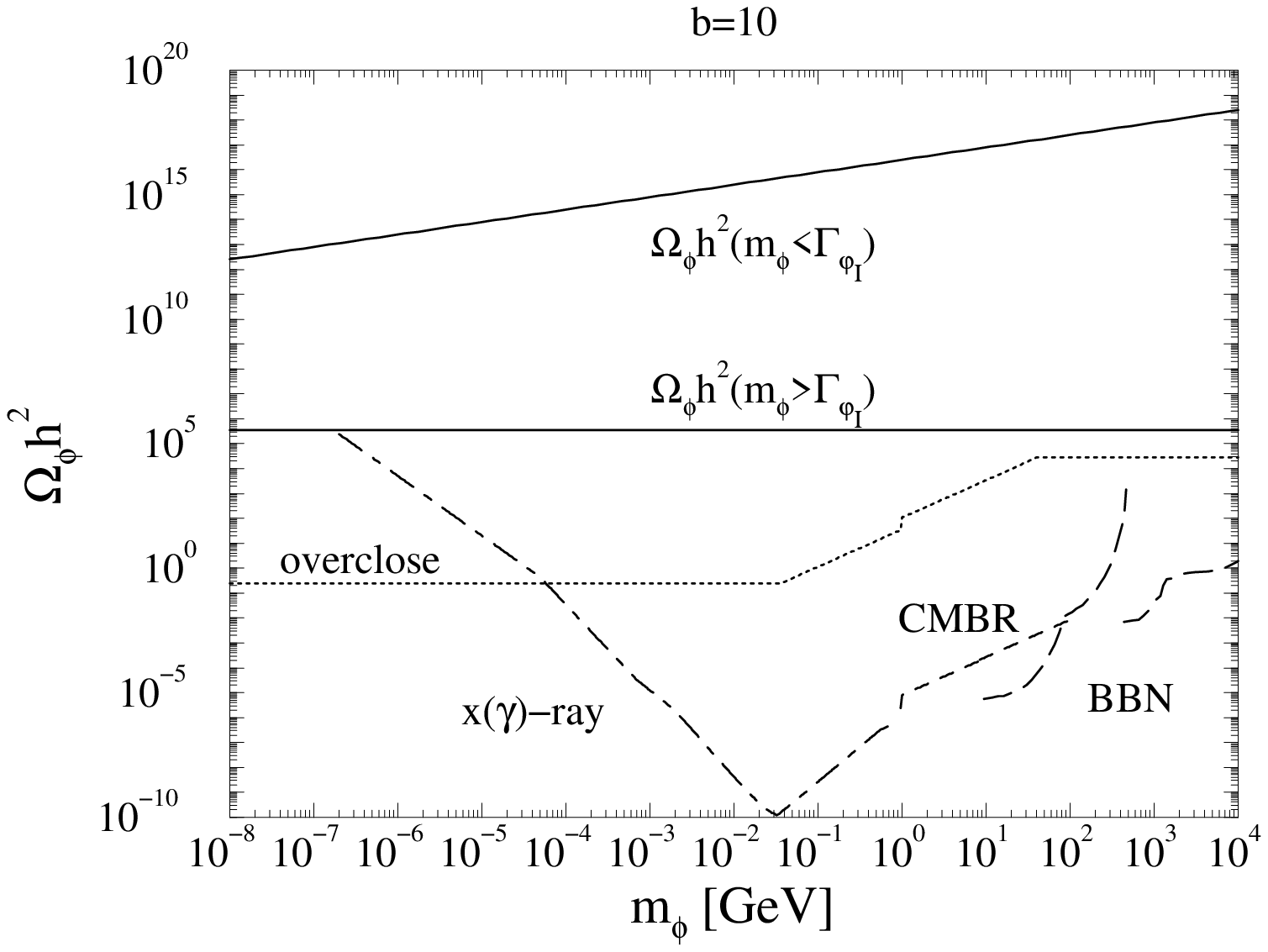,width=12cm}}
    \caption{
    Same figure as Fig.\ref{fig:ompb1} except for the value of the
    modulus coupling $b=10$.}
    \label{fig:ompb10}
\end{figure}

We show these upper bounds on the modulus abundance
$\Omega_\phi$%
\footnote{
For the stable modulus of mass $m_\phi \lesssim 100$ MeV,
$\Omega_\phi$ is given by $\Omega_\phi 
= (\rho_\phi)_0 / \rho_{cr}$ with the present energy density 
of the modulus $(\rho_\phi)_0$.
On the other hand, 
for the modulus of mass $m_\phi \gtrsim 100$ MeV,
$\Omega_\phi$ is regarded as the ratio,
$(\rho_\phi / s )_D /(\rho_{cr}/s_0)$,
where $(\rho_\phi / s )_D$ denotes the ratio of the energy density
of the modulus to the entropy density when the modulus decays.
}
in Figs. \ref{fig:ompb1}, \ref{fig:ompb01} and \ref{fig:ompb10}
taking the modulus coupling as $b =$ 1, 0.1 and 10, respectively.  
The predicted
modulus abundances Eqs.~(\ref{Rp1}) and (\ref{Rp2}) with $T_{RI}=10$
MeV are also found in those figures.  One can easily see that the
string modulus with a mass from $10$ eV to 10 TeV is excluded by the
various cosmological observations.  Since $m_\phi \simeq m_{3/2}$, the
whole gravitino mass region typically predicted by both GMSB and
HSSB models is not cosmologically allowed.  This difficulty is often
referred as ``cosmological moduli problem''.  Here we would like to
stress that this problem could not be solved by choosing the model of
the primordial inflation, i.e., even if one assumes the extremely low
reheating temperature $T_{RI} \sim$ 10 MeV.  This is very different
from the gravitino problem.  Therefore, we required some extra
mechanism to dilute the modulus mass density sufficiently other than the
primordial inflation.  In the following, we consider the thermal
inflation model proposed by Lyth and Stewart \cite{Lyth-Stewart} as
such a dilution mechanism.
\clearpage
\section{Original Thermal Inflation Model}
\label{sec-oti}

In this section we review an original thermal inflation model proposed
by Lyth and Stewart \cite{Lyth-Stewart}.  The thermal inflation is
caused by a scalar field (called ``flaton''), and the flaton
supermultiplet $X$ which is a singlet under the standard model
gauge groups has the following superpotential:
\begin{eqnarray}
    \label{sp-X-oti}
    W = 
    \sum_{k=1}^{\infty} 
    \frac{ \lambda_k }{ (n + 3)k } \frac{ X^{(n+3)k} }
    {M_\ast^{(n+3)k - 3} } + C,
\end{eqnarray}
where $\lambda_k$ denotes the coupling constant and we take $\lambda_1
=1$.  Here we impose the discrete $Z_{n+3}$-symmetry to guarantee the
flatness of the potential.  In order to cancel out the cosmological
constant the constant term $C$ should satisfy
\begin{eqnarray}
    \label{const-c}
    |C| \simeq m_{3/2} M_G^2.
\end{eqnarray}
The cutoff scale of the model is denoted by $M_\ast$ in
Eq. (\ref{sp-X-oti}).\footnote{
As the cutoff scale of the theory it is natural to choose the
gravitational scale.  Thus the interaction term of the superpotential
(\ref{sp-X-oti}) should be considered as
\begin{eqnarray}
    W_{int} =
    \frac{ 1}{ n+3 } \frac{ X^{n+3} }{ M_\ast^{n} }
    =
    \lambda \frac{  X^{n+3} }{ M_G^{n} },
\end{eqnarray}
Here $\lambda$ is introduced as a coupling which should be $\lambda
\lesssim {\cal O}(1)$, thus $M_\ast \gtrsim M_G$ is naturally
expected.  Although we will take it as a free parameter in the
following analysis, we will also mention about this lower bound on
$M_{\ast}$.}
Note that we only have to consider the leading term of $k=1$
since the higher power terms are highly suppressed.
 
Then we obtain the following effective potential at low energy ($|X|
\ll M_G, M_{\ast}$) as
\begin{eqnarray}
    \label{veff-oti}
    V_{\mbox{eff}}(X) 
    = V_0 - m_0^2 |X|^2 +
    \frac{n}{n+3} \frac{C}{M_G^2} \frac{ X^{n+3} +
    {X^\ast}^{n+3}}{M_\ast^n}
    +
    \frac{ |X|^{2n+4} }{ M_\ast^{2n} },
\end{eqnarray}
where we use the same letter $X$ for the scalar component of
the superfield.
In this potential 
we have assumed the negative mass squared $-m_0^2$ at the origin
which is induced by the SUSY breaking effects and 
$m_0$ is of the order of the electroweak scale 
$\Lambda_{EW} \sim$ 100 GeV.  Then one finds the vacuum
expectation value (vev) of the flaton ($\langle X \rangle \equiv M$)
as
\begin{eqnarray}
    \label{vev-X-oti}
    \frac{ M^{n+1} }{ M_\ast^n }
    &=&
    \frac{ 1 }{ 2(n+2 ) }
    \left[ ~ 
        n m_{3/2} + \sqrt{ n^2 m_{3/2}^2 + 4 (n+2) m_0^2 }  ~
    \right].
\end{eqnarray}
Therefore, the flaton is found to have a very large vev $M \gg
\Lambda_{EW}$ since $M_\ast \gg \Lambda_{EW}$.  
Here notice that we assumed the soft SUSY breaking mass ($m_0$)
for the flaton, and hence the potential (\ref{veff-oti}) is 
applicable only for the scale below masses of messenger 
fields when we consider the light gravitino (modulus) mass region
in the GMSB models.
Therefore, the vev of the flaton $M$ should be smaller than
their masses. However, in the original thermal inflation model,
this constraint is always satisfied 
when we estimate the minimum of the modulus abundance 
in the next section.

The vacuum energy of
the flaton $V_0$ is written as
\begin{eqnarray}
    V_0 
    &=&
    \frac{ n+1 }{ n+2 } M^2
    \left[ 
        m_0^2 +
        \frac{ n m_{3/2} }{ 2 (n+2) (n+3) }
        \left( 
            n m_{3/2} + 
            \sqrt{ n^2 m_{3/2}^2 + 4 (n+2 ) m_0^2 }
        \right)    
    \right].
\end{eqnarray}
The masses of the scalar particles associated with $X$ can be
estimated by using the variables $\chi$ and $a$ which is defined as
\begin{eqnarray}
    X = \left( M + \frac{ \chi }{ \sqrt{2} } \right)
    \mbox{exp} \left[  \frac{ i a }{ \sqrt{2} M }  \right].
\end{eqnarray}
Here $\chi$ is the flaton particle which causes the thermal inflation.
On the other hand, we call the imaginary part of $X$, $a$, as an
``$R$-axion''.  Since the superpotential (\ref{sp-X-oti}) posses an
approximate $U(1)_R$ symmetry which explicitly breaks down to
$Z_{n+3}$ symmetry by the constant term $C$, $a$ is considered as a NG
boson with an explicit breaking mass proportional to $C$.  The masses
of the flaton and the $R$-axion are estimated as
\begin{eqnarray}
    m_\chi^2 
    &=&
    2(n+1) m_0^2 +
    \frac{ n (n+1) m_{3/2} }{ 2 (n+2 ) }
    \left[ ~ 
        n m_{3/2} + \sqrt{ n^2 m_{3/2}^2 + 4 (n+2) m_0^2 }  ~
    \right], 
    \label{mx}
    \\
    m_a^2 
    &=&
    \frac{ n(n+3) m_{3/2} }{ 2 (n+2) }
    \left[ ~ 
        n m_{3/2} + \sqrt{ n^2 m_{3/2}^2 + 4 (n+2) m_0^2 }  ~
    \right].
    \label{ma}
\end{eqnarray}
Notice that if one takes $m_{0}$ as
\begin{eqnarray}
    \label{m0-Raopen0}
    m_0^2 &>&  \frac{ n^2 (3n+11) (5n+13) }{ 4 (n+1)^2 (n+2) }
               m_{3/2}^2, \nonumber \\
          &=&  \frac{21}{4} m_{3/2}^2 \for n=1,    
\end{eqnarray}
the flaton decay into two $R$-axions is kinematically allowed.
Therefore, in the GMSB models where the
gravitino mass is in the range $m_{3/2} \sim$ 10 eV--1 GeV, the decay
process $\chi \rightarrow 2 a$ seems almost to be open since 
$m_{3/2} \ll m_0 \sim \Lambda_{EW}$.  
On the other hand, in the HSSB models, 
we expect $m_{0} \sim m_{3/2} \sim \Lambda_{EW}$ and the
flaton might not decay into $R$-axions.  
Although the decay channel of the flaton is crucial
to estimate the dilution factor of the string modulus energy, 
we assume that the decay process $\chi \rightarrow 2 a$ 
be always open in the whole gravitino mass region 
and that $m_{0}$ always satisfy Eq.~(\ref{m0-Raopen0}) 
in the original model.
Furthermore, if $m_{0} \gg m_{3/2}$, the vev of the flaton $M$ is
determined independently on $m_{3/2}$ as
\begin{eqnarray}
    \label{M-otiap}
    M \simeq  \left( \frac{ 1 }{ n+2 } \right)^{ \frac{1}{2(n+1)} }
    \left( m_0 M_\ast^n \right)^{ \frac{ 1 }{ n+1 } },
\end{eqnarray}
and we also obtain simple expressions for $V_0$, $m_{\chi}$ and $m_a$
as
\begin{eqnarray}
    \label{V0-simple}
    V_0 &\simeq& \frac{ n+1 }{ n+2 } m_0^2 M^2,
    \\
    \label{mx-simple}
    m_\chi^2 &\simeq& 2 (n+1) m_0^2,
    \\
    \label{ma-simple}
    m_a^2 &\simeq& \frac{ n (n+3) }{ \sqrt{n+2} } m_{3/2} m_0.
\end{eqnarray}

We are now at the point to see 
how the flaton with the potential (\ref{veff-oti}) 
causes a mini-inflation ($=$ thermal inflation) at
the late time of the evolution of the universe.
For the thermal inflation to work, the
flaton does not sit at the true minimum (\ref{vev-X-oti}), but sits
around the origin due to the finite temperature effects in the early
universe.  To realize it, the flaton has to interact rapidly with
fields in the thermal bath of the universe.  The Yukawa interaction
\begin{eqnarray}
    \label{sp-Xxixi}
    W = g_\xi X \xi \overline{\xi},
\end{eqnarray}
with $\xi$ and $\overline{\xi}$ in the thermal bath is sufficient.
Here $g_\xi$ is the coupling.  When the flaton sits at the true
minimum, the fields $\xi$ and $\overline{\xi}$ obtain heavy masses
$m_\xi \simeq g_\xi M$.  However, for the case that the flaton is near
the origin, $\xi$ and $\overline{\xi}$ are almost massless and can be
in the thermal bath if they couple strongly to the fields in the
thermal bath.  This condition is satisfied if, for example, $\xi$ and
$\overline{\xi}$ are 5 and $5^\ast$ in SU(5) gauge group respecting the
unification of the gauge coupling constants.  Then the interaction
(\ref{sp-Xxixi}) gives an additional mass to the flaton in the early
universe as
\begin{eqnarray}
    V_{\mbox{eff}} (X) =
    V_0 +
    \left( c T^2 - m_0^2 \right) |X|^2
    - \frac{ n }{ n+3 } \frac{ m_{3/2} }{M_\ast^n}
    \left( X^{n+3} + {X^\ast}^{n+3} \right)
    + \frac{ |X|^{2n+4} }{M_\ast^{2n} },
\end{eqnarray}
where $c$ is the order one coupling
and we take $c = 1$ in the following analysis.\footnote{
In the case $\xi$ and $\overline{\xi}$ are 5 and $5^\ast$ in SU(5),
the constant $c$ takes a value $c = 5 g_\xi^2 /3$.  }
Thus for the cosmic temperature $ T \gtrsim T_c ( \simeq m_0)$ the flaton
sits at the origin.

Here it should be noted that the Yukawa interaction (\ref{sp-Xxixi})
with $g_{\xi} \sim 1$ is also required from another reason.  In fact,
the interaction (\ref{sp-Xxixi}) with a large $g_\xi$ can induce the
negative mass squared at the origin 
in the potential (\ref{veff-oti}) 
by the renormalization group effects.

During the time that the flaton is trapped at the origin, its vacuum
energy $V_0$ becomes dominant and the universe expands exponentially,
i.e., the thermal inflation takes place \cite{Lyth-Stewart}. There are
three possibilities for the form of the energy which dominates the
universe before the thermal inflation occurs.  If the radiation energy
dominates the universe before thermal inflation, the thermal inflation
starts when $V_0$ becomes comparable to the energy of the radiation at $T =
T_{STI} \sim V_0^{1/4}$.  Then the thermal inflation lasts for
$T_{STI} > T > T_c$.  On the other hand, if the universe
before the thermal inflation is dominated by the energy of the modulus
oscillation, the temperature at the beginning of the thermal inflation
is estimated as $T_{STI} \sim ( V_0^2/(m_\phi M_G) )^{1/6}$ with the
modulus mass $m_\phi$.  The universe before the thermal inflation
might be dominated by the oscillating energy of the inflaton of the
primordial inflation.  In this case, $T_{STI} \sim V_0^{1/8} M_G^{1/4}
\Gamma_{\varphi_I}^{1/4}$, where $\Gamma_{\varphi_I}$ is the decay
rate of the inflaton $\varphi_I$ [see Eq. (\ref{T_STI})
in Appendix \ref{Ap-2}].  Anyway, the flaton causes the
thermal inflation for $T_{STI} > T > T_c$ and it occurs just before
the electroweak phase transition since $T_{c} = m_{0} \sim
\Lambda_{EW}$.

We turn to the thermal history after the thermal inflation ends and
see how the modulus abundance is diluted.  When the cosmic temperature
becomes smaller than $T_{c}$, the flaton rolls down to its true
minimum and oscillates around it.  And the flaton decay occurs when
the Hubble parameter becomes comparable to the total width
$\Gamma_\chi$ of the flaton.  By the flaton decay and the successive
decay of the $R$-axion, the vacuum energy of the thermal inflation is
transferred into the thermal bath and the universe is reheated.  At
this epoch the tremendous entropy is produced and the string moduli is
diluted significantly.  

Since we have assumed in the original thermal inflation model that
$m_0$ satisfy Eq. (\ref{m0-Raopen0}), the flaton always decays into
$R$-axions.  This decay rate is given by
\begin{eqnarray}
    \Gamma (\chi \rightarrow 2 a)
    = \frac{ 1 }{ 64 \pi}
    \frac{ m_\chi^3 }{ M^2},
\end{eqnarray}
where we have neglected the mass of the $R$-axion.  In addition, the
flaton decay into two photons occurs through the Yukawa interaction
(\ref{sp-Xxixi}) via one-loop diagrams of $\xi$ and $\overline{\xi}$.
For example, $\xi$ and $\overline{\xi}$ are 5 and $5^\ast$ in SU(5),
its rate is estimated as
\begin{eqnarray}
    \Gamma (\chi \rightarrow 2 \gamma) 
    = \frac{ 2 }{ 9 \pi}
    \left( \frac{ \alpha_{em} }{ 4 \pi } \right)^2
    \frac{ m_\chi^3 }{ M^2 }.
\end{eqnarray}
Here we have neglected the effects of the SUSY breaking.  In a similar
way, the flaton can decay into two gluons if kinematically allowed,
i.e., $m_{\chi} \gtrsim$ 1 GeV, and the decay rate is given by
\begin{eqnarray}
    \label{Gam-2g}
    \Gamma (\chi \rightarrow 2 g)
    = \frac{ 1 }{ 4 \pi }
    \left( \frac{ \alpha_{s} }{ 4 \pi } \right)^2
    \frac{ m_\chi^3 }{ M^2 }.
\end{eqnarray}
Furthermore, the flaton might also decay into the gaugino pair.
However, we forbid them since such a decay overproduces the LSPs and
hence is cosmologically dangerous.

On the other hand, the flaton has a chance to couple directly to the
SUSY standard model particles.  Within the renormalizable interactions
the suitable charges of the Higgs supermultiplets $H$ and
$\overline{H}$ under the $Z_{n+3}$-symmetry allow the superpotential
\begin{eqnarray}
    \label{sp-Xhh}
    W = \lambda_\mu X H \overline{H},
\end{eqnarray}
where $\lambda_\mu$ is a dimensionless coupling.  In this case the
ordinary ``$\mu$ term'' is forbidden but is induced by the flaton vev
as $\mu = \lambda_\mu M$, and hence the coupling constant
$\lambda_\mu$ should be extremely small so that $\mu$ becomes the
electroweak scale.\footnote{
The electroweak scale of $\mu$ can be naturally understood by the
nonrenormalizable interaction~{\protect
\cite{Stewart-Kawasaki-Yanagida}} as
\begin{eqnarray}
     W \sim \frac{ X^2 H \overline{H}}{ M_{G} },
\end{eqnarray}
other than Eq. ({\protect \ref{sp-Xhh}}).  However, the following
discussion is almost same in both cases.  }
Through the interaction Eq. (\ref{sp-Xhh})
the flaton can decay into Higgs bosons or Higgsinos
if allowed. Their decay rate can be written as
\begin{eqnarray}
    \label{gam-xhh}
    \Gamma (\chi \rightarrow 2 h)
    = C_h \frac{1}{16 \pi} \frac{ m_\chi^3 }{M^2},
\end{eqnarray}
where we have ignored the masses of final states.  Here $C_h$ is a
constant parameter and $C_h = ( \lambda_\mu M/ m_\chi )^4 = (\mu /
m_\chi)^4$ for the decay into the Higgs bosons  
and $C_h = ( \lambda_\mu /m_\chi )^2$ 
for the decay into Higgsino pair.  In order that $\mu$
does not exceeds the electroweak scale $\mu \lesssim \Lambda_{EW} \sim
m_\chi$, it should be $C_h \lesssim 1$.  Since Higgsinos produced by
the flaton decay might also lead to the overclosure of the LSP, we
only consider the flaton decay into Higgs bosons.
Note that when we take $\lambda_\mu \neq 0$, the charged Higgs and
Higgsino also contribute to the flaton decay into two photons and its
rate becomes
\begin{eqnarray}
    \Gamma (\chi \rightarrow 2 \gamma )
    &=& \frac{ 49 }{72 \pi }
    \left( \frac{ \alpha_{em} }{ 4 \pi } \right)^{2}
    \frac{ m_\chi^3 }{ M^2 }.
\end{eqnarray}

The $R$-axion produced by the flaton decay has the decay processes
similar to the flaton.  The $R$-axion can decay into two photons with
the rate
\begin{eqnarray}
    \label{gam-agamgam}
    \Gamma (a \rightarrow 2 \gamma) 
    = \frac{ 2 }{ 9 \pi}
    \left( \frac{ \alpha_{em} }{ 4 \pi } \right)^2
    \frac{ m_a^3 }{ M^2 },
\end{eqnarray}
for $\lambda_\mu = 0$ and 
\begin{eqnarray}
    \label{gam-agamgam1}
    \Gamma (a \rightarrow 2 \gamma) 
    = \frac{ 49 }{ 72 \pi}
    \left( \frac{ \alpha_{em} }{ 4 \pi } \right)^2
    \frac{ m_a^3 }{ M^2 },
\end{eqnarray}
for $\lambda_\mu \neq 0$.  In addition, it also decays into two gluons
for $m_a \gtrsim 1$ GeV with the rate
\begin{eqnarray}
    \label{gam-agg}
    \Gamma (a \rightarrow 2 g)
    = \frac{ 1 }{ 4 \pi }
    \left( \frac{ \alpha_{s} }{ 4 \pi } \right)^2
    \frac{ m_a^3 }{ M^2 }.
\end{eqnarray}
Note that the $R$-axion can not decay into Higgs boson pair even for
$\lambda_\mu \neq 0$.  Although it might decay into Higgsino pair,
this process is assumed to be forbidden in the same way as the flaton
decay.  Therefore, the R-axion is assumed to have only the radiative
decay modes.

Now we are ready to estimate how an entropy is produced 
by the decays of the flaton and the $R$-axion.  
When the Hubble parameter becomes
comparable to the flaton's total width $\Gamma_{\chi}$, the flaton
decays into both the SM particles and the $R$-axions.  Since the
$R$-axion only has the interaction with the thermal bath suppressed by
$1/M$, the flaton energy transferred into the $R$-axion could not
reheat the universe at this time.  Then the only the energy
transferred into the SM particles reheats the universe at $T = T_{SM}$
by the flaton decay.  The ratio of the entropy densities just before
to after the flaton decay is estimated as
\begin{eqnarray}
    \Delta_{SM} =
    1 + 
    \left( 1 - \epsilon_a \right)~
    \frac{ 4 }{ 3 }
    \frac{ V_0 }{ ( 2 \pi^2 / 45 ) g_\ast T_c^3 T_{SM} },
\end{eqnarray}
where $\epsilon_a$ denotes the branching ratio of the 
flaton decay into two $R$-axions.

After that, when the Hubble parameter becomes comparable to the total
width of the $R$-axion, the $R$-axion decays into the SM particles
occurs and reheats the universe at $ T = T_R$.  At this time the
entropy of the universe increases by the factor $\Delta_a$:
\begin{eqnarray}
    \Delta_a =
    1 + \epsilon_a ~\frac{4}{3}~
    \frac{ V_0 }{ ( 2 \pi^2 / 45 ) g_\ast T_c^3 T_R } 
    \left( \frac{ 2 m_a }{ m_\chi } \right)
    \frac{1}{ \Delta_{SM} }.
\end{eqnarray}
Here note that the energy transferred into the $R$-axions is diluted
at the rate $R^{-4}$ for $T > m_{a}$ while diluted at $R^{-3}$ for $T
< m_a$.

Then, all of the vacuum energy of the flaton $V_0$ is released into
the thermal bath and the thermal inflation does increase the entropy
of the universe by the factor $\Delta$:
\begin{eqnarray}
    \label{Delta-OTI}
    \Delta &=& \Delta_{SM} \times \Delta_a,
    \nonumber \\
    &=&
    1 +
    \left( 1 - \epsilon_a \right)~
    \frac{ 4 }{ 3 }
    \frac{ V_0 }{ ( 2 \pi^2 / 45 ) g_\ast T_c^3 T_{SM} }
    ~+~
    \epsilon_a ~\frac{4}{3}~
    \frac{ V_0 }{ ( 2 \pi^2 / 45 ) g_\ast T_c^3 T_R } 
    \left( \frac{ 2 m_a }{ m_\chi } \right).
\end{eqnarray}
Here the branching ratio of $\chi \rightarrow 2 a$ is $\epsilon_a
\simeq 1$ for the case that the flaton can not decay into Higgs
bosons, because the radiative decay channels
of the flaton are only induced by one-loop
diagrams and their rates are significantly suppressed.  On the
other hand, if the flaton decay into Higgs bosons is allowed, 
$\epsilon_a \simeq 1/( 1 + C_h/4)$.  
Even in this case, however, $\Delta$ takes its
maximum value when $\epsilon_a \simeq 1$.  This is because the
$R$-axion decay rate is much smaller than the flaton decay rate and
hence $m_\chi T_R/( 2 m_a T_{SM}) \ll 1$.  Therefore in order to
obtain the maximum entropy production, the flaton decay into
Higgs bosons should be suppressed.  In the following, in order to make
a conservative analysis we take $\epsilon_a = 1$ and use
\begin{eqnarray}
    \label{Delta-Raopen}
    \Delta ~\simeq~ \frac{4}{3} 
    \frac{ V_0 }{ ( 2 \pi^2 / 45 ) g_\ast T_c^3 T_R } 
    \left( \frac{ 2 m_a }{ m_\chi } \right).
\end{eqnarray}
For this (maximum) entropy production the reheating temperature $T_R$
is determined by the total decay width of the $R$-axion which can be
written as
\begin{eqnarray}
         \Gamma_{a} ~= ~ C_{a} \frac{ m_{a}^3 }{ M^2 },
\end{eqnarray}
where the parameter $C_{a}$ depends on the decay mode and is given by
[see Eqs. (\ref{gam-agamgam}), (\ref{gam-agamgam1}), and (\ref{gam-agg})]
\begin{eqnarray}
    C_a =
    \left\{
    \begin{array}{l l}
        \frac{2}{9 \pi}
        \left( \frac{ \alpha_{em} }{ 4 \pi } \right)^2 
        & \mbox{for}~ \lambda_\mu = 0\\[1em]
        \frac{49}{72 \pi}
        \left( \frac{ \alpha_{em} }{ 4 \pi } \right)^2 
        & \mbox{for}~ \lambda_\mu \neq 0
    \end{array}
    \right. ,
\end{eqnarray}
for the case that the $R$-axion decays only into photons
($m_{a} \lesssim$ 1 GeV), 
and since a heavier $R$-axion dominately decays into two gluons
\begin{eqnarray}
    C_a \simeq \frac{1}{4 \pi}
    \left( \frac{ \alpha_{s} }{ 4 \pi } \right)^2,
\end{eqnarray}
for $m_a \gtrsim 1$ GeV.
Then, the reheating temperature $T_R$  is obtained as
\begin{eqnarray}
    T_R ~\simeq~
    0.96 ~ \sqrt{ \Gamma_a M_G } 
    ~\simeq~
    0.96 ~
    C_a^{1/2} \frac{ m_a^{3/2} M_G^{1/2} }{ M }.
\end{eqnarray}
From Eqs. (\ref{V0-simple}), (\ref{mx-simple}), and (\ref{ma-simple})
we can write the vev and the vacuum energy of the flaton as
\begin{eqnarray}
    \label{M-TRa}
    M &\simeq& 0.96 ~C_a^{1/2} ~
    \frac{ n^{3/4} (n+3)^{3/4} }{ (n+2)^{3/8} }
    \frac{ m_0^{3/4} m_{3/2}^{3/4} M_G^{1/2} }{ T_R},
\end{eqnarray}
\begin{eqnarray}
    \label{V0-m0TR}
    V_0 &\simeq& 0.92 ~C_a~ 
    \frac{ n^{3/2} (n+1) (n+3)^{3/2} }{ (n+2)^{7/4} } ~
    \frac{ m_0^{7/2} m_{3/2}^{3/2} M_G }{ T_R^2 }.
\end{eqnarray}
The entropy production factor (\ref{Delta-Raopen}), therefore,
is given by
\begin{eqnarray}
    \Delta 
    ~\simeq~
    2.0 \times 10^{-2} ~
    \frac{ n^2 (n+1)^{1/2} (n+3)^2 }{ (n+2)^2 }
    C_a \TC^{-3}
    \frac{ m_{3/2}^2 M_G }{ T_R^3 }.
\end{eqnarray}
Note that $\Delta$ is independent on $m_0$ since $m_0 \simeq T_c$.  It
can be seen that the lowest reheating temperature $T_R \sim 10$ MeV
gives the maximum entropy production as
\begin{eqnarray}
    \Delta
    ~\simeq~
    8.4 \times 10^{17}
    \TC^{-3}
    \left( \frac{ m_{3/2} }{1\GeV} \right)^2
    \left( \frac{ T_R }{ 10\MeV} \right)^{-3},
\end{eqnarray}
for the case $n=1$ and 
the $R$-axion decays dominately decays into gluons ($m_a \gtrsim$ 1 GeV).
Therefore, the thermal inflation is found to produce a
tremendous entropy at late time of the universe and can dilute all of
the unwanted particles which are long-lived like the string moduli.

To end this section, we briefly discuss about the initial condition of
the thermal inflation.  In order to realize the thermal inflation, the
flaton should be trapped at the origin of the potential by the
thermal effects.  Thus we have required the Yukawa interaction
(\ref{sp-Xxixi}).  If the flaton sits around the origin just after the
primordial inflation, $\xi$ and $\overline{\xi}$ become massless and
give the flaton a mass comparable to $T$.  Therefore the key point of
the initial condition of the thermal inflation is where the flaton
sits just after the primordial inflation.

One may think that the flaton sits at the true minimum ($\langle X
\rangle = M$) just after the primordial inflation.  In this case,
since $\xi$ and $\overline{\xi}$ obtain very heavy masses, the maximum
temperature $T_{MAX}$ achieved after the inflation should be higher
than their mass, i.e., $T_{MAX} \gtrsim m_{\xi, \overline{\xi}} \simeq
M$ in order to thermalize them.  The maximum temperature is estimated
as [see Appendix \ref{Ap-1}]
\begin{eqnarray}
    T_{MAX} \simeq 0.702 ~ g_{\ast}(T_{MAX})^{-1/4} ~
    g_{\ast}(T_{RI})^{1/8} ~ T_{RI}^{1/2} V_I^{1/8},
\end{eqnarray}
where the reheating temperature of the primordial inflation 
is $T_{RI}$ and its vacuum energy is $V_I$.
Requiring $T_{MAX} \gtrsim m_{\xi, \overline{\xi}}$, we obtain 
\begin{eqnarray}
    V_I^{1/4} &\gtrsim& 2.03 ~ g_\ast (T_{MAX})^{1/2} ~
    g_\ast (T_{RI})^{-1/4} ~  M^2 T_{RI}^{-1},
    \nonumber \\
    &\simeq&
    7.64 \times 10^{17}~\mbox{GeV}~
    \left( \frac{ M }{ 10^{10}~\mbox{GeV} } \right)^2
    \left( \frac{ T_{RI} }{1~\mbox{TeV} } \right)^{-1}.
\end{eqnarray}
Thus the inflation model with very low $T_{RI}$ should have
very large vacuum energy.

However, one can avoid this difficulty by considering
the effects of the supergravity.  
The flaton can sit at the origin due to the
additional SUSY breaking effect from the large vacuum energy of the
inflaton.  In this case the flaton starts to roll down toward 
its true minimum when $H \simeq m_0 \sim \Lambda_{EW}$.  
Then the flaton can be trapped at the origin, if the
cosmic temperature at $H \simeq m_0$ is higher than the negative
curvature at the origin, i.e., $T( H \simeq m_0 ) \gtrsim m_0$.  This
gives the lower bound on the reheating temperature $T_{RI}$ as
\begin{eqnarray}
    T_{RI} 
    &\gtrsim&
    7.10 ~ \frac{ m_0^{3/2} }{ M_G^{1/2} },
    \nonumber \\
    &\simeq&
    4.59 \times 10^{-6}~\mbox{GeV}~ 
    \left( \frac{ m_0 }{100 ~\mbox{GeV} } \right)^{3/2}.
\end{eqnarray}
This condition is always hold since $T_{RI}$ should be
higher than about 10 MeV from the BBN observations.
Therefore, the initial condition of the thermal inflation 
can be explained naturally by the
supergravity effects.

\section{Moduli Problem with Original Thermal Inflation Model}
\label{sec-mpoti}
In this section, we estimate the modulus abundance 
in the presence of the original thermal inflation 
explained in the previous section and examine whether it could solve
the cosmological moduli problem or not.

If the Hubble parameter during the thermal inflation, $H_{TI}$%
\footnote{
The Hubble parameter during the thermal inflation $H_{TI}$
is estimated as $H_{TI} = V_0^{1/2}/(\sqrt{3}M_G)$ with the vacuum energy
of the flaton $V_0$.},
is larger than the modulus mass $m_\phi$, the
modulus oscillation starts after the end of the thermal inflation.
The ratio of the energy density of this oscillation to the entropy
density $s$ after the reheating process of the thermal inflation is
estimated as
\begin{eqnarray}
    \label{Rp-ATI}
    \frac{ \rho_\phi }{ s } \simeq \frac{1}{8} T_R
    \left( \frac{ m_\chi }{ 2 m_a } \right)
    \left( \frac{ \phi_0 }{ M_G } \right)^{2} 
    \gtrsim 1.25 \times 10^{-3} ~\mbox{GeV}
    \left( \frac{ \phi_0 }{ M_G } \right)^{2}.
\end{eqnarray}
where $T_R$ is the final reheating temperature of the thermal
inflation. Here we have used the fact that $T_R \gtrsim$ 10 MeV from
the BBN observations and $m_\chi > 2 m_a$ since the
flaton always decays into to the R-axions [see
Eq.~(\ref{m0-Raopen0})].  We have also assumed that the modulus
oscillation starts at least before the reheating of the thermal
inflation, i.e., we have assumed the modulus mass as
\begin{eqnarray}
    m_\phi > 1.1 \frac{ T_R^2 }{ M_G }
    \simeq 4.5 \times 10^{-23} ~\mbox{GeV}~
    \left( \frac{ T_R}{10 ~\mbox{MeV} } \right)^2.
\end{eqnarray}
The abundance (\ref{Rp-ATI}) becomes larger than the result
Eq.(\ref{Rp2}) for $T_{RI} = T_R$.  Therefore, 
the cosmological moduli problem could not solved in this case 
and we should consider the case $H_{TI} < m_\phi$.

In the following we will discuss the present modulus abundance 
in the presence of the original thermal inflation model 
with $H_{TI} < m_\phi$.  
Most of the previous works assumed that the oscillation of the
modulus start after the reheating process of the primordial inflation
completes.  Thus the thermal inflation should occur after the end of
the reheating of the primordial inflation.  However, various models of
the primordial inflation do not meet this assumption and give
completely different consequences.  In the present article, we will
consider more general cases.  The key time scales of the discussion
are the following three scales: (I)$H \simeq m_\phi$ : the cosmic time
when the modulus starts to oscillate.  (II)$H \simeq
\Gamma_{\varphi_I}$: the cosmic time when the reheating process of the
primordial inflation ends. Here $\Gamma_{\phi_I}$ denotes the decay
rate of the inflaton $\varphi_I$ of the primordial inflation.  (III)$H
\simeq H_{TI}$: the cosmic time at which the thermal inflation occurs.
Since we are considering the case $m_{\phi} > H_{TI}$ as mentioned
above, the thermal history of the universe is classified
into the following three cases in general.
\begin{itemize}
\item Case I: $\Gamma_{\varphi_I} \ge m_\phi \ge H_{TI}$.
\item Case II: $m_\phi \ge \Gamma_{\varphi_I} \ge H_{TI}$.
\item Case III: $m_\phi \ge H_{TI} \ge \Gamma_{\varphi_I}$.
\end{itemize}
For these three cases we will estimate the modulus abundance
with the original thermal inflation model.\footnote{
The modulus oscillation is considered to start after the primordial
inflation, i.e., $m_\phi < H_{I} = V_I^{1/2}/(\sqrt{3} M_G)$.}

\subsection{Case I: For the case $\Gamma_{\varphi_I} \ge m_\phi \ge H_{TI}$}
\label{subsec-oti1}

First of all, we consider the case $\Gamma_{\varphi_I} \ge m_\phi \ge
H_{TI}$ where a modulus field $\phi$\footnote{
Here we also assume one modulus field with $m_\phi \simeq m_{3/2}$
to make a conservative analysis.
}
starts to oscillate after the reheating process of the primordial
inflation completes.  In this case, the reheating temperature of the
primordial inflation, $T_{RI}$, which can be written as Eq.
(\ref{TRI}), should be
\begin{eqnarray}
    \label{condition-case1}
    T_{RI} \ge T_{\phi} \simeq 
    7.2 \times 10^{8} ~\mbox{GeV}~
    \left( \frac{m_\phi}{1~\mbox{GeV}} \right)^{1/2},
\end{eqnarray}
where $T_\phi$ denotes the cosmic temperature when the modulus starts
to oscillate [see Eq. (\ref{T-phi})].  Therefore, a relatively higher
reheating temperature is required and in some models of the inflation,
such as chaotic or hybrid inflation, we can easily obtain such high
$T_{RI}$ \cite{Linde}.

At $T=T_\phi$ the ratio between the energy density of the modulus
oscillation to the entropy density is given by Eq. (\ref{Rp1}). After
that, the energy density of the modulus is carried by the coherent
oscillation.  Hereafter, we call this modulus as ``big-bang modulus''.
In the presence of the original thermal inflation
the present abundance\footnote{
For the unstable modulus of mass $m_\phi \gtrsim 100$ MeV,
it corresponds to the abundance just before the modulus decay.
}
of this big-bang modulus is diluted by the entropy production factor
$\Delta$ [see Eq. (\ref{Delta-Raopen})] and becomes
\begin{eqnarray}
    \label{RBB-case1-oti}
    \left( \frac{ \rho_\phi }{ s } \right)_{BB}
    &=&
    \frac{ \frac{1}{2} m_\phi^2 \phi_0^2 }
         { \frac{ 2 \pi^2 }{ 45 } g_\ast T_\phi^3 }
    \times \frac{ 1 }{ \Delta }.
\end{eqnarray}

Moreover, it should be noted that the modulus energy is also
produced after the thermal inflation.  During the thermal
inflation the modulus dose not sit at the true minimum but is
displaced from it by an amount $\delta \phi_0 \sim (V_0/m_\phi^2
M_G^2) \phi_0$, and this causes the secondary oscillation of the
modulus \cite{Lyth-Stewart}.  We call this modulus as the 
``thermal-inflation'' modulus.  The abundance of the thermal-inflation modulus
is given by
\begin{eqnarray}
    \label{RTI-case1-oti}
    \left( \frac{ \rho_\phi }{ s } \right)_{TI}
    &=& \frac{ \frac{1}{2} m_\phi^2 \delta \phi_0^2 }
             { \frac{ 2 \pi^2}{ 45 } g_\ast T_c^3 }
        \times \frac{ 1 }{ \Delta }
    =
    \frac{ \frac{1}{2} V_0^2  }
         { \frac{ 2 \pi^2}{ 45 } g_\ast T_c^3  m_\phi^2 M_G^2 }
    \left( \frac{ \phi_0 }{ M_G } \right)^2
    \times \frac{ 1 }{ \Delta }.
\end{eqnarray}
Here notice that the thermal-inflation modulus starts to
oscillate before the entropy production of the thermal inflation takes
place and can be diluted.  From Eq.(\ref{Delta-Raopen}) the both
modulus abundances are expressed as
\begin{eqnarray}
    \RBB &\simeq& 3.8 
    \frac{ m_\phi^{1/2} m_0^{3} M_G^{1/2} T_R}{ V_0 }
    \TC^3
    \left( \frac{ m_\chi }{ 2 m_a } \right)
    \phiz^2,\\
    \RTI &\simeq& 0.38
    \frac{ V_0 T_R }{ m_\phi^2 M_G^2 }
    \left( \frac{ m_\chi }{ 2 m_a } \right)
    \phiz^2.
\end{eqnarray}
Therefore, the total abundance of the modulus is given by
\begin{eqnarray}
    \label{Rtot}
    \left( \frac{ \rho_\phi }{ s } \right)_0
    \simeq
    \mbox{MAX} \left[
        \left( \frac{ \rho_\phi }{ s } \right)_{BB},
        \left( \frac{ \rho_\phi }{ s } \right)_{TI}
        \right].
\end{eqnarray}

We turn to estimate the lower bound on this total abundance
of the modulus and compare with various cosmological constraints.  The
original thermal inflation model is parameterized by two mass scales
$m_{0}$ and $M_{\ast}$ besides the gravitino mass $m_{3/2} (\simeq
m_{\phi}$). 
Here we take $m_{0}$ and $T_{R}$ as two free parameters
since $M_{\ast}$ is the function of $m_0, T_R$ and $m_{3/2}$. 
These two parameters are constrained as Eq. (\ref{m0-Raopen0})
for $m_0$ since the flaton is assumed to decay into $R$-axions
in the original thermal inflation model,
and $T_{R} \gtrsim 10$ MeV from the BBN observations.

The lower bound on the total abundance of the modulus (\ref{Rtot}) 
can be estimated by using the fact
\begin{eqnarray}
    \Rz 
    &\ge& 
    \sqrt{ \RBB \RTI }, \nonumber 
    \\
    &\simeq&
    0.84 ~
    \frac{ (n+1)^{1/2} (n+2)^{1/4} }{ n^{1/2} (n+3)^{1/2} }
    \frac{ m_0^2 T_R }{ m_\phi^{5/4} M_G^{3/4} }
    \mgra^{-1/2} 
    \TC^{3/2}
    \phiz^2,
\end{eqnarray}
where the equality holds when $\RBB = \RTI$, i.e.,
\begin{eqnarray}
    \label{m0eq-oti1}
    m_0 
    &\simeq&
    \frac{ 1.9 }{ C_a^{1/2} } ~
    \frac{ (n+2)^{7/8} }{ n^{3/4} (n+1)^{1/2} (n+3)^{3/4} }
    \frac{ T_R M_G^{1/8} }{ m_\phi^{1/8} }
    \mgra^{-3/4}
    \left( \frac{ T_c }{ m_0 } \right)^{3/4}.
\end{eqnarray}
Therefore, we obtain 
\begin{eqnarray}
    \label{Rmin-oti1}
    \Rz
    &\gtrsim&
    \frac{ 2.9 }{ C_a } ~
    \frac{ (n+2)^2 }{ n^2 (n+1)^{1/2} (n+3)^2 } ~
    \frac{ T_R^3 }{ m_\phi^{3/2} M_G^{1/2} }
    \mgra^{-2} 
    \TC^{3}
    \phiz^2.
\end{eqnarray}
and the lowest reheating temperature $T_R \simeq 10$ MeV gives the
lower bound on the total abundance of the modulus.  
It should be noted that the dependence of the index $n$ 
in the flaton superpotential (\ref{sp-X-oti}) appears only
in the pre-factor.
For $n=1$ we write
down it in terms of the density parameter as
\begin{eqnarray}
    \label{OM-oti1g}
    \Omega_\phi h^2 
    &\gtrsim &
    3.0 \times 10^{-2} ~
    \left( \frac{ m_\phi }{ 1~\mbox{GeV} } \right)^{-3/2}
    \left( \frac{ m_{3/2} }{ m_\phi } \right)^{-2}
    \left( \frac{ T_c }{ m_0 } \right)^3
    \left( \frac{ \phi_0 }{ M_G } \right)^2,
\end{eqnarray}
for the case that a $R$-axion can decay into gluons
($m_\phi \gtrsim 10$ keV), and 
\begin{eqnarray}
    \Omega_\phi h^2 
    &\gtrsim&
    \left\{
        \begin{array}{ll}
            \displaystyle{
            8.7 \times 10^{9} ~
            \left( \frac{ m_\phi }{ 1~\mbox{keV} } \right)^{-3/2}
            \left( \frac{ m_{3/2} }{ m_\phi } \right)^{-2}
            \left( \frac{ T_c }{ m_0 } \right)^3
            \left( \frac{ \phi_0 }{ M_G } \right)^2 }&
            \mbox{for}~\lambda_\mu = 0\\[1em]
            \displaystyle{
            2.8 \times 10^{9} ~
            \left( \frac{ m_\phi }{ 1~\mbox{keV} } \right)^{-3/2}
            \left( \frac{ m_{3/2} }{ m_\phi } \right)^{-2}
            \left( \frac{ T_c }{ m_0 } \right)^3
            \left( \frac{ \phi_0 }{ M_G } \right)^2 }&
            \mbox{for}~\lambda_\mu \neq 0
        \end{array}
    \right..
\end{eqnarray}
for the case that a $R$-axion can only decay into photons
($m_\phi \lesssim 10$ keV).

However, the above estimation should be changed 
when the modulus mass becomes
\begin{eqnarray}
    \label{mp-oti1-cut}
        m_{\phi} &\gtrsim&
        1.2 \TeV~
        \frac{ (n+1)^{4/9} (n+2)^{11/9} }
             { n^{14/9} (n+3)^{2/3} (3n+11)^{4/9} (5n+13)^{4/9} }
        \mgra^{-14/9}
        \TC^{2/3},
        \nonumber \\
        &\simeq&
        200 \GeV~
        \mgra^{-14/9}
        \TC^{2/3} \fn,
\end{eqnarray}
because $m_0$ in Eq.~(\ref{m0eq-oti1}) lies outside of the region
(\ref{m0-Raopen0}) if one takes $T_R \simeq 10$ MeV.  In such modulus
mass region the reheating temperature which gives the lower bound on
$\Rz$ should be higher than $10$ MeV and is given by
\begin{eqnarray}
    T_R 
    &\simeq&
    7.1 \times 10^{-4} ~
    \frac{ n^{7/4} (n+3)^{3/4} (3n+11)^{1/2} (5n+13)^{1/2} }
         { (n+1)^{1/2} (n+2)^{11/8} }
    \frac{ m_\phi^{9/8} }{ M_G^{1/8} }
    \mgra^{7/4}
    \TC^{-3/4}, \nonumber \\
    &\simeq&
    60 \MeV
    \left( \frac{ m_\phi }{ 1\TeV} \right)^{9/8}
    \mgra^{7/4}
    \TC^{-3/4} \fn,
\end{eqnarray}
and then we obtain from Eq. (\ref{Rmin-oti1})
\begin{eqnarray}
    \label{OM-oti1Mst}
    \Omega_\phi h^2
    &\gtrsim&
    2.0 \times 10^{-4}
    \left( \frac{ m_\phi }{1\TeV} \right)^{15/8}
    \mgra^{13/4}
    \TC^{3/4}
    \phiz^2 \fn.
\end{eqnarray}
Here note that the $R$-axion is heavy enough to decay into gluons in
the modulus mass region (\ref{mp-oti1-cut}).

\begin{figure}[t]
    \centerline{\psfig{figure=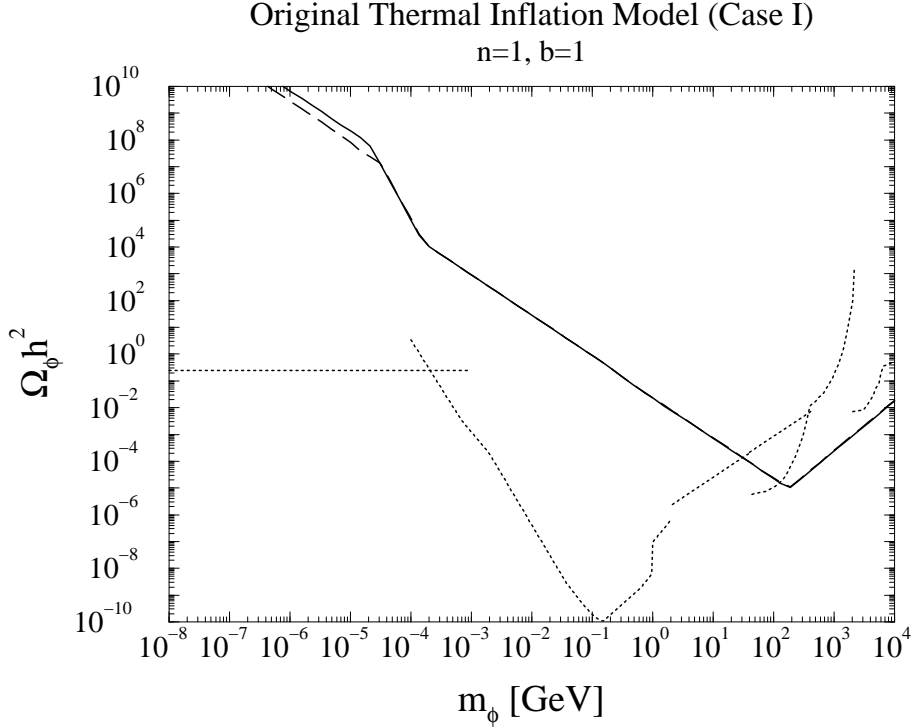,width=12cm}}
\caption{
The lower bounds on the modulus abundance
in the presence of the original thermal inflation
for the case I: $\Gamma_{\varphi_I} \ge m_\phi \ge H_{TI}$.
We assume $m_{\phi}=m_{3/2}$, $T_c=m_0$ and $\phi_{0}=M_{G}$
and take $n=1$ and $b=1$.
The solid (dashed) line denotes the lower bound 
when $\lambda_\mu = 0$ ($\lambda_\mu \neq 0$).
Upper bounds from various cosmological constraints
are all shown by the dotted lines.
}
    \label{fig:oti_case1}
\end{figure}

We show in Fig. \ref{fig:oti_case1} the lower bound on the total
modulus abundance.  It is found that only the modulus with mass
$m_\phi (\simeq m_{3/2} ) \gtrsim 100$ GeV is cosmologically allowed.
Thus the moduli problem in the HSSB models can
be solved by the original thermal inflation model.
However, the light modulus $m_{\phi} \lesssim 1$ GeV
predicted by the GMSB models is still faced with 
serious cosmological difficulties even if one assumes
the original thermal inflation in this case I:
$\Gamma_{\varphi_I} \ge m_\phi \ge H_{TI}$ 
\cite{Asaka-Hashiba-Kawasaki-Yanagida1}.

%
%
\begin{figure}[t]
    \centerline{\psfig{figure=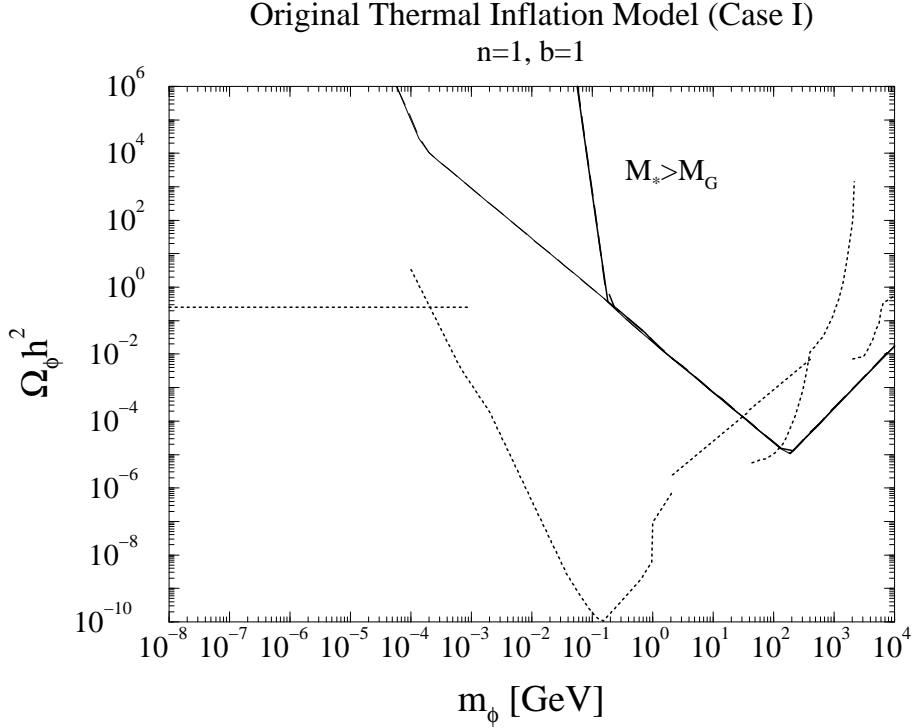,width=12cm}}
\caption{
Same figure as Fig. {\protect \ref{fig:oti_case1}}
except for $M_{\ast} > M_{G}$.
We also show the results in Fig. {\protect \ref{fig:oti_case1}}
by the thin lines.
}
    \label{fig:oti_case1cut}
\end{figure}

We have considered the parameters of the original thermal inflation
model, $m_0$ and $M_\ast$, (i.e., $m_{0}$ and $T_{R}$), as free
parameters.  Here we discuss how the lower bound on the modulus
abundance obtained here is changed, when the cutoff scale of the
original thermal inflation model is bounded from below as $M_\ast >
M_{cr} \sim M_G$ [see the footnote in Sec. \ref{sec-oti}].  This lower
bound on $M_\ast$ leads to the lower bound on $m_0$ as
\begin{eqnarray}
    m_0 > (n+2)^{1/2}
    \left[ \frac{ 1.2 }{ C_a^2 n^3 (n+3)^3 } 
    \right]^{ \frac{n+1}{3n-1} } ~
    T_R^{ \frac{ 4(n+1) }{ 3n-1 } } ~
    m_{3/2}^{ - \frac{ 3(n+1) }{ 3n-1} }
    M_G^{ \frac{ 2(n-1) }{3n-1} }
    \left( \frac{ M_{cr} }{ M_G } \right)^{\frac{ 4n }{ 3n-1 } }.
\end{eqnarray}
Since this lower bound becomes more stringent as $n$ becomes large,
which results in the larger abundance of the modulus, we take $n=1$ in
the following:
\begin{eqnarray}
    \label{const-m0cut}
    m_0 > \frac{ 3.2 \times 10^{-2} }{ C_a^{2} }
    \frac{ T_R^4 }{ m_{3/2}^3 }
    \left( \frac{ M_{cr} }{ M_G } \right)^2.
\end{eqnarray}

Due to this lower bound on $m_{0}$
we find that in the modulus mass region 
\begin{eqnarray}
    \label{const-mpcut}
    m_\phi < 0.18 ~\mbox{GeV} ~
    \left( \frac{ T_R }{ 10 ~\mbox{MeV} } \right)^{24/23}
    \left( \frac{ m_{3/2} }{ m_\phi } \right)^{-18/23}
    \left( \frac{ T_c }{ m_0 } \right)^{-6/23}
    \left( \frac{ M_{cr} }{ M_G } \right)^{16/23}, 
\end{eqnarray}
the abundance of the thermal-inflation modulus 
becomes always larger than that of the big-bang modulus.
Therefore the minimum of the total abundance of the modulus 
is given by using Eq. (\ref{const-m0cut}) as
\begin{eqnarray}
    \Rz = \RTI 
    &\gtrsim&
    \frac{ 4.9 \times 10^{-7} }{ C_a^{7 } } ~
    \frac{ T_R^{15} }{ m_\phi^{13} M_G }
    \left( \frac{ m_{3/2} }{ m_\phi } \right)^{-11}
    \left( \frac{ M_{cr} }{ M_G } \right)^{8}
    \left( \frac{ \phi_0 }{ M_G } \right)^2,
\end{eqnarray}
and the lowest reheating temperature $T_R$ = 10 MeV leads to 
\begin{eqnarray}
    \label{OMmin-oti1-cut}
    \Omega_\phi h^2 
    \gtrsim 680 ~
    \left( \frac{ m_\phi }{ 100~\mbox{MeV} } \right)^{-13}
    \left( \frac{ m_{3/2} }{ m_\phi } \right)^{-11}
    \left( \frac{ M_{cr} }{ M_G } \right)^{8}
    \left( \frac{ \phi_0 }{ M_G } \right)^2,
\end{eqnarray}
for $m_a \gtrsim$ 1 GeV.  On the other hand, for the heavier mass
region, the minimum of the modulus abundance is same as the previous
one Eqs.(\ref{OM-oti1g}) and (\ref{OM-oti1Mst}), even if one takes
$M_\ast \gtrsim M_{cr} \sim M_G$.

We show the result in Fig. \ref{fig:oti_case1cut}.  One finds that the
minimum of the modulus abundance becomes much larger than the previous
results in the mass region given by Eq. (\ref{const-mpcut}).  On the
other hand, the allowed region of the modulus mass $m_\phi \sim$ 100
GeV still survives even if one takes $M_\ast \gtrsim M_{cr} \sim M_G$.
Thus, for the case I: $\Gamma_{\varphi} \ge m_\phi \ge H_{TI}$ 
the original thermal inflation 
can naturally solve the cosmological difficulties of the string
moduli particles if the gravitino mass is $m_{3/2} (\simeq m_\phi)
\gtrsim$ 100 GeV which is predicted in the HSSB models, 
while cannot solve the moduli problem 
in the lighter gravitino mass region of the GMSB models.
%
%
\subsection{Case II: 
For the case $ m_\phi \ge \Gamma_{\varphi_I} \ge H_{TI}$}
\label{subsec-oti2}

Next we consider the case that $ m_\phi \ge \Gamma_{\varphi_I} \ge
H_{TI}$.  In this case, the big-bang modulus starts to oscillate
before the reheating process of the primordial inflation completes.
Thus the energy density of the modulus is diluted by the primordial
inflation as Eq. (\ref{Rp2}).  Then the original thermal inflation
takes place after that, and the mass density of the big-bang modulus
is further reduced by the thermal inflation as
\begin{eqnarray}
    \label{RBB-case2}
    \left( \frac{ \rho_\phi }{ s } \right)_{BB}
    &\simeq&
    \frac{1}{8} T_{RI}
    \left( \frac{ \phi_0 }{ M_G } \right)^{2}
    \times 
    \frac{ 1 }{ \Delta }
    ~\simeq~
    5.8 \times 10^{-2} ~
    \Gamma_{\varphi_I}^{1/2} M_G^{1/2}
    \left( \frac{ \phi_0 }{ M_G } \right)^{2}
    \times 
    \frac{ 1 }{ \Delta },
\end{eqnarray}
with the reheating temperature
\begin{eqnarray}
        T_{\phi} \ge T_{RI} \ge 0.35 V_{0}^{1/4},
\end{eqnarray}
where $T_{\phi}$ is the temperature when the big-bang modulus
starts to oscillate and is given by Eq. (\ref{T-phi}).
This abundance takes its minimum value 
when $\Gamma_{\varphi_I} = H_{TI}$ as
\begin{eqnarray}
    \left( \frac{ \rho_\phi }{ s } \right)_{BB}
    ~\ge~
    \left( \frac{ \rho_\phi }{ s } \right)_{BBm}
    &\equiv&
    4.4 \times 10^{-2}
    V_0^{1/4}
    \left( \frac{ \phi_0 }{ M_G } \right)^{2}
    \times 
    \frac{ 1 }{ \Delta }.
\end{eqnarray}
On the other hand, the abundance of the thermal-inflation modulus
is the same as Eq. (\ref{RTI-case1-oti}) in the previous case I.

Let us estimate the lower bound on the total abundance of the modulus
(\ref{Rtot}) in this case.  One can write $\RBBm$ and $\RTI$ in terms
of $m_0$ and $T_R$ as
\begin{eqnarray}
    \label{RBBm-oti2}
    \left( \frac{ \rho_\phi }{ s } \right)_{BBm}
    &\simeq&
    \frac{ 2.2 }{ C_a^{3/4} } ~
    \frac{ (n+2)^{25/16} }{ n^{13/8} (n+1)^{1/4} (n+3)^{13/8} } ~
    \frac{ m_0^{7/8} T_R^{5/2} }{ m_{3/2}^{13/8} M_G^{3/4} }
    \left( \frac{ T_c }{ m_0 } \right)^{3}
    \left( \frac{ \phi_0 }{ M_G } \right)^2,
    \\
    \label{RTI-oti2}
    \left( \frac{ \rho_\phi }{ s } \right)_{TI}
    &\simeq&
    0.24  ~C_a ~
    \frac{ n (n+1)^{3/2} (n+3) }{ (n+2)^{3/2} } ~
    \frac{ m_0^4 m_{3/2} }{ m_\phi^2 M_G T_R }
    \left( \frac{ \phi_0 }{ M_G } \right)^2.
\end{eqnarray}
Both abundances become larger as $m_0$ becomes larger,
and $\RBBm = \RTI$ is achieved when
\begin{eqnarray}
    m_0 = (m_0)_{eq} 
    &\simeq&
    \frac{ 2.01 }{ C_a^{14/25} }~
    \frac{ (n+2)^{49/50} }{ n^{21/25} (n+1)^{14/25} (n+3)^{21/25} }
    \frac{ m_\phi^{16/25} M_G^{2/25} T_R^{28/25} }{ m_{3/2}^{21/25} }
    \left( \frac{ T_c }{ m_0 } \right)^{24/25}. ~~
\end{eqnarray}
Therefore the minimum of the total modulus abundance is given by
$\RBBm$ with the lowest values of $T_R$ and $m_0 (\le
(m_0)_{eq})$.  The lower bound on $m_0$ Eq. (\ref{m0-Raopen0}) gives
\begin{eqnarray}
    \label{RBBm-oti2min}
    \Rz &=&
    \left( \frac{ \rho_\phi }{ s } \right)_{BBm}
    \nonumber \\
    &\gtrsim&
    \frac{ 1.2 }{ C_a^{3/4} } ~
    \frac{ (n+2)^{9/8} (3n+11)^{7/16} (5n+13)^{7/16} }
         { n^{3/4} (n+1)^{9/8} (n+3)^{13/8} } ~
    \frac{ T_R^{5/2} }{ m_{3/2}^{3/4} M_G^{3/4} }
    \left( \frac{ T_c }{ m_0 } \right)^{3}
    \left( \frac{ \phi_0 }{ M_G } \right)^2.~~~
\end{eqnarray}
For $n=1$ the lowest reheating temperature $T_R = 10$ MeV leads to
\begin{eqnarray}
    \Omega_\phi h^2 
    ~\gtrsim ~
    7.3 \times 10^{-7}~
    \left( \frac{ m_\phi }{ 1~\mbox{GeV} } \right)^{-3/4}
    \left( \frac{ m_{3/2} }{ m_\phi } \right)^{-3/4}
    \left( \frac{ T_c }{ m_0 } \right)^{3}
    \left( \frac{ \phi_0 }{ M_G } \right)^2,
\end{eqnarray}
for the case that the $R$-axion can decay into gluons 
($m_\phi \gtrsim 100$ MeV), and 
\begin{eqnarray}
    \label{oti2-min}
    \Omega_\phi h^2 
    ~\gtrsim ~
    \left\{
        \begin{array}{ll}
            5.2 \times 10^{-2} ~
            {\displaystyle
            \left( \frac{ m_\phi }{ 100 ~\mbox{keV} } \right)^{-3/4}
            \left( \frac{ m_{3/2} }{ m_\phi } \right)^{-3/4}
            \left( \frac{ T_c }{ m_0 } \right)^{3}
            \left( \frac{ \phi_0 }{ M_G } \right)^2 }&
            \mbox{for}~\lambda_\mu = 0 \\
            2.3 \times 10^{-2} ~
            {\displaystyle
            \left( \frac{ m_\phi }{ 100 ~\mbox{keV} } \right)^{-3/4}
            \left( \frac{ m_{3/2} }{ m_\phi } \right)^{-3/4}
            \left( \frac{ T_c }{ m_0 } \right)^{3}
            \left( \frac{ \phi_0 }{ M_G } \right)^2 }&
            \mbox{for}~\lambda_\mu \neq 0 
        \end{array}
        \right.,
\end{eqnarray}
for the case that the $R$-axion cannot decay into gluons
($m_\phi \lesssim 100$ MeV).

However, for the modulus mass region $m_\phi \lesssim 10^{-5}$ GeV,
when the total abundance takes its minimum value Eq. (\ref{oti2-min}),
$m_0$ becomes so small that the vacuum energy of the original thermal
inflation $V_0$ is less than the energy of the radiation at $T = T_R$.
To avoid this failure, $m_0$ should be
\begin{eqnarray}
    \label{const-m02}
    m_0 > \frac{ 1.5 }{ C_a^{2/7} } ~
    \frac{ (n+2)^{1/2} }{ n^{3/7} (n+1)^{2/7} (n+3)^{3/7} } ~
    \frac{ T_R^{12/7} }{ m_\phi^{3/7} M_G^{2/7} }
    \left( \frac{ m_{3/2} }{ m_\phi } \right)^{-3/7}.
\end{eqnarray}
Therefore the minimum of the total abundance is given by
\begin{eqnarray}
    \left( \frac{ \rho_\phi }{ s } \right)_{0}
    &\gtrsim&
    \frac{ 3.0 }{ C_a } ~
    \frac{ (n+2)^2 }{ n^2 (n+1)^{1/2} (n+3)^{2} } ~
    \frac{ T_R^4 }{ m_{3/2}^2 M_G }
    \left( \frac{ T_c }{ m_0 } \right)^{3}
    \left( \frac{ \phi_0 }{ M_G } \right)^2.
\end{eqnarray}
For $n=1$  we obtain 
\begin{eqnarray}
    \Omega_\phi h^2 
    \gtrsim 
    \left\{
        \begin{array}{ll}
            59 ~
            {\displaystyle
            \left( \frac{ m_\phi }{ 1 ~\mbox{keV} } \right)^{-2}
            \left( \frac{ m_{3/2} }{ m_\phi } \right)^{-2}
            \left( \frac{ T_c }{ m_0 } \right)^{3}
            \left( \frac{ \phi_0 }{ M_G } \right)^2 }&
            \mbox{for}~\lambda_\mu = 0 \\
            19 ~
            {\displaystyle
            \left( \frac{ m_\phi }{ 1 ~\mbox{keV} } \right)^{-2}
            \left( \frac{ m_{3/2} }{ m_\phi } \right)^{-2}
            \left( \frac{ T_c }{ m_0 } \right)^{3}
            \left( \frac{ \phi_0 }{ M_G } \right)^2 }&
            \mbox{for}~\lambda_\mu \neq 0 
        \end{array}
        \right. .
\end{eqnarray}
\begin{figure}[t]
    \centerline{\psfig{figure=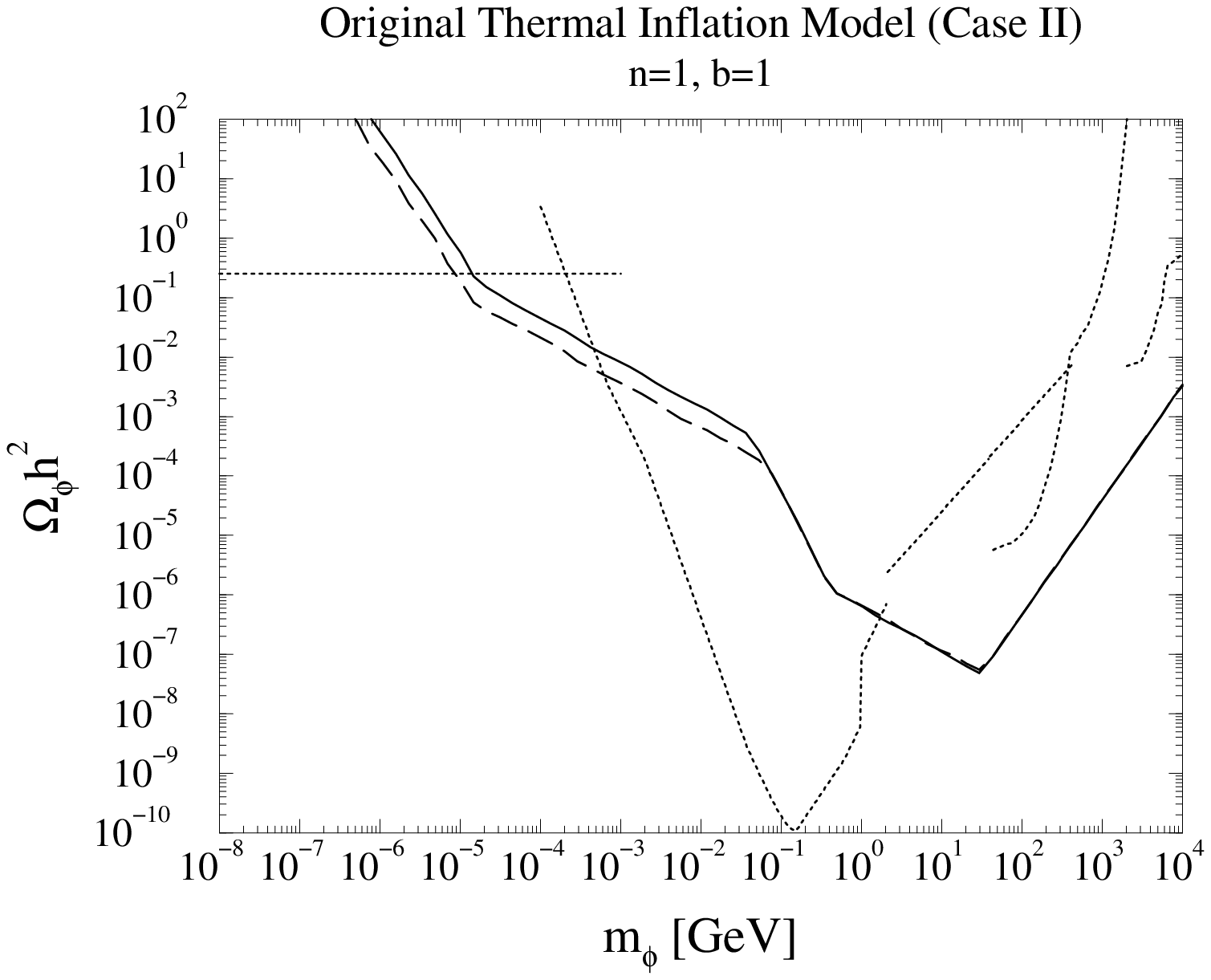,width=12cm}}
\caption{
Same figure as Fig. {\protect \ref{fig:oti_case1}}
for the case II: $m_\phi \ge \Gamma_{\varphi_I} \ge H_{TI}$.
}
    \label{fig:oti_case2}
\end{figure}
\begin{figure}[htb]
    \centerline{\psfig{figure=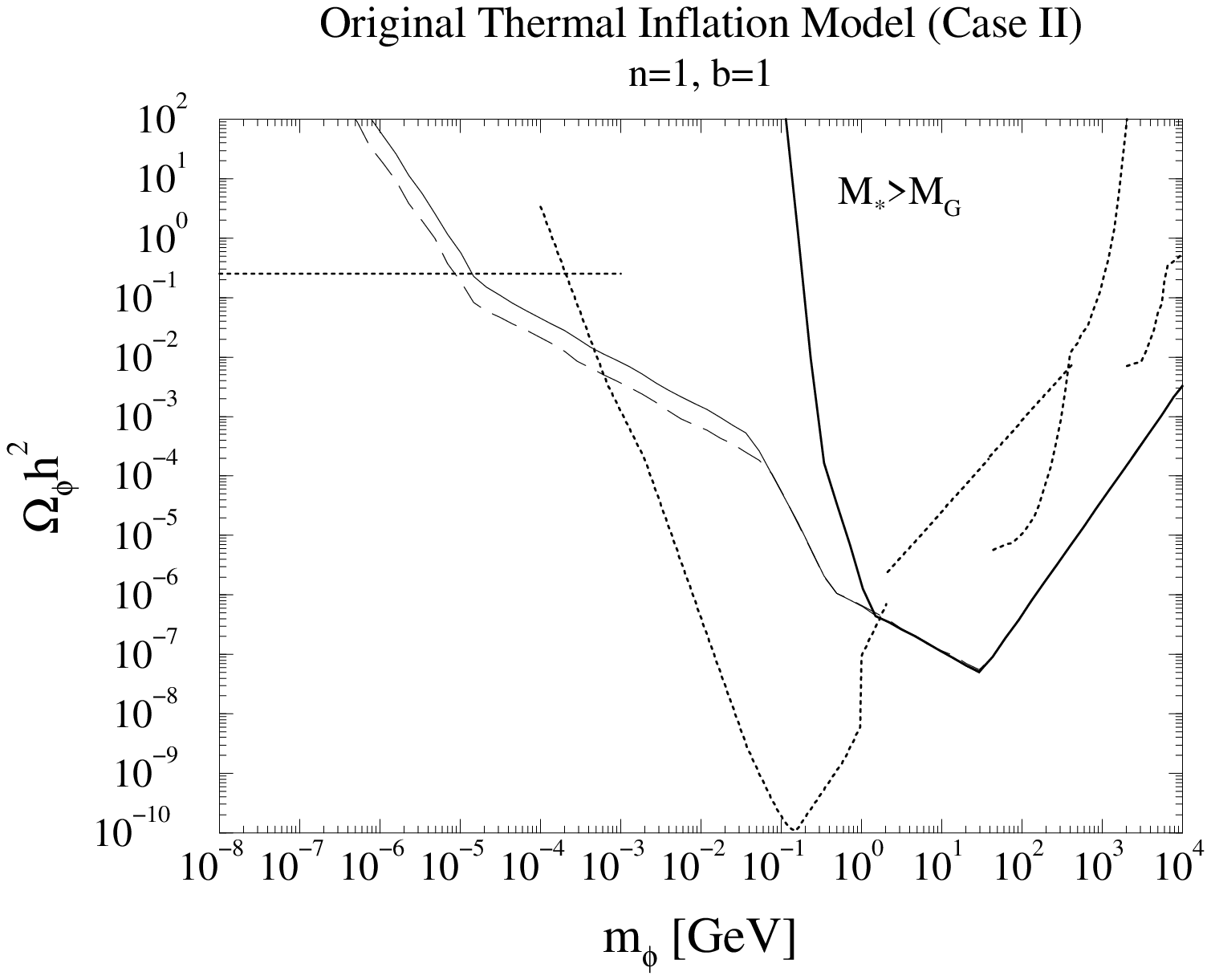,width=12cm}}
\caption{
Same figure as Fig. {\protect \ref{fig:oti_case2}}
except for $M_\ast > M_G$.
We also show the results in Fig. {\protect \ref{fig:oti_case2}}
by the thin lines.
}
    \label{fig:oti_case2cut}
\end{figure}

On the other hand, for the modulus mass region $m_\phi \gtrsim$ 10 GeV, in
order that $m_0$ satisfies both $m_0 \le (m_0)_{eq}$ and 
Eq. (\ref{m0-Raopen0}), 
the reheating temperature should be higher than 10 MeV as
\begin{eqnarray}
    T_R 
    &\gtrsim&
    7.7\times 10^{-4} ~
    \frac{ n^{23/14} (n+3)^{3/4} (3n+11)^{25/56} (5n+13)^{25/56} }
         { (n+1)^{11/28} (n+2)^{37/28} } ~
    \frac{ m_\phi^{15/14} }{ M_G^{1/14} }
    \left( \frac{ m_{3/2} }{ m_\phi } \right)^{23/14}
    \left( \frac{ T_c }{ m_0 } \right)^{-6/7},
    \nonumber \\
    &\simeq&
    31 ~\mbox{MeV}~
    \left( \frac{ m_\phi }{ 100~\mbox{GeV} } \right)^{15/14}
    \left( \frac{ m_{3/2} }{ m_\phi } \right)^{23/14}
    \left( \frac{ T_c }{ m_0 } \right)^{-6/7}
    ~~~~\mbox{for}~n=1.
\end{eqnarray}
Therefore we obtain from Eq. (\ref{RBBm-oti2min})
\begin{eqnarray}
    \label{OM-oti2-hm}
    \Omega_\phi h^2 \gtrsim
    4.0 \times 10^{-7} ~
    \left( \frac{ m_\phi }{ 100~\mbox{GeV} } \right)^{27/14}
    \left( \frac{ m_{3/2} }{ m_\phi } \right)^{47/14}
    \left( \frac{ T_c }{ m_0 } \right)^{6/7}
    \left( \frac{ \phi_0 }{ M_G } \right)^2 \fn.
\end{eqnarray}

We show the obtained lower bound on the total abundance of the modulus
in Fig. \ref{fig:oti_case2}.  It is found that since the primordial
inflation does dilute the energy of the big-bang modulus, the lower
bound becomes weaker than the previous case I and the allowed modulus
mass regions are $m_\phi \gtrsim 1$ GeV and $m_\phi \simeq$ 10 keV--1
MeV.  
Therefore, the gravitino mass region predicted by the HSSB
scenario can be cosmologically viable and it
should be noted that small window for the gravitino mass range 
in the GMSB models does appear.  
This feature is crucially different from the previous results.

However, when we take the cutoff scale of the original thermal
inflation model as $M_\ast > M_{cr} \sim M_G$, this new window is
disappeared as shown in Fig. \ref{fig:oti_case2cut}.  The lower bound
on the total abundance of the modulus becomes more stringent for
$m_\phi \lesssim 1$ GeV where the abundance of the thermal-inflation
modulus is always larger than the big-bang modulus one and then the
limit is the same as Eq. (\ref{OMmin-oti1-cut}) in the previous case.
Therefore, in order that the original thermal inflation 
dilutes sufficiently the light modulus of mass 
$m_\phi \simeq$ 10 keV--1 MeV in GMSB models,
the extremely low cut off scale as $M_{\ast} \sim 10^6$--$10^{10}$ GeV 
is required.
On the other hand, the allowed region for $m_\phi \gtrsim$ 1 GeV 
still exists even for the case $M_\ast \gtrsim M_G$.

\subsection{Case III: 
For the case $ m_\phi \ge H_{TI} \ge \Gamma_{\varphi_I}$}
\label{subsec-oti3}

Finally we consider the case $ m_\phi \ge H_{TI} \ge \Gamma_{\varphi_I}$
where the reheating process of the primordial inflation completes
after the thermal inflation ends and its reheating temperature becomes
extremely low.  In this case the present abundance of the big-bang
modulus is given by [see Appendix \ref{Ap-2}]
\begin{eqnarray}
    \left( \frac{ \rho_\phi }{ s } \right)_{BB}
    \simeq 4.8 
    \frac{ m_0^4 T_R }{ \Gamma_{\varphi_I} V_0^{1/2} M_G }
    \left( \frac{ m_\chi }{ 2 m_a } \right)
    \left( \frac{ T_c }{ m_0 } \right)^4
    \left( \frac{ \phi_0 }{ M_G } \right)^2.
\end{eqnarray}
This ratio takes its minimum value when $\Gamma_{\varphi_I} = H_{TI}$ as
\begin{eqnarray}
    \left( \frac{ \rho_\phi }{ s } \right)_{BB} 
    ~\gtrsim ~
    \left( \frac{ \rho_\phi }{ s } \right)_{BBm} 
    &\equiv&
    8.2 ~
    \frac{ m_0^4 T_R }{ V_0 }
    \left( \frac{ m_\chi }{ 2 m_a } \right)
    \left( \frac{ T_c }{ m_0 } \right)^4
    \left( \frac{ \phi_0 }{ M_G } \right)^2.
\end{eqnarray}
As well as the previous cases we can rewrite in terms of $m_0$ and
$T_R$ as
\begin{eqnarray}
    \label{RBBm-oti3}
    \left( \frac{ \rho_\phi }{ s } \right)_{BBm} 
    &\simeq&
    \frac{ 6.3 }{ C_a } ~
    \frac{ (n+2)^2 }{ n^2 (n+1)^{1/2} (n+3)^2 } ~
    \frac{ m_0 T_R^3 }{ m_{3/2}^2 M_G }
    \left( \frac{ T_c }{ m_0 } \right)^4
    \left( \frac{ \phi_0 }{ M_G } \right)^2.
\end{eqnarray}
On the other hand the abundance of the thermal-inflation modulus 
is the same as the previous two cases [see Eq. (\ref{RTI-oti2})].
Now the condition $\left( \frac{\rho_\phi}{s} \right)_{BBm}$=
$\left( \frac{\rho_\phi}{s} \right)_{TI}$ holds when
\begin{eqnarray}
    m_0 = (m_0)_{eq} 
    &\simeq&
    \frac{ 3.0 }{ C_a^{2/3} } ~
    \frac{ (n+2)^{7/6} }{ n (n+1)^{2/3} (n+3) }~
    \frac{ m_\phi^{2/3} T_R^{4/3} }{ m_{3/2} }
    \left( \frac{ T_c }{ m_0 } \right)^{4/3}.
\end{eqnarray}
Therefore the abundance of the big-bang modulus (\ref{RBBm-oti3}) with
the lowest values of $T_R$ and $m_0 (\le (m_0)_{eq})$ gives the lower
bound on the total modulus abundance (\ref{Rtot}).  From the lower
bound on $m_0$ [Eq. (\ref{m0-Raopen0})] in order to 
allow the flaton decay into $R$-axions
we find
\begin{eqnarray}
    \label{RBBm-oti3min}
    \Rz &=&
    \left( \frac{ \rho_\phi }{ s } \right)_{BBm}
    \nonumber \\
    &\gtrsim&
    \frac{ 3.2 }{ C_a } ~
    \frac{ (n+2)^{3/2} (3n+11)^{1/2} (5n+13)^{1/2} }
         { n (n+1)^{3/2} (n+3)^2 } ~
    \frac{ T_R^3 }{ m_{3/2} M_G }
    \left( \frac{ T_c }{ m_0 } \right)^{4}
    \left( \frac{ \phi_0 }{ M_G } \right)^2.
\end{eqnarray}
For $n=1$ the lowest reheating temperature $T_R $ = 10 MeV gives 
\begin{eqnarray}
    \Omega_\phi h^2 
    \gtrsim 
    0.96 \times 10^{-10}~
    \left( \frac{ m_\phi }{ 1 ~\mbox{GeV} } \right)^{-1}
    \left( \frac{ m_{3/2} }{ m_\phi } \right)^{-1}
    \left( \frac{ T_c }{ m_0 } \right)^{4}
    \left( \frac{ \phi_0 }{ M_G } \right)^2,
\end{eqnarray}
when the $R$-axion can decay into gluons ($m_\phi \gtrsim$ 100 MeV), 
and
\begin{eqnarray}
    \Omega_\phi h^2 
    \gtrsim 
    \left\{
        \begin{array}{ll}
            {\displaystyle 
            2.8 \times 10^{-5}  ~
            \left( \frac{ m_\phi }{ 1 ~\mbox{MeV} } \right)^{-1}
            \left( \frac{ m_{3/2} }{ m_\phi } \right)^{-1}
            \left( \frac{ T_c }{ m_0 } \right)^{4}
            \left( \frac{ \phi_0 }{ M_G } \right)^2 } &
            \mbox{for}~\lambda_\mu = 0 \\
            {\displaystyle 
            0.92 \times 10^{-5}  ~
            \left( \frac{ m_\phi }{ 1 ~\mbox{MeV} } \right)^{-1}
            \left( \frac{ m_{3/2} }{ m_\phi } \right)^{-1}
            \left( \frac{ T_c }{ m_0 } \right)^{4}
            \left( \frac{ \phi_0 }{ M_G } \right)^2 } &
            \mbox{for}~\lambda_\mu \neq 0
        \end{array}
    \right.,    
\end{eqnarray}
when the $R$-axion cannot decay into two gluons ($m_\phi \lesssim$ 100 MeV).

In the same way as the previous case II, however, 
for the modulus mass $m_\phi \lesssim 10$ keV we find 
from the lower bound on $m_0$ [Eq. (\ref{const-m02})]
\begin{eqnarray}
    \left( \frac{ \rho_\phi }{ s } \right)_{0}
    &\gtrsim&
    \frac{ 9.3 }{ C_a^{9/7} } ~
    \frac{ (n+2)^{5/2} }{ n^{17/7} (n+1)^{11/14} (n+3)^{17/7} } ~
    \frac{ T_R^{33/7} }{ m_{3/2}^{17/7} M_G^{9/7} }
    \left( \frac{ T_c }{ m_0 } \right)^{4}
    \left( \frac{ \phi_0 }{ M_G } \right)^2,
\end{eqnarray}
and taking $T_R$ = 10 MeV we obtain for $n=1$
\begin{eqnarray}
    \Omega_\phi h^2
    \gtrsim
    \left\{
        \begin{array}{ll}
            {\displaystyle 
            1.6 ~
            \left( \frac{ m_\phi }{1~\mbox{keV}} \right)^{-17/7}
            \left( \frac{ m_{3/2} }{ m_\phi } \right)^{-17/7}
            \left( \frac{ T_c }{ m_0 } \right)^{4}
            \left( \frac{ \phi_0 }{ M_G } \right)^2 } &
            \mbox{for}~\lambda_\mu = 0 \\
            {\displaystyle 
            0.39 ~
            \left( \frac{ m_\phi }{1~\mbox{keV}} \right)^{-17/7}
            \left( \frac{ m_{3/2} }{ m_\phi } \right)^{-17/7}
            \left( \frac{ T_c }{ m_0 } \right)^{4}
            \left( \frac{ \phi_0 }{ M_G } \right)^2 } &
            \mbox{for}~\lambda_\mu \neq 0 
        \end{array}
   \right..     
\end{eqnarray}

When the total abundance takes its minimum value,
the reheating temperature becomes higher than 10 MeV
for $m_\phi \gtrsim$ 3 GeV as
\begin{eqnarray}
    T_R &\gtrsim& 7.0\times 10^{-4} ~
    \frac{ n^{3/2} (n+3)^{3/4} (3n+11)^{3/8} (5n+13)^{3/8} }
         { (n+1)^{1/4} (n+2)^{5/4} } ~
    \frac{ m_{3/2}^{3/2} }{ m_\phi^{1/2} }
    \left( \frac{ T_c }{ m_0 } \right)^{-1},
    \nonumber \\
    &\simeq&
    3.4 ~\mbox{GeV}~
    \left( \frac{m_\phi}{1 ~\mbox{TeV} }\right)
    \left( \frac{ m_{3/2} }{ m_\phi^{3/2} } \right)^{3/2}
    \left( \frac{ T_c }{ m_0 } \right)^{-1}
    ~~~~\mbox{for}~n=1,
\end{eqnarray}
and from Eq. (\ref{RBBm-oti3min}) we obtain for $n=1$
\begin{eqnarray}
    \Omega_\phi h^2 
    \gtrsim
    3.7 \times 10^{-6}
    \left( \frac{ m_\phi }{ 1 ~ \mbox{TeV} } \right)^{2}
    \left( \frac{ m_{3/2} }{ m_\phi } \right)^{7/2}
    \left( \frac{ T_c }{ m_0 } \right)
    \left( \frac{ \phi_0 }{ M_G } \right)^2,
\end{eqnarray}
where the flaton dominately decays into gluons.

\begin{figure}[t]
    \centerline{\psfig{figure=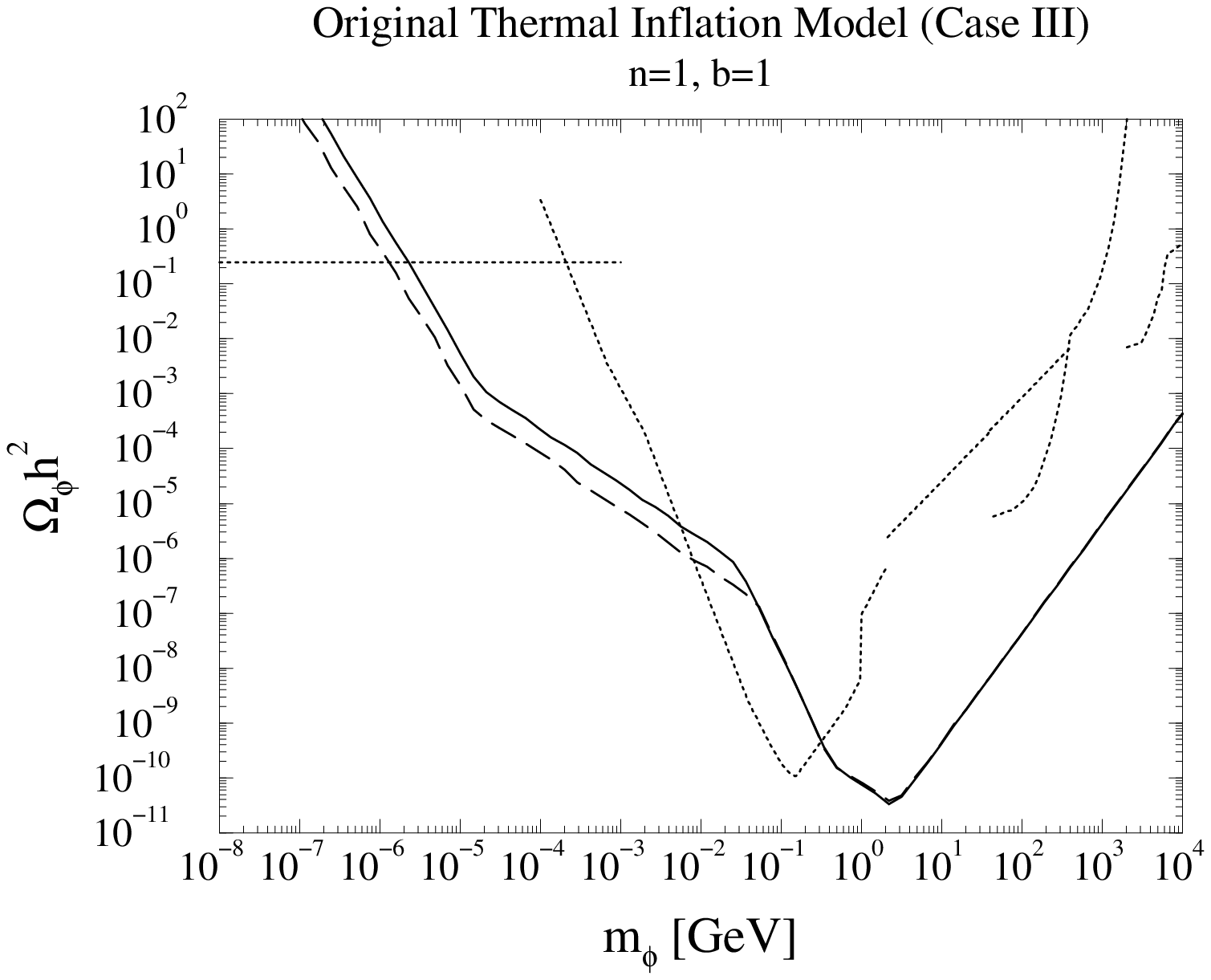,width=12cm}}
\caption{
Same figure as Fig. {\protect \ref{fig:oti_case1}}
for the case III: $m_\phi \ge H_{TI} \ge \Gamma_{\varphi_I}$.
}
    \label{fig:oti_case3}
\end{figure}
\begin{figure}[t]
    \centerline{\psfig{figure=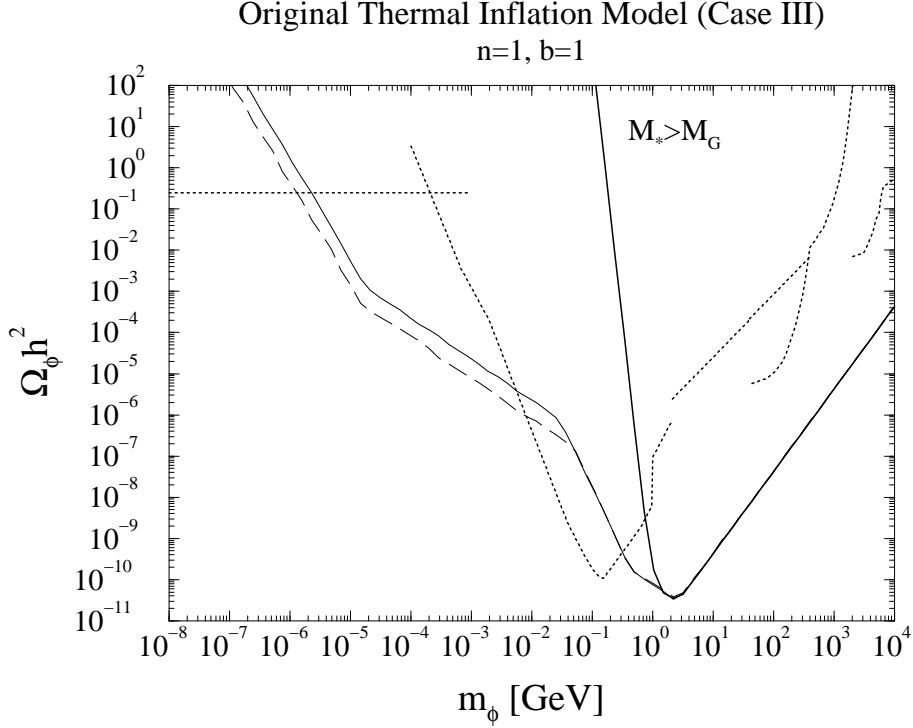,width=12cm}}
\caption{
Same figure as Fig. {\protect \ref{fig:oti_case3}}
except for $M_\ast > M_G$.
We also show the results in Fig. {\protect \ref{fig:oti_case3}}
by the thin lines.}
    \label{fig:oti_case3cut}
\end{figure}

We show the lower bound on the total abundance of the modulus in Fig.
\ref{fig:oti_case3}.  It is found that the lower bound becomes smaller
than those in the previous two cases, because the big-bang modulus is
also diluted by the primordial inflation during the thermal inflation.
In this case the allowed regions for the modulus mass are $m_\phi
\simeq$ 1 keV--10 MeV and $m_\phi \gtrsim$ 300 MeV, and the allowed
mass region in the GMSB models
extends wider than the case II.

In these allowed modulus mass regions,
in order to obtain the lowest modulus abundance
the required reheating temperature 
of the ``primordial inflation'' is 
$T_{RI} \simeq$ 10 MeV for $m_\phi \simeq$ 1 keV--10 keV,
and 
\begin{eqnarray}
    \label{TRI-oti3}
    T_{RI} ~\simeq~
    \left\{
        \begin{array}{ll}
            1.6 \GeV ~
            {\displaystyle
            \left( \frac{ m_\phi }{ 1~\mbox{MeV} } \right)^{5/4}
            \left( \frac{ m_{3/2} }{ m_\phi } \right)^{5/4} }
            &
            \mbox{for}~\lambda_\mu = 0 \\
            2.1 \GeV ~
            {\displaystyle
            \left( \frac{ m_\phi }{ 1 ~\mbox{MeV} } \right)^{5/4}
            \left( \frac{ m_{3/2} }{ m_\phi } \right)^{5/4} }
            &
            \mbox{for}~\lambda_\mu \neq 0 
        \end{array}
        \right. ,
\end{eqnarray}
for $m_\phi \simeq$ 10 keV--10 MeV.
On the other hand, for the heavier modulus of mass 
$m_\phi \gtrsim 3$ GeV, $T_{RI}$ is given by 
\begin{eqnarray}
    T_{RI} ~\simeq~
    5.5 \times 10^{6} \GeV~
    \left( \frac{ m_\phi }{ 1~\mbox{TeV} } \right)^{3/4}
    \mgra^{1/2}
    \TC^{1/2}.
\end{eqnarray}
Therefore, extremely low reheating temperature 
$T_{RI} \sim 10$ MeV--10 GeV is required to dilute sufficiently 
the light modulus of mass $m_\phi \simeq$ 1 keV--10 MeV, 
even if one assume the original thermal inflation model.

Furthermore, if one takes the cutoff scale as $M_\ast \gtrsim M_G$,
the allowed region for the modulus mass
$m_\phi \simeq$ 1 keV--10 MeV vanishes as shown in 
Fig. \ref{fig:oti_case3cut}, and only the modulus of mass
$m_\phi \gtrsim 1$ GeV is allowed.  
As well as the previous cases I and II,
the lower bound on the total modulus abundance when $M_\ast \gtrsim
M_G$ is given by Eq. (\ref{OMmin-oti1-cut}) for 
$m_\phi \lesssim 1$ GeV.
Therefore, in order to dilute sufficiently 
the light modulus,
the thermal inflation model with the extremely low cut-off 
scale $M_\ast \sim 10^{5}$--$10^{12}$ GeV 
is required for $m_\phi \sim$ 1 keV--10 MeV,
as well as the low reheating temperature of the primordial inflation
as $T_{RI} \sim 10$ MeV--10 GeV.
\section{Modified Thermal Inflation Model}
\label{sec-mti}
As showed in the previous section, the original thermal inflation
can dilute the relic abundance of the string modulus
significantly, and the modulus mass regions $m_\phi (\simeq m_{3/2})
\sim$ 1 keV--10 MeV, and $m_\phi \gtrsim 300$ MeV survive the various
cosmological constraints if we consider the model of the primordial
inflation with an extremely low reheating temperature.  However, if we
take the gravitational scale as the natural cutoff scale of the
thermal inflation model, i.e. $M_\ast \gtrsim M_G$, the former allowed
region vanishes and only the modulus mass region $m_\phi \gtrsim$ 1
GeV predicted by the HSSB models is allowed.
To dilute sufficiently the light modulus particle predicted by the
GMSB models, one needs the relatively low
cutoff scale $M_\ast \sim 10^{5}$ GeV--$10^{12}$ GeV for $m_\phi \sim$
1 keV--10 MeV.  This weak point of the original model comes from the
fact that the flaton can always decay into $R$-axions.  The vacuum
energy of the thermal inflation is mostly transferred into
relativistic $R$-axions by the flaton decay and the $R$-axions lose
their energy faster than non-relativistic particles as the universe
expands, which leads to much less entropy production (i.e., less
dilution of the modulus density).  Therefore, one might expect to
dilute the modulus density more effectively by the thermal inflation
model which the decay process $\chi \rightarrow 2 a$ is not
kinematically allowed. 

Furthermore, the original model is faced with another serious
difficulty. Since the potential of the flaton possesses the exact discreet
$Z_{n+3}$ symmetry to ensure the flatness of the potential, the
degenerate minima leads to the domain wall problem.  In order to avoid
this problem one has to introduce a term which breaks the symmetry
explicitly in the potential.

One of economical modifications of the original thermal inflation
model, which solves above two difficulties simultaneously, has been
proposed by Ref. \cite{Asaka-Hashiba-Kawasaki-Yanagida1}.  
A linear term which breaks the $Z_{n+3}$ completely 
is added to the original 
superpotential of the flaton Eq. (\ref{sp-X-oti}) as
\begin{eqnarray}
    \label{sp-X-mti}
    W \simeq \frac{1}{(n+3)} \frac{ X^{n+3} }{ M_\ast^n} 
    + C + \alpha X,
\end{eqnarray}
with the dimensionfull parameter $\alpha$ which 
is required to be
\begin{eqnarray}
    \label{alpha-DW}
    \left| \alpha \right|
    ~\gtrsim~
    \frac{ m_{3/2}^2 m_\chi M }{ M_{pl}^2 }
\end{eqnarray}
in order to eliminate the domain walls \cite{Vilenkin}.  Then this
superpotential gives the low energy potential to the flaton as
\begin{eqnarray}
    \label{veff-mti}
    V_{\mbox{eff}}(X)
    &=& V_0 - \frac{ 2 \alpha C }{ M_G^2 } ( X + X^{\ast} )
    - m_0^2 |X|^2 
    + \frac{ \alpha }{ M_\ast^n } ( X^{n+2} + X^{\ast n+2} ) 
    \nonumber \\
    &&+ \frac{ n }{ n + 3 } ~
    \frac{ C }{ M_G^2 M_\ast^{n} } ( X^{n+3} + X^{\ast n+3} )
    + \frac{ |X|^{2n+4} }{ M_\ast^{2n} }.
\end{eqnarray}
The vev of the flaton is estimated as
\begin{eqnarray}
    \label{vev-mti}
    \langle X \rangle = M 
    \simeq 
    \left[ \frac{ 1 }{ (n+2)(1-x) } \right]^\frac{1}{2(n+1)}
    \left( m_0 M_\ast^n \right)^\frac{1}{n+1},
\end{eqnarray}
and the vacuum energy $V_0$ is
\begin{eqnarray}
    \label{V0-mti}
    V_0 \simeq
    \frac{ n(1-x) + 1 }{ (n+2) (1-x) } m_0^2 M^2.
\end{eqnarray}
with the dimensionless parameter $x < 1$ which is defined as $\alpha =
- x M^{n+2} / M_\ast^n$.  Here we assumed $m_0 \gg m_{3/2}$.  It
should be noted that even in the presence of the explicit breaking
term in Eq. (\ref{sp-X-mti}) the dynamics of the thermal inflation is
shown not to change much if $m_0 \gg m_{3/2}$ and $x$ is not so close
to one \cite{Asaka-Hashiba-Kawasaki-Yanagida1}.  

In the flaton potential (\ref{veff-mti}) 
we assume the SUSY breaking mass $m_0$ at the origin 
as well as in the original model.
Therefore, when we work in the light gravitino mass region
of the GMSB models,
we obtain the constraint that 
the vev of the flaton $M$ should be smaller than the 
masses of the messenger multiplets.
In the modified model, this constraint 
plays a significant role when we take the cutoff scale as
$M_\ast \gtrsim M_G$ [see the discussion in Sec. \ref{sec-mpmti}].

Then, masses of the
flaton and the $R$-axion are given by
\begin{eqnarray}
    \label{mx-mti}
    m_\chi^2 &\simeq& \frac{ 2 (n+1) - n x }{ 1 - x } m_0^2,\\
    m_a^2 &\simeq& \frac{ (n+2) x }{ 1 - x } m_0^2.
\end{eqnarray}
Therefore, if $x$ takes a value 
\begin{eqnarray}
    \label{x-Raclose}
    \left( 1 > \right) x \ge \frac{ 2(n+1) }{ 5n + 8 } \equiv x_{min},
\end{eqnarray}
the flaton decay into two $R$-axions is kinematically
forbidden.\footnote{
In this region of $x$, the domain wall problem is solved,
i.e., $\alpha$ satisfies Eq. ({\protect \ref{alpha-DW}}).}
Therefore, this modified inflation model with $m_0 \gg m_{3/2}$ and
$x$ in the region (\ref{x-Raclose}) gives one way to forbid the flaton
decay into $R$-axions and as well as to solve the domain wall problem.
In the following we take $x = x_{min}$ for simplicity and $m_0$ as
\begin{eqnarray}
    \label{m0-Raopen2}
    m_0^2 &>&  \frac{ n^2 (3n+11) (5n+13) }{ 4 (n+1)^2 (n+2) }
               m_{3/2}^2
          ~=~  \frac{21}{4} m_{3/2}^2 \for n=1.
\end{eqnarray}
for the purpose of the comparison with the previous results
in the original model.  This is
the same condition for $m_0$  as  assumed in the original thermal
inflation model [Eq. (\ref{m0-Raopen0})].

Then we turn to discuss the thermal history after the modified thermal
inflation ends.  In the modified model the vacuum energy of the flaton
is completely transferred into the thermal bath when the Hubble
parameter becomes comparable to the total width of the flaton
$\Gamma_\chi$ and reheats the universe at the temperature $T = T_R$.
As well as the case of the $R$-axion,
the total width $\Gamma_\chi$ can be written as
[see Sec. \ref{sec-oti}]
\begin{eqnarray}
    \Gamma_\chi ~=~
    C_\chi \frac{ m_\chi^3 }{ M^2 }.
\end{eqnarray}
Here $C_\chi$ is given by
\begin{eqnarray}
    C_\chi =
    \left\{
    \begin{array}[]{ll}
        \displaystyle{
        \frac{ 2 }{ 9 \pi } \left(  \frac{ \alpha_{em} }{ 4 \pi }
        \right)^2 } & ~~\mbox{for}~\lambda_\mu = 0\\
        &\\
        \displaystyle{
        \frac{ 49 }{ 72 \pi } \left(  \frac{ \alpha_{em} }{ 4 \pi }
        \right)^2 } & ~~\mbox{for}~\lambda_\mu \neq 0 
    \end{array}
    \right. ,
\end{eqnarray}
for $m_\chi \lesssim 1$ GeV,
\begin{eqnarray}
    C_\chi \simeq 
    \frac{ 1 }{ 4 \pi } \left(  \frac{ \alpha_{s} }{ 4 \pi } \right)^2,
\end{eqnarray}
for 1 GeV $ \lesssim m_\chi \le 2 m_h$ 
($m_h$ is the Higgs boson mass.%
\footnote{
It should be regarded as the lightest Higgs boson mass
and we take $m_h \simeq 70$ GeV.
}), and
\begin{eqnarray}
    C_\chi \simeq \frac{ C_h }{ 16 \pi } +
    \frac{ 1 }{ 4 \pi } \left(  \frac{ \alpha_{em} }{ 4 \pi } \right)^2,
\end{eqnarray}
for $m_\chi > 2 m_h$.
Then the reheating temperature is obtained as
\begin{eqnarray}
    \label{TR-mti}
    T_R = 0.96 \sqrt{ \Gamma_\chi M_G }.
\end{eqnarray}
By using this reheating temperature, the modified thermal inflation
model increases the entropy of the universe by a factor
\begin{eqnarray}
    \label{Delta-mti}
    \Delta &\simeq& 
    1 +
    \frac{ 4 }{ 3 }
    \frac{ V_0 }{ ( 2 \pi^2 / 45 ) g_\ast T_c^3 T_R },
\end{eqnarray}
which is obtained by putting $\epsilon_a = 0$ and substituting $T_R$
for $T_{R,SM}$ in Eq. (\ref{Delta-OTI}).  Comparing with the entropy
production factor Eq. (\ref{Delta-Raopen}) in the original thermal
inflation model you can see that the suppression factor $(2
m_a/m_\chi)$ is dropped off so that the relic density of the string
modulus is expected to be diluted more effectively.

\section{Moduli Problem with Modified Thermal Inflation Model}
\label{sec-mpmti}

In this section we examine whether the modified thermal inflation
model could solve the cosmological moduli problem or not.  Crucial
difference from the original model is that the flaton decay into two
$R$-axions is kinematically forbidden.  Then the reheating temperature
$T_R$ and also the entropy production factor $\Delta$ of the modified
thermal inflation are determined by the flaton decay.

\subsection{Case I: 
For the case $\Gamma_{\varphi_I} \ge m_\phi \ge H_{TI}$}
\label{subsec-mti1}

First, we consider the case I: $\Gamma_{\varphi_I} \ge m_\phi \ge
H_{TI}$.  In the presence of the modified thermal inflation model with
the entropy production factor (\ref{Delta-mti}), the abundances of the
big-bang modulus and the thermal inflation modulus are given from
Eqs. (\ref{RBB-case1-oti}) and (\ref{RTI-case1-oti}) as
\begin{eqnarray}
    \left( \frac{ \rho_\phi }{ s } \right)_{BB}
    &\simeq&
    3.8 \frac{ m_\phi^{1/2} M_G^{1/2} m_0^3 T_R }{ V_0 }
    \left( \frac{ T_c }{ m_0 } \right)^{3}
    \left( \frac{ \phi_0 }{ M_G } \right)^{2},
    \\
    \left( \frac{ \rho_\phi }{ s } \right)_{TI}
    &\simeq&
    0.38 \frac{ V_0 T_R }{ m_\phi^2 M_G^2 }
    \left( \frac{ \phi_0 }{ M_G } \right)^2.
    \label{RTI-mti-case10}
\end{eqnarray}
Then, as well as for the original thermal inflation model, we find the
minimum of the total modulus abundance given by Eq.~(\ref{Rtot})
in the parameter space of $m_0$ and $T_R$.

The vacuum energy of the modified thermal inflation
(\ref{V0-mti}) can be written in terms of $m_0$ and $T_R$ as
\begin{eqnarray}
    \label{V0-mti2}
    V_0 \simeq C_{V0} \frac{ m_0^5 M_G }{ T_R^2 }.
\end{eqnarray}
Here $C_{V0}$ is defined as
\begin{eqnarray}
    C_{V0} &=& 0.92 ~C_\chi 
    \left[ \frac{ 2(n+1) - nx }{ 1 - x } \right]^{3/2}
    \frac{ n(1-x) + 1 }{ (n+2)(1-x) },
    \nonumber \\
    &\simeq&
    9.3 ~C_\chi ~~\mbox{for}~n=1, x = x_{min}.
\end{eqnarray}
From Eq. (\ref{V0-mti2}) we can express both abundances 
in term of $m_0$ and $T_R$ as
\begin{eqnarray}
    \left( \frac{ \rho_\phi }{ s } \right)_{BB}
    &\simeq&
    \frac{3.8}{C_{V0}}~
    \frac{ m_\phi^{1/2} T_R^{3} }{ m_0^2 M_G^{1/2} }
    \left( \frac{ T_c }{ m_0 } \right)^{3}
    \left( \frac{ \phi_0 }{ M_G } \right)^{2},
    \\
    \left( \frac{ \rho_\phi }{ s } \right)_{TI}
    &\simeq&
    0.38~C_{V0}~ \frac{ m_0^5 }{T_R m_\phi^2 M_G }
    \left( \frac{ \phi_0 }{ M_G } \right)^2.
    \label{RTI-mti-case1}
\end{eqnarray}
The dependence on $m_0$ tells that the total abundance takes its
minimum value when $\RBB$ = $\RTI$ is achieved, i.e.,
\begin{eqnarray}
    \label{m0eq-mti-case1}
    m_0 ~=~ \frac{ 1.4 }{ C_{V0}^{2/7} } ~
    m_\phi^{5/14} T_R^{4/7} M_G^{1/14} 
    \TC^{3/7},
\end{eqnarray}
and we obtain
\begin{eqnarray}
    \label{R0-mti-case1}
    \Rz ~\ge~ \sqrt{ \RBB \RTI } 
    &\simeq&
    1.2 \frac{ m_0^{3/2} T_R }{ m_\phi^{3/4} M_G^{3/4} }
    \TC^{3/2} \phiz^2,
    \\
    &\simeq&
    \frac{ 2.0 }{ C_{V0}^{3/7} } ~
    \frac{ T_R^{13/7} }{ m_\phi^{3/14} M_G^{9/14} }
    \TC^{15/7} \phiz^2.
\end{eqnarray}
Therefore the lowest reheating temperature $T_R = 10$ MeV gives for
$n=1$
\begin{eqnarray}
    \Omega_\phi h^2 ~ \gtrsim ~
    \left\{
        \begin{array}{ll}
            6.0 \times 10^{-3 }
            \displaystyle{
            \left( \frac{ m_\phi }{10^{-8}\GeV} \right)^{-3/14}
            \TC^{15/7} \phiz^2 } &
            ~~\for~\lambda_\mu = 0\\
            3.7 \times 10^{-3 }
            \displaystyle{
            \left( \frac{ m_\phi }{10^{-8}\GeV} \right)^{-3/14}
            \TC^{15/7} \phiz^2 } &
            ~~\for~\lambda_\mu \neq 0
        \end{array}
    \right.,
\end{eqnarray}
for $m_\chi \lesssim 1$ GeV (the flaton decays only into two photons)
and
\begin{eqnarray}
    \Omega_\phi h^2 \gtrsim
    1.01 \times 10^{-5} ~
    \left( \frac{ m_\phi }{1\GeV} \right)^{-3/14}
    \TC^{15/7} \phiz^2 , 
\end{eqnarray}
for $2 m_h > m_\chi \gtrsim 1$ GeV (the flaton decays dominately
into two gluons).

\begin{figure}[t]
    \centerline{\psfig{figure=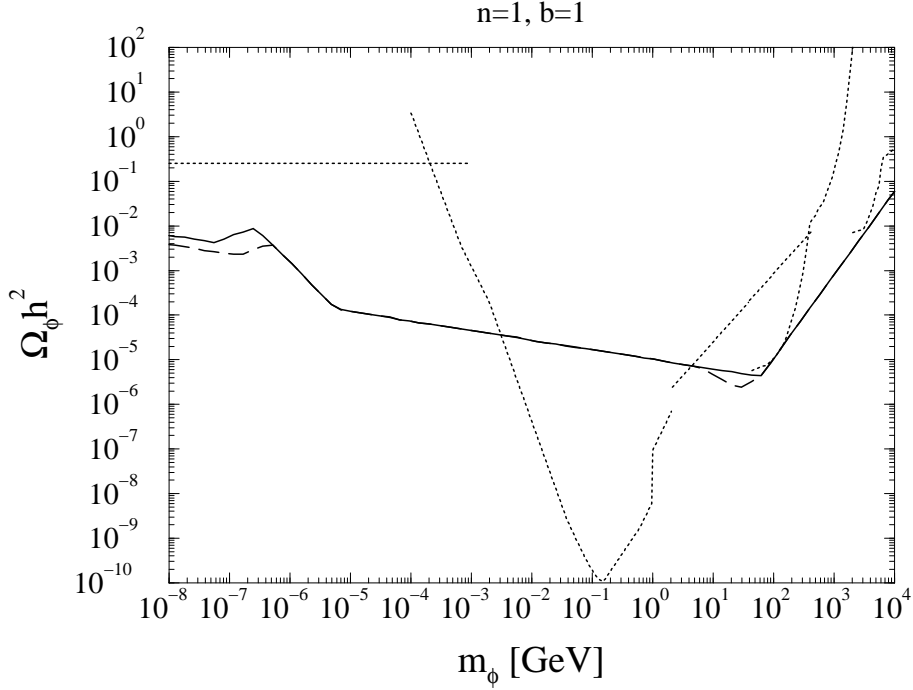,width=12cm}}
\caption{
The lower bounds on the modulus abundance
in the presence of the modified thermal inflation
for the case I: $\Gamma_{\varphi_I} \ge m_\phi \ge H_{TI}$.
We assume $m_{\phi}=m_{3/2}$, $T_c = m_0$ and $\phi_{0}=M_{G}$
and take $n=1$ and $b=1$.
The solid (dashed) line denotes the lower bound 
when $\lambda_\mu = 0$ ($\lambda_\mu \neq 0$).
Upper bounds from various cosmological constraints
are all shown by the dotted lines.
}
    \label{fig:mti_case1}
\end{figure}

On the other hand, when the flaton can decay
into Higgs bosons ($m_\chi \ge 2 m_h$), 
one more free parameter $C_h$ appears in the
flaton decay width and the condition $\RBB$ = $\RTI$, i.e., Eq.
(\ref{m0eq-mti-case1}), could be achieved by taking a moderate $C_h$.
Then we find from Eq. (\ref{R0-mti-case1}) that the lowest values of
$m_0$ which corresponds to $m_\chi = 2 m_h$ and $T_R = 10$ MeV give
the minimum of the total abundance as
\begin{eqnarray}
    \Omega_\phi h^2 \gtrsim 
    4.6 \times 10^{-6} ~
    \left( \frac{ m_\phi }{ 10\GeV} \right)^{-3/4}
    \left( \frac{ m_h }{70\GeV} \right)^{3/2}
    \TC^{3/2} \phiz^2.
\end{eqnarray}
for $n=1$.  In fact, this result gives the absolute minimum abundance
for the modulus with a mass $m_\phi \simeq$ 6 -- 26 GeV.  (See Fig.
\ref{fig:mti_case1}.)

However, for the modulus mass region $m_\phi \gtrsim$ 60 GeV, the
reheating temperature $T_R = 10$ MeV could not satisfy $\RBB = \RTI$
because of the lower bound on $m_0$ [Eq. (\ref{m0-Raopen2})].  
From Eq. (\ref{m0eq-mti-case1}) $T_R$ should be
\begin{eqnarray}
    T_R \gtrsim 18\MeV~
    \left( \frac{ m_\phi}{100\GeV} \right)^{9/8}
    \TC^{-3/4} \mgra^{7/4} \fn.
\end{eqnarray}
Then from Eq. (\ref{R0-mti-case1})
this gives the minimum abundance as
\begin{eqnarray}
    \Omega_\phi h^2 \gtrsim 1.1 \times 10^{-5} ~
    \left( \frac{ m_\phi }{100\GeV} \right)^{15/8}
    \TC^{3/4}  \mgra^{13/4} \phiz^2 \fn.
\end{eqnarray}

Fig. \ref{fig:mti_case1} shows the lower bound on the total abundance
of the modulus.  We find that the modified thermal inflation 
can dilute extensively the modulus density than the original one in
the light modulus mass region $m_\phi \lesssim 100$ GeV.  On the other
hand, the lower bound is almost the same
as the original model for $m_\phi \gtrsim 100$ GeV\footnote{
The difference of the order one factor comes from the difference of
the pre-factor in the expression of $m_\chi$ [see
Eqs. (\ref{mx-simple}) and (\ref{mx-mti})] and so on.}
[see Fig. \ref{fig:oti_case1}].
We obtain the allowed regions for the modulus mass:
$m_\phi \lesssim 3$ MeV and $m_\phi \gtrsim 4$ GeV.
Therefore, the window for the gravitino mass
predicted by the GMSB models does exist 
even for the case I: $\Gamma_{\varphi_I} \ge m_\phi \ge H_{TI}$
and this point is crucially different from 
the original thermal inflation model.
%
\begin{figure}[t]
    \centerline{\psfig{figure=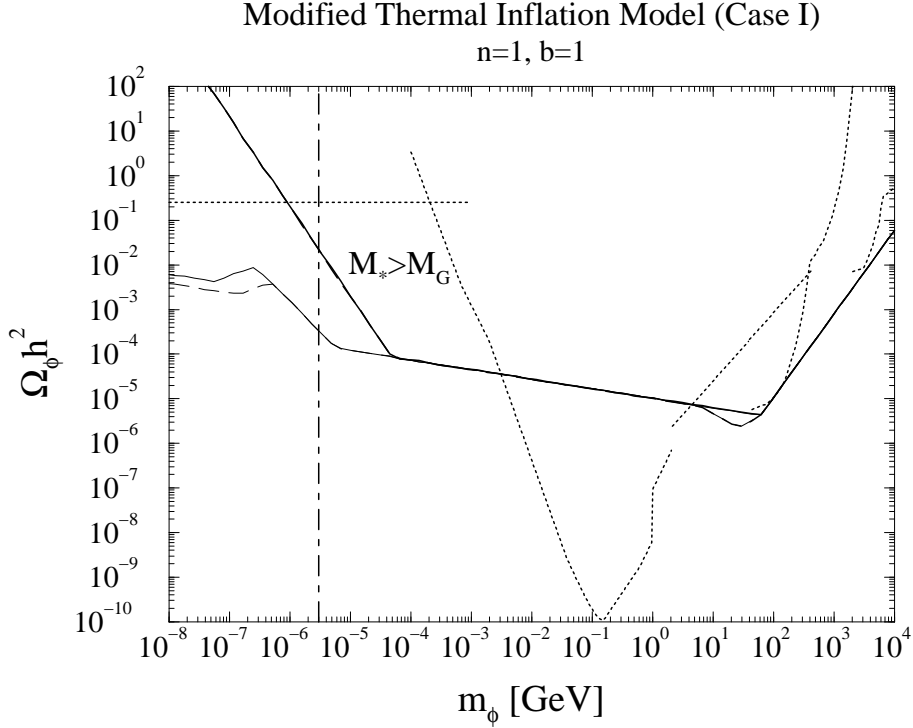,width=12cm}}
\caption{
Same figure as Fig. \ref{fig:mti_case1} except for $M_\ast > M_G$.
We also show the results in Fig. \ref{fig:mti_case1}
by the thin lines.
The lower bound on the modulus mass [Eq. (\ref{const-mp-messenger})]
is represented by the dot-dashed line.
}
    \label{fig:mti_case1cut}
\end{figure}

Let us discuss the effect of the lower bound on the cutoff scale
$M_\ast > M_{cr} \sim M_G$.  In the same way as the original thermal
inflation model, it leads to the lower bound on $m_0$:
\begin{eqnarray}
    \label{const-m0-mti}
    m_0 > \frac{ 0.25 }{ C_\chi^{1/2} } ~ T_R
    \left( \frac{ M_{cr} }{ M_G } \right)^{1/2},
\end{eqnarray}
for $n=1$. Here we consider only the case $n=1$ since
larger $n$ induces more abundant mass density of the modulus.
This lower bound on $m_0$ does change the above analysis
for the modulus mass region
\begin{eqnarray}
    \label{mpc-mti-case1}
    m_\phi < 4.9 \times 10^{-5} \GeV~
    \left( \frac{ T_R }{10\MeV} \right)^{6/5}
    \TC^{-6/5}
    \left( \frac{ M_{cr} }{ M_G } \right)^{7/5},
\end{eqnarray}
because the condition (\ref{m0eq-mti-case1}) could not be satisfied
and $\RTI$ be always larger than $\RBB$.  In this case
we find from Eqs.(\ref{RTI-mti-case1}) and (\ref{const-m0-mti})
\begin{eqnarray}
    \Rz ~=~ \RTI
    ~\gtrsim~ \frac{ 3.2 \times 10^{-3} }{ C_\chi^{3/2} }
    \frac{ T_R^{4 } }{ m_\phi^2 M_G }
    \left( \frac{ M_{cr} }{ M_G } \right)^{5/2}
    \phiz^2,
\end{eqnarray}
and the lowest reheating temperature $T_R$ = 10 MeV gives 
\begin{eqnarray}
    \label{OM-mti-cut}
    \Omega_\phi h^2 \gtrsim
    0.20 ~
    \left( \frac{ m_\phi }{ 1\keV} \right)^2
    \left( \frac{ M_{cr} }{ M_G } \right)^{5/2}
    \phiz^2,
\end{eqnarray}
for $m_\chi \gtrsim 1$ GeV.
We show the lower bound on the modulus
abundance in Fig. \ref{fig:mti_case1cut} when $M_\ast > M_G$.  

Here we neglect the modulus mass region $m_\phi \lesssim 3$ keV,
since the vev of the flaton becomes always larger than 
masses of messenger fields to obtain the reheating temperature
$T_R \gtrsim$ 10 MeV.
In the GMSB models masses of messenger fields $m_{mess}$ 
can be written as $m_{mess} \simeq \langle F_{mess} \rangle /
\Lambda$ with the $F$-component vev $\langle F_{mess} \rangle$
in the messenger sector and $\Lambda \sim 10^4$--$10^{5}$ GeV
\cite{GMSB}, and we obtain the upper bound on 
masses of messenger fields as
\begin{eqnarray}
    m_{mess} \lesssim 
    \frac{ \sqrt{3} m_{3/2} M_G }{ \Lambda }.
\end{eqnarray}
On the other hand, the vev of the flaton can be written 
from Eqs. (\ref{mx-mti}) and (\ref{TR-mti}) as
\begin{eqnarray}
    M = 3.4 C_\chi^{1/2} \frac{ m_0^{3/2} M_G^{1/2} }{ T_R},
\end{eqnarray}
for $n=1$. Therefore, when one consider the GMSB models, 
in order that the vev of the flaton should be smaller than 
$m_{mess}$ we obtain 
\begin{eqnarray}
    m_0 ~\lesssim~ \frac{ 0.64 }{ C_\chi^{1/3} }
    \frac{ m_{3/2}^{2/3} M_G^{1/3} T_R^{2/3} }{ \Lambda^{2/3} }.
\end{eqnarray}
For the case $M_\ast > M_{cr} \sim M_G$ 
the upper bound on the reheating temperature 
can be obtained from Eq. (\ref{const-m0-mti}) as
\begin{eqnarray}
    T_R \lesssim 18 C_\chi^{1/2} \frac{ m_{3/2}^2 M_G }{ \Lambda^2 }
    \left( \frac{ M_{cr} }{ M_G } \right)^{-3/2}.
\end{eqnarray}
Then, the reheating temperature $T_R \gtrsim 10$ MeV is achieved only when
\begin{eqnarray}
    \label{const-mp-messenger}
    m_\phi \simeq m_{3/2} \gtrsim 
    3.0 \keV 
    \left( \frac{ \Lambda}{10^4 \GeV} \right)
    \left( \frac{ M_{cr} }{ M_G } \right)^{3/4},
\end{eqnarray}
where we consider the flaton dominately decays into gluons.

From Fig. \ref{fig:mti_case1cut} we see
that the dilution of the modulus density becomes less effective
for the modulus mass region Eq. (\ref{mpc-mti-case1}) and the modulus
with a mass $m_\phi \lesssim 3$ keV is excluded.  However, the allowed
region of the modulus (gravitino) mass does exist in the region where
the GMSB models predicts, even if we take
$M_\ast \gtrsim M_G$ in the modified thermal inflation model, while it
does not exist in the original model even for the case III ($ m_\phi >
H_{TI}> \Gamma_{\varphi_I}$) [see Figs. \ref{fig:oti_case1cut},
\ref{fig:oti_case2cut} and \ref{fig:oti_case3cut}].

\subsection{Case II: For the case $m_\phi \ge \Gamma_{\varphi_I} \ge H_{TI}$}
\label{subsec-mti2}
%
Next we consider the case $m_\phi \ge \Gamma_{\varphi_I} \ge H_{TI}$.
From Eq. (\ref{RBB-case2}) the modified thermal inflation model with
the entropy production factor (\ref{Delta-mti}) gives the abundance of
the big-bang modulus as
\begin{eqnarray}
    \RBB 
    &\simeq&
    3.8 ~ \frac{ \Gamma_{\varphi_I}^{1/2} M_G^{1/2} m_0^3 T_R }{ V_0 }
    \TC^3
    \phiz^2.
\end{eqnarray}
On the other hand, the abundance of the thermal-inflation modulus is the
same as Eqs. (\ref{RTI-mti-case10}) and (\ref{RTI-mti-case1}) in the
previous case. Now the abundance of the big-bang modulus takes its
minimum value when $\Gamma_{\varphi_I} = H_{TI}$ as
\begin{eqnarray}
    \label{RBBm-mti-case2}
    \RBB ~\ge~ \RBBm
    &\simeq&
    2.9 ~\frac{ m_0^3 T_R }{ V_0^{3/4} }
    \TC^{3} \phiz^2.
\end{eqnarray}
Then the lower bound on the total abundance Eq. (\ref{Rtot}) 
is given by
\begin{eqnarray}
    \Rz ~\ge ~\sqrt{ \RBBm \RTI} 
    ~\simeq ~
    \frac{ 2.2 }{ C_{V0}^{12/23} }
    \frac{ T_R^{47/23} }{ m_\phi^{6/23 } M_G^{18/23} }
    \TC^{60/23} \phiz^2,
\end{eqnarray}
with
\begin{eqnarray}
    m_0 ~=~ \frac{ 1.4}{ C_{V0}^{7/23} } 
    m_\phi^{8/23} M_G^{1/23} T_R^{14/23}
    \TC^{12/23}.
\end{eqnarray}
Therefore, we obtain for $n=1$ with
the lowest reheating temperature $T_R = 10$ MeV:
\begin{eqnarray}
    \Omega_\phi h^2 ~ \gtrsim ~
    \left\{
        \begin{array}{ll}
            7.6 \times 10^{-5}
            \displaystyle{
            \left( \frac{ m_\phi }{10^{-8}\GeV} \right)^{-6/23}
            \TC^{60/23} \phiz^2 } &
            ~~\for~\lambda_\mu = 0\\
            4.3 \times 10^{-5}
            \displaystyle{
            \left( \frac{ m_\phi }{10^{-8}\GeV} \right)^{-6/23}
            \TC^{60/23} \phiz^2 } &
            ~~\for~\lambda_\mu \neq 0
        \end{array}
    \right.,
\end{eqnarray}
for $m_\chi \lesssim 1$ GeV and
\begin{eqnarray}
    \Omega_\phi h^2 \gtrsim
    3.2 \times 10^{-8} ~
    \left( \frac{ m_\phi }{1\GeV} \right)^{-6/23}
    \TC^{60/23} \phiz^2 , 
\end{eqnarray}
for $2 m_h > m_\chi \gtrsim 1$ GeV.  

\begin{figure}[t]
    \centerline{\psfig{figure=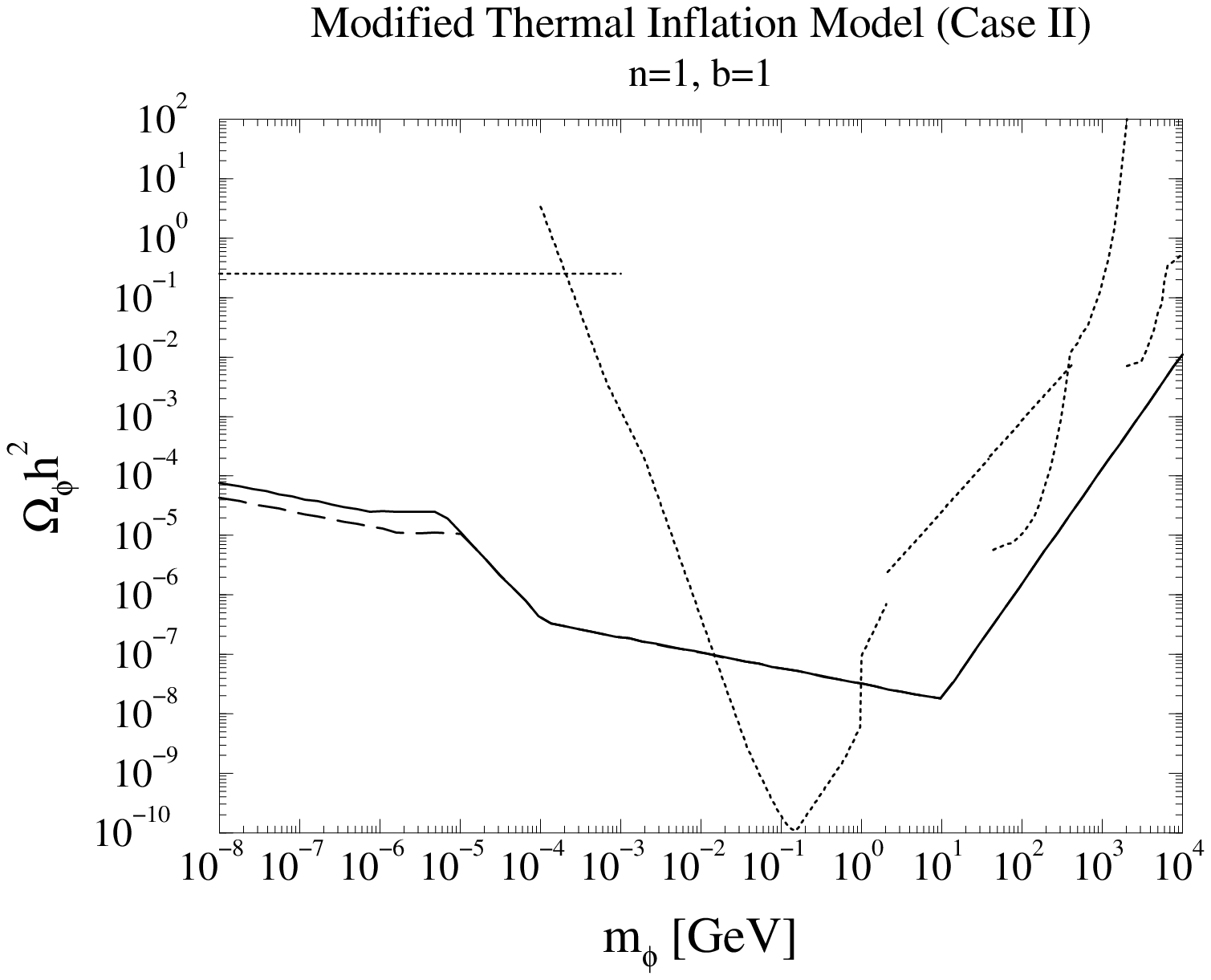,width=12cm}}
\caption{
Same figure as Fig. {\protect \ref{fig:mti_case1}}
for the case II: $m_\phi \ge \Gamma_{\varphi_I} \ge H_{TI}$.
}
    \label{fig:mti_case2}
\end{figure}
\begin{figure}[t]
    \centerline{\psfig{figure=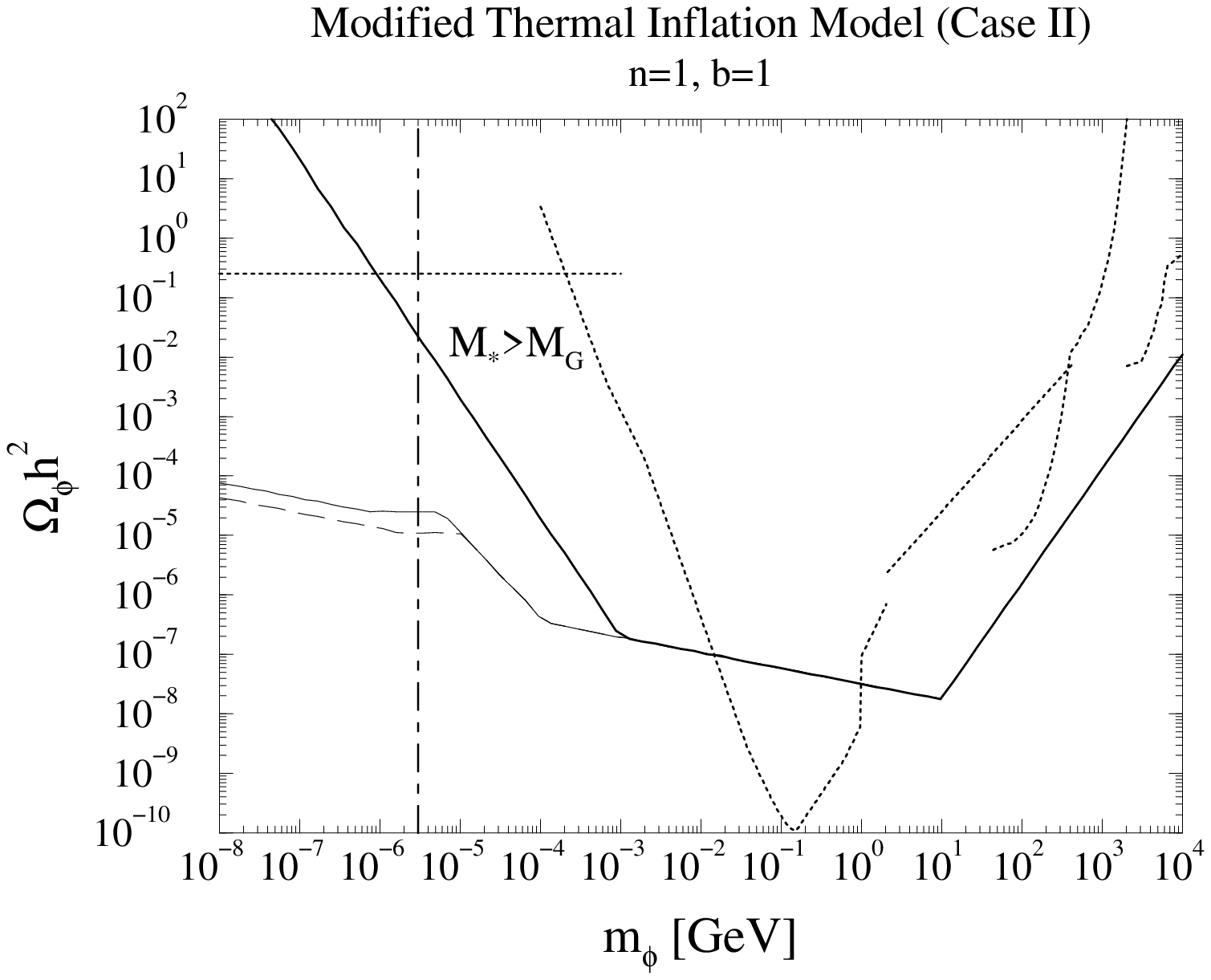,width=12cm}}
\caption{
Same figure as Fig. {\protect \ref{fig:mti_case2}}
except for $M_\ast > M_G$.
We also show the results in Fig. {\protect \ref{fig:mti_case2}}
by the thin lines.
The lower bound on the modulus mass [Eq. (\ref{const-mp-messenger})]
is represented by the dot-dashed line.
}
    \label{fig:mti_case2cut}
\end{figure}

Furthermore, if the modulus mass becomes larger than about 10 GeV,
the total abundance takes its minimum value when 
the reheating temperature is not 10 MeV but
\begin{eqnarray}
    T_R \simeq 120 \MeV~
    \left( \frac{ m_\phi }{100 \GeV } \right)^{15/14}
    \mgra^{23/14}
    \TC^{-6/7} \fn,
\end{eqnarray}
due to the lower bound on $m_0$ [Eq. (\ref{m0-Raopen2})], which leads to
\begin{eqnarray}
    \Omega_\phi h^2 \gtrsim 1.6 \times 10^{-6} ~
    \left( \frac{ m_\phi }{100 \GeV} \right)^{27/14}
    \TC^{6/7} \mgra^{47/14} \phiz^2 \fn.
\end{eqnarray}
This result is the same as Eq. (\ref{OM-oti2-hm}) 
in the original thermal inflation model
if we neglect the order one factor.
We find that the dilution becomes ineffective
when the flaton decays into Higgs bosons,
and then the lower bounds listed above are the absolute ones.

In Fig. \ref{fig:mti_case2} the lower bound on the total abundance is
displayed.  Comparing the result in the original thermal inflation
model showed in Fig. \ref{fig:oti_case2}, the lower bound on the
modulus abundance is extensively reduced for the lighter modulus mass
region $m_\phi \lesssim 10$ GeV, while it is almost the same for
$m_\phi \gtrsim 100$ GeV.  In particular, the lower bound becomes much
weaker than the previous case I for the lighter modulus mass region
$m_\phi \lesssim 1$ GeV, so that only the modulus mass region $m_\phi
(\simeq m_{3/2}) \sim$ 10 MeV--1 GeV is excluded by the stringent
constraint from the cosmic $X( \gamma )$-ray backgrounds.

Moreover, when we take the cutoff scale as $M_\ast > M_{cr} \sim M_G$,
the lower bound behaves as shown in Fig. \ref{fig:mti_case2cut}.
This changes the lower bound for $m_\phi \lesssim$ 1 MeV 
and the bound is given by Eq. (\ref{OM-mti-cut}) in the previous
case I.  
Therefore, the modulus whose mass is $m_\phi \lesssim$ 3 keV is
excluded as well as the previous case I.

\subsection{Case III: 
For the case $m_\phi \ge H_{TI} \ge \Gamma_{\varphi_I}$ }
\label{subsec-mti3}
Finally we consider the case $m_\phi \ge H_{TI} \ge
\Gamma_{\varphi_I}$, where the abundance of the big-bang modulus in
the presence of the modified thermal inflation model is given by 
[see Appendix \ref{Ap-2}]
\begin{eqnarray}
    \RBB \simeq 4.8 
    \frac{ m_0^4 T_R }{ \Gamma_{\varphi_I} V_0^{1/2} M_G }
    \TC^4 \phiz^2.
\end{eqnarray}
This ratio takes its minimum value when $\Gamma_{\varphi_I} = H_{TI}$ as
\begin{eqnarray}
    \RBB ~\ge~ \RBBm ~=~
    8.2 ~\frac{ m_0^4 T_R }{ V_0 } 
    \TC^4 \phiz^2.
\end{eqnarray}
While the abundance of the thermal-inflation modulus 
is the same as the previous two cases and given by 
Eqs. (\ref{RTI-mti-case10}) and (\ref{RTI-mti-case1}).
Since $\RBBm$ becomes equal to $\RTI$ when
\begin{eqnarray}
    m_0 ~=~
    \frac{ 1.7 }{ C_{V0}^{1/3} } 
    m_\phi^{1/3} T_R^{2/3}  
    \TC^{2/3},
\end{eqnarray}
the lower bound on the total abundance (\ref{Rtot}) is given as
\begin{eqnarray}
    \Rz &\gtrsim&
    \frac{ 4.9}{ C_{V0}^{2/3} } 
    \frac{ T_R^{7/3} }{ m_\phi^{1/3} M_G }
    \TC^{10/3} \phiz^2.
\end{eqnarray}
%
\begin{figure}[t]
    \centerline{\psfig{figure=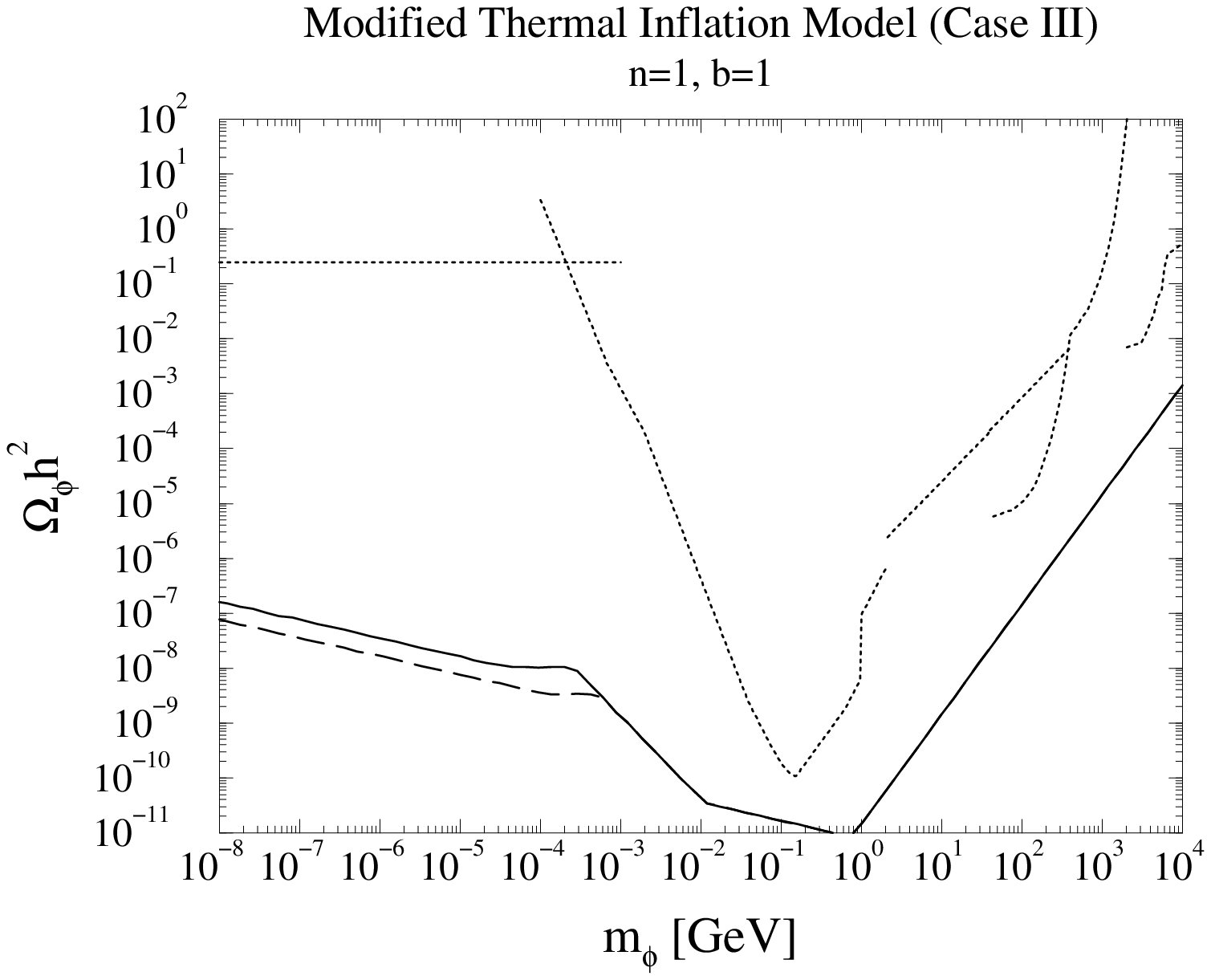,width=12cm}}
\caption{
Same figure as Fig. {\protect \ref{fig:mti_case1}}
for the case III: $m_\phi \ge H_{TI} \ge \Gamma_{\varphi_I}$.
}
    \label{fig:mti_case3}
\end{figure}
%
\begin{figure}[t]
    \centerline{\psfig{figure=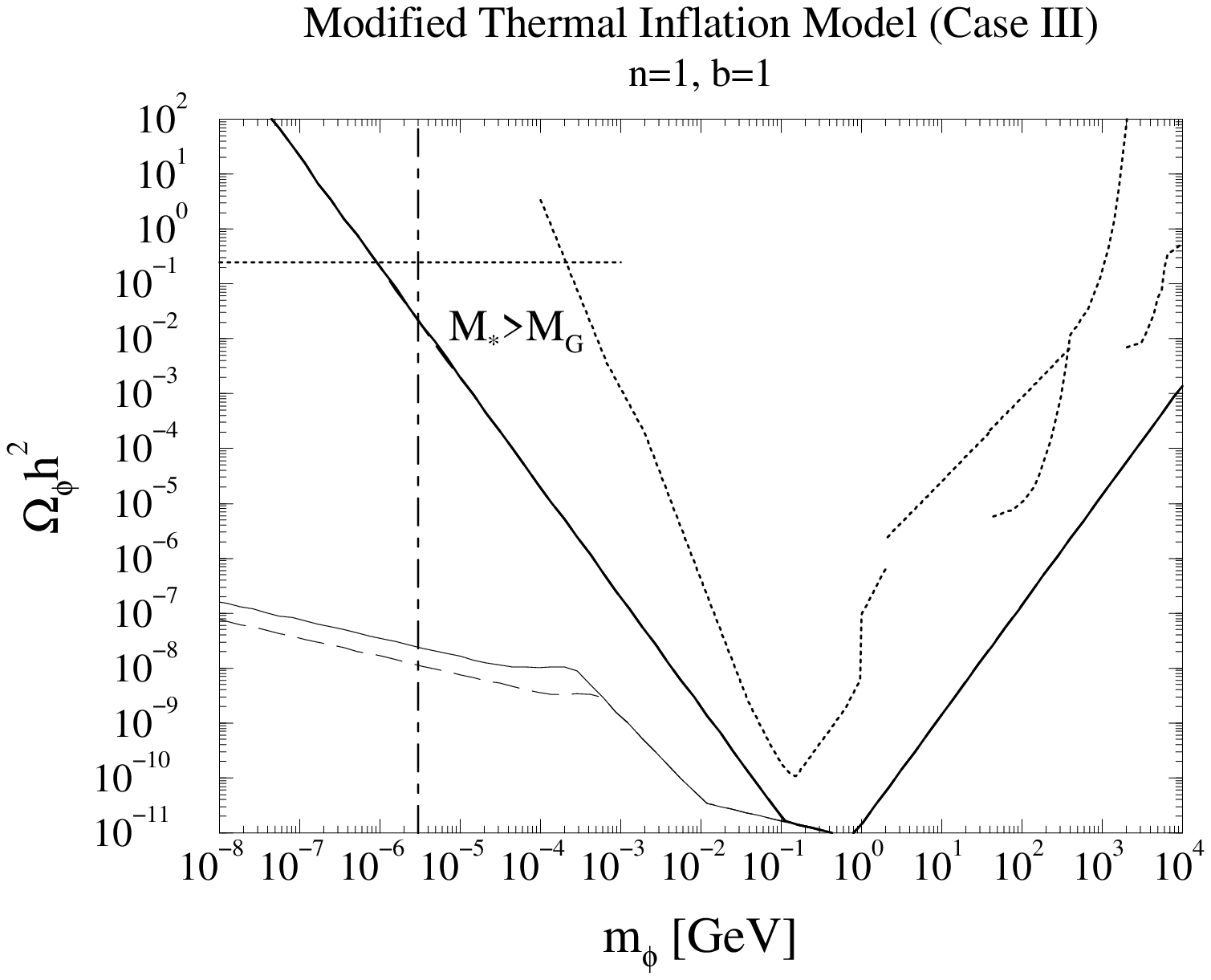,width=12cm}}
\caption{
Same figure as Fig. \ref{fig:mti_case3} except for $M_\ast > M_G$.
We also show the results in Fig. \ref{fig:mti_case3}
by the thin lines.
The lower bound on the modulus mass [Eq. (\ref{const-mp-messenger})]
is represented by the dot-dashed line.
}
    \label{fig:mti_case3cut}
\end{figure}
Therefore, the lowest reheating temperature $T_R = 10$ MeV
leads to for $n=1$
\begin{eqnarray}
    \Omega_\phi h^2 ~ \gtrsim ~
    \left\{
        \begin{array}{ll}
            3.4 \times 10^{-8}
            \displaystyle{
            \left( \frac{ m_\phi }{1\keV} \right)^{-1/3}
            \TC^{10/3} \phiz^2 } &
            ~~\for~\lambda_\mu = 0\\
            1.6 \times 10^{-8}
            \displaystyle{
            \left( \frac{ m_\phi }{1\keV} \right)^{-1/3}
            \TC^{10/3} \phiz^2 } &
            ~~\for~\lambda_\mu \neq 0
        \end{array}
    \right.,
\end{eqnarray}
for the flaton can only decay into photons ($m_\phi \lesssim 1$ MeV),
and
\begin{eqnarray}
    \Omega_\phi h^2 \gtrsim
    1.6 \times 10^{-11} ~
    \left( \frac{ m_\phi }{100\MeV} \right)^{-1/3}
    \TC^{10/3} \phiz^2 , 
\end{eqnarray}
for the flaton dominately decays into two gluons 
($m_\phi \gtrsim 1$ MeV).
In the case III the flaton decay into Higgs bosons
makes dilution ineffective and gives no absolute minimum
of the total modulus abundance.

We show the lower bound on the total modulus abundance in Fig.
\ref{fig:mti_case3}.  It is seen that in the case III
the modulus mass density is sufficiently diluted 
so that the whole modulus (gravitino) mass region 
$10$ eV--10 TeV  predicted by both models of the GMSB and the HSSB 
survives the various cosmological constraints.  
This is a crucial result when one considers the primordial
inflation model with quite low reheating temperature.
When the total abundance takes its minimum value, 
the required reheating temperature of the primordial inflation is
estimated as, for the modulus mass region $m_\phi \lesssim 1$ GeV,
\begin{eqnarray}
    T_{RI} ~ \simeq ~
    \left\{
        \begin{array}{ll}
            230 \GeV~
            \displaystyle{
            \left( \frac{ m_\phi }{1\keV} \right)^{5/12}
            \TC^{5/6} } &
            ~~\for~\lambda_\mu = 0\\
            190 \GeV~
            \displaystyle{
            \left( \frac{ m_\phi }{1\keV} \right)^{5/12}
            \TC^{5/6}  } &
            ~~\for~\lambda_\mu \neq 0
        \end{array}
    \right.,
\end{eqnarray}
when the flaton decays only into two photons, and
\begin{eqnarray}
    T_{RI} ~\simeq~
    1.1 \times 10^{4} \GeV~
    \left( \frac{ m_\phi }{100 \MeV} \right)^{5/12}
    \left( \frac{ T_c }{ m_0 } \right)^{5/6},
\end{eqnarray}
when the flaton dominately decays into two gluons.
On the other hand, for the mass region $m_\phi \gtrsim 1$ GeV, 
the required reheating temperature is given by
\begin{eqnarray}
    T_{RI} ~\simeq~
    5.5 \times 10^{6} \GeV
    \left( \frac{ m_\phi }{ 1\TeV} \right)^{3/4}
    \TC^{1/2}
    \mgra^{1/2},
\end{eqnarray}
where the flaton dominately decays into two gluons.
Therefore, in order to sufficiently dilute the light modulus predicted by 
the GMSB models,
the required reheating temperature of the primordial inflation is
not so extremely low as that required 
in the original thermal inflation model [see Eq. (\ref{TRI-oti3})].

Furthermore, we show in Fig. \ref{fig:mti_case3cut} the lower bound on
the modulus abundance for the case $M_\ast > M_{cr} \sim M_G$.  The
lower bound becomes more stringent for the modulus mass $m_\phi
\lesssim 100$ MeV where the bound is given by Eq. (\ref{OM-oti2-hm})
which is the same as the previous two cases.  
However, we find that in the wide region of the modulus mass
$m_\phi \sim$ 3 keV--10 TeV 
the cosmological moduli problem can be solved naturally
by the modified thermal inflation,
if the reheating temperature of the primordial inflation 
is low enough.
\section{Discussion}
\label{sec-dis}

The thermal inflation, 
whether it is original one or modified one,
can dilute significantly unwanted long-lived particles (i.e., the
string moduli).  
Especially, as shown in Subsec. \ref{subsec-mti3}, the modified model
can gives a solution to the cosmological moduli problem
and the whole modulus mass region 
$m_\phi (\simeq m_{3/2}) \sim 10$ eV--$10^4$ GeV 
predicted by both the GMSB and HSSB models 
survives from various cosmological constraints
[see Fig. \ref{fig:mti_case3}],
if the reheating temperature of the primordial inflation 
is low enough for its reheating process to finish
after the thermal inflation ends.
Furthermore, in this case, we find that even if one takes 
the gravitational scale as
the cutoff scale of the modified thermal inflation model 
($M_\ast \gtrsim M_G$),
the modulus mass region $m_\phi (\simeq m_{3/2}) \sim$ 3 keV--$10^4$ GeV
is allowed [see Fig. \ref{fig:mti_case3cut}].
Note that gravitino is also diluted sufficiently and
we have no gravitino problem.

However, since the abundances of all relic particles are also
diluted by the thermal inflation, one might be faced with the
following problems; how generate the present observed baryon asymmetry
and what is the dark matter of our universe.  Here we give some
possible solutions to them.

We first discuss the baryon asymmetry of the universe.
The generation of the observed baryon asymmetry $Y_B \equiv
(n_B/s)_0 \sim 10^{-10}$--$10^{-11}$ is one of the challenging question
in particle cosmology.  Furthermore, if one assume the thermal
inflation in the history of the universe, the situation becomes worse.
Because the primordial baryon asymmetry is also diluted after the
thermal inflation and the temperature at that epoch is $T \sim $ 10
MeV in order to dilute the moduli abundance maximally, the GUT
baryogenesis and the electroweak baryogenesis do not work in this
case.

However, as pointed out by Ref. \cite{Gouvea-Moroi-Murayama}, enough
baryon number could be produced by the Affleck-Dine mechanism
\cite{Affleck-Dine}.  The generated asymmetry is related with the
present modulus abundance as \cite{Gouvea-Moroi-Murayama}
\begin{eqnarray}
    Y_B ~\lesssim ~
    \frac{ 1 }{ m_{3/2} } 
    \left( \frac{ \rho_\phi }{ s } \right)_{BB}
    = 
    3.6 \times 10^{-9} ~
    ( \Omega_{\phi_{BB}} h^2 )
    \left( \frac{ m_\phi }{1\GeV}  \right)^{-1}
    \mgra^{-1},
\end{eqnarray}
where $\Omega_{\phi_{BB}} = \rho_{\phi_{BB}}/ \rho_{cr}$.
Therefore, a tremendous baryon number is produced primordially 
(i.e., $H \sim m_{3/2}$) 
and enough asymmetry could be left even after the entropy
production by the thermal inflation.  Then the lower bound on the
modulus abundance can be obtained from the present value of $Y_B$ as
\begin{eqnarray}
    \label{OM_AD}
    \Omega_\phi h^2  ~\ge~
    \Omega_{\phi_{BB} } h^2 
    ~\gtrsim
    2.8 \times 10^{-3} ~
    \left( \frac{ Y_B}{ 10^{-11} } \right)
     \left( \frac{ m_\phi }{1\GeV}  \right)
    \mgra.
\end{eqnarray}

\begin{figure}[t]
    \centerline{\psfig{figure=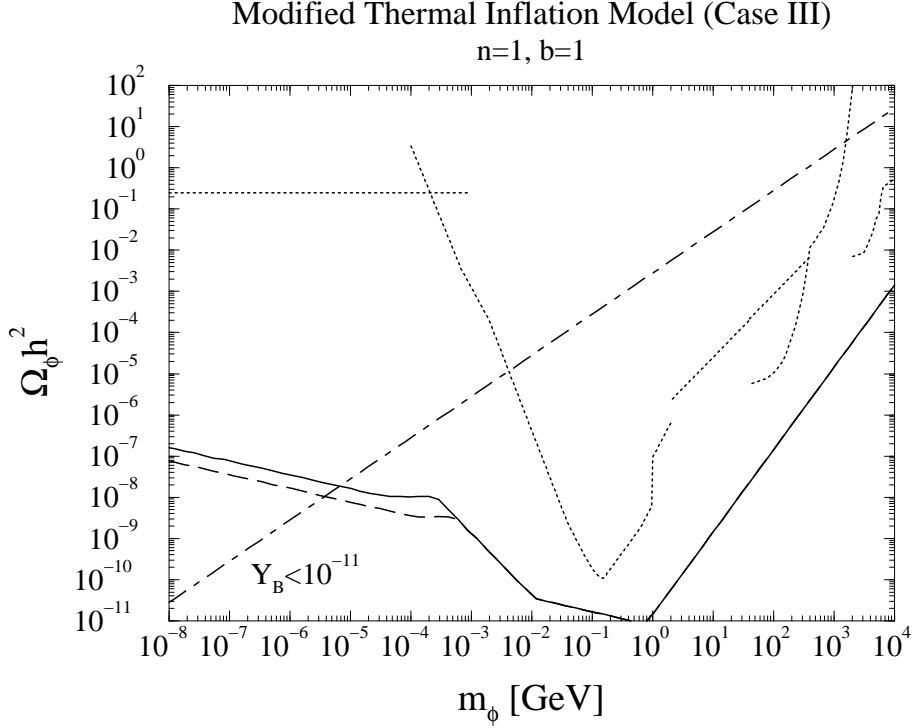,width=12cm}}
\caption{
Same figure as Fig. {\protect{\ref{fig:mti_case3}}}.
We also show by the dot-dashed line
the lower bound on the modulus abundance 
from the present baryon asymmetry $Y_B = 10^{-11}$ 
which is generated by the Affleck-Dine mechanism.
}
    \label{fig:mti_case3AD}
\end{figure}
In Fig. \ref{fig:mti_case3AD} we show this bound together with the
result obtained in Subsec. \ref{subsec-mti3} 
[Fig. \ref{fig:mti_case3}].  You can see
that this lower bound (\ref{OM_AD}) lies above various upper bounds on
the modulus abundance for $m_\phi \gtrsim 4$ MeV.  Therefore the
Affleck-Dine mechanism can generate enough baryon asymmetry in the
lighter modulus mass region with $m_\phi \lesssim $ 4 MeV
\cite{Kawasaki-Yanagida} even in the presence of the thermal
inflation, and, on the other hand, we require other mechanism of
baryogenesis in the heavier modulus mass region.
 
In Ref. \cite{Stewart-Kawasaki-Yanagida} a variant type of the
Affleck-Dine mechanism was proposed, where an $D$-flat direction,
$LH_u$, could produce the required baryon asymmetry after the thermal
inflation ends.  They assumed that $m_{H_u}^2 + m_L^2$ be negative
($m_{H_u}$ and $m_L$ are the soft SUSY breaking mass of a $H_u$ and a
scalar lepton doublet $L$) in order that the $LH_u$ condensate rolls
away from the origin. However, in this case the potential along a
specific direction is unbounded from below \cite{Komatsu}.  Therefore,
this interesting mechanism of baryogenesis is not phenomenologically
viable.%
\footnote{
We thank T. Yanagida for informing us of this point.
}

%
\begin{figure}[t]
    \centerline{\psfig{figure=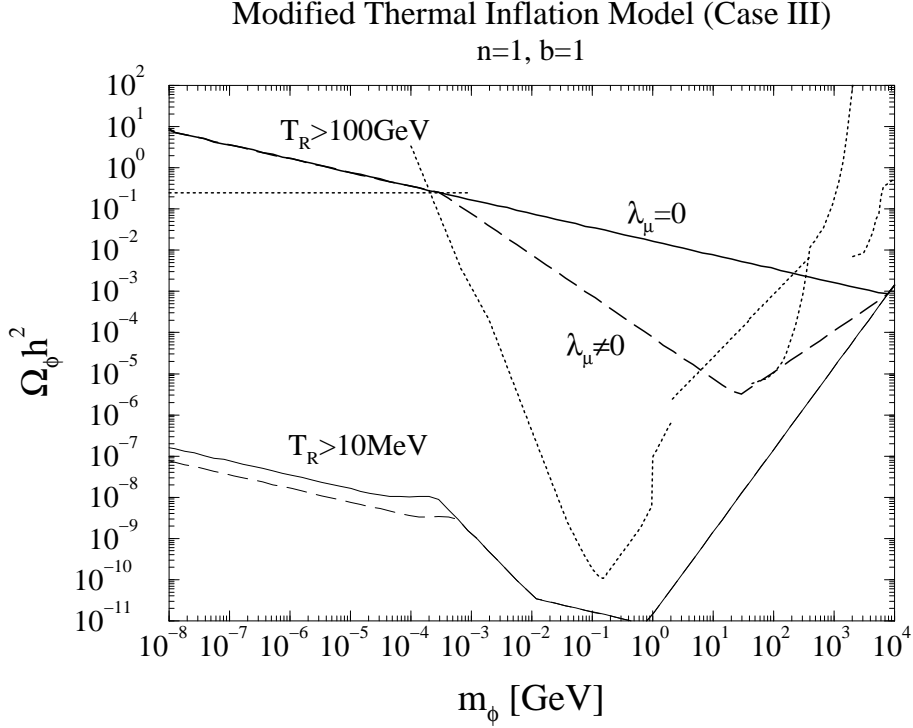,width=12cm}}
\caption{
The minimum abundance of the modulus in the presence of the modified 
thermal inflation for the case III: 
$m_\phi \ge H_{TI} \ge \Gamma_{\varphi_I}$.
The thick solid (dashed) line denotes the minimum abundance for the case
$\lambda_\mu = 0$ ($\lambda_\mu \neq 0$) when $T_R > 100$ GeV and 
the thin solid (dashed) line denotes that when $T_R > 10$ MeV.
Upper bounds from various cosmological constraints are all shown
by the dotted lines.
}
    \label{fig:mti_case3TR100}
\end{figure}

Although we have assumed the lowest reheating temperature $T_R \sim
10$ MeV of the thermal inflation to dilute the modulus density
most effectively, one might have a chance to generate appropriate baryon
asymmetry by the electroweak baryogenesis mechanism if one takes a
reheating temperature as high as $\sim$ 100 GeV.  In Fig.
\ref{fig:mti_case3TR100}, we show the lower bound on the modulus
abundance in the modified thermal inflation with $m_\phi \ge H_{TI}
\ge \Gamma_{\varphi_I}$ and $T_R \gtrsim 100$ GeV.  We find that the
moduli with $m_\phi \gtrsim 100$ GeV is cosmologically allowed even if
$T_R \gtrsim 100$ GeV for $\lambda_\mu =0$.  
Furthermore, for the case $\lambda_\mu \neq 0$, 
the allowed region becomes wider as $m_\phi
\gtrsim $ 5 GeV by the flaton decay into Higgs bosons.  Therefore,
in the heavy modulus (gravitino) 
mass region predicted by the HSSB models, 
one had a possibility to obtain the enough baryon
asymmetry even in the presence of the thermal inflation, if the
electroweak baryogenesis would work.

Therefore, in the presence of the thermal inflation,
we have two possibilities, so far,
to produce enough baryon asymmetry 
as well as to dilute the mass density of the moduli sufficiently:
(i) $m_\phi \lesssim$ 4 MeV by the Affleck-Dine mechanism and
(ii) $m_\phi \gtrsim$ 5 GeV by the electroweak baryogenesis mechanism.

Finally, we would like to mention about the dark matter
candidates under the tremendous entropy production by the thermal
inflation.  Note that a dark matter of our universe is diluted
away as well as the string moduli.  One possibility is the
stable modulus itself, if its mass is less than about 100 keV, since
in this mass region we can obtain $\Omega_\phi \sim 1$ 
without conflicting  with the $x(\gamma)$-ray background
constraints.  In this case, as shown in Ref.
\cite{Asaka-Hashiba-Kawasaki-Yanagida2}, the moduli dark matter with
$m_\phi \sim$ 100 keV will be tested by the future x-ray background
experiments with high energy resolution.
The axion is another candidate for the dark matter.
If its decay constant is high
enough as $f_{PQ} \sim 10^{15}$--$10^{16}$ GeV, its energy becomes
comparable to the whole energy of the present universe 
for the case that the reheating temperature 
is $T_R \sim$ 10 MeV \cite{AXIONDM}.  
Recently, the author and Yanagida proposed another candidate
for the dark matter in the presence of the thermal inflation 
\cite{Asaka-Kawasaki-Yanagida}.
The superheavy particle of mass $10^{12}$--$10^{14}$ GeV,
which was primordially in the thermal equilibrium,
could be the dark matter, if its lifetime was longer than 
the age of the universe.
%
\section*{Acknowledgements}
We would like to thank T. Yanagida for various suggestions 
and stimulating discussions.
This work was partially supported by the Japan Society for the
Promotion of Science (TA) and ``Priority Area:
Supersymmetry and Unified Theory of Elementary Particles
($\sharp$707)''(MK).
\appendix
\section{Evolution of energy densities of
inflaton and radiation after the primordial inflation}
\label{Ap-1}

Here we briefly estimate the evolution of energy densities of the
inflaton $\varphi_I$ and the radiation after the primordial inflation.
The model of the primordial inflation is basically fixed by two
parameters: the vacuum energy of the inflation $V_I$ and
the decay width of the inflation $\Gamma_{\varphi_I}$.

When the inflation ends at $t=t_{EI}$,
the energy densities of the inflation $\rho_{\varphi_I}$ 
and the radiation $\rho_R$ are given by
\begin{eqnarray}
    &&\rho_{\varphi_I}(t_{EI}) = V_I, \label{EI-1}\\
    &&\rho_R (t_{EI}) \simeq 0. \label{EI-2}
\end{eqnarray}
The Hubble parameter of the universe is given by
\begin{eqnarray}
    H(t_{EI}) \equiv H_{EI} 
    = \frac{ V_I^{1/2} }{ \sqrt{3} M_G }.
\end{eqnarray}
Then the inflaton causes a coherent oscillation for $H_{EI} > H(t) >
\Gamma_{\varphi_I}$.  During this period the energy of the oscillation
dominates the density of universe and the Hubble parameter is given as
\begin{eqnarray}
    H(t) \simeq \frac{ \rho_{\varphi_I}(t)^{1/2} }{ \sqrt{3} M_G }.
\end{eqnarray}
The evolution of $\rho_{\varphi_I}$ and $\rho_R$ is 
governed by the following equations:
\begin{eqnarray}
    &&
    {\dot \rho}_{\varphi_I} + 3 H \rho_{\varphi_I}
    = - \Gamma_{\varphi_I} \rho_{\varphi_I},\\
    &&
    {\dot \rho}_{\varphi_R} + 4 H \rho_{\varphi_R}
    = \Gamma_{\varphi_I} \rho_{\varphi_I},
\end{eqnarray}
where the dot represents the derivative with respect to the cosmic
time.  When $t \gg \Gamma_{\varphi_I}$ the solutions of them with the
initial conditions (\ref{EI-1}) and (\ref{EI-2}) are given by
\begin{eqnarray}
    \rho_{\varphi_I}(t)
    &\simeq&
    V_I \left( \frac{ R(t_{EI}) }{ R(t) } \right)^3,
    \\
    \rho_R (t)
    &\simeq&
    \frac{ 2 \sqrt{3} }{ 5 }
    \Gamma_{\varphi_I} M_G V_I^{1/2}
    \left( \frac{ R(t_{EI}) }{ R(t) } \right)^{3/2}
    \left[ 1 - 
       \left( \frac{ R(t_{EI}) }{ R(t) } \right)^{5/2} 
    \right], 
\end{eqnarray}
where $R(t)$ is the scale factor of the universe at the time $t$.
Thus the radiation energy density for $R(t) \gg R(t_{EI})$ is diluted
by a rate $R^{-3/2}$ (not $R^{-4}$) as the universe expands.  Note
that $\rho_R$ takes its maximum value
\begin{eqnarray}
    \left. \rho_R \right|_{MAX}
    \simeq 0.24 \Gamma_{\varphi_I} M_G V_I^{1/2},
\end{eqnarray}
at $R(t) \simeq 1.5 R(t_{EI})$. 
Then the maximum temperature achieved after the primordial inflation is
estimated as
\begin{eqnarray}
    T_{MAX} 
    &\simeq&
    0.93 ~
    g_\ast (T_{MAX})^{- 1/4}
    \Gamma_{\varphi_I}^{ 1/4 } M_G^{1/4} V_I^{1/8},\\
    &\simeq&
    0.70 ~
    g_\ast (T_{MAX})^{-1/4} g_\ast(T_{RI})^{1/8}
    T_{RI}^{1/2} V_I^{1/8},
\end{eqnarray}
where $T_{RI}$ denotes the reheating temperature of the 
primordial inflation and is given by
\begin{eqnarray}
    T_{RI} \simeq
    1.7 g_\ast (T_{RI})^{-1/4} \sqrt{ \Gamma_{\varphi_I} M_G }.
\end{eqnarray}

\section{Estimation of the abundance of the big-bang modulus
for the case III: $m_\phi \ge H_{TI} \ge \Gamma_{\varphi_I}$}
\label{Ap-2}

Here we derive the abundance of the big-bang modulus for the case III:
$m_\phi \ge H_{TI} \ge \Gamma_{\varphi_I}$, where the reheating
process of the primordial inflation completes after the thermal
inflation ends and its reheating temperature becomes extremely low.
The mass density of the modulus oscillation is diluted not only by the
thermal inflation, but also by the primordial inflation even during
the thermal inflation.  Therefore, we expect the abundance of the
modulus is extensively reduced than case I and II.

The big-bang modulus starts to oscillate with the initial amplitude
$\phi_0 \sim M_G$ at $t = t_{BB}$ when $H \simeq m_\phi$.  At this
time, the energy densities of the modulus, the inflaton, and the
radiation are estimated as
\begin{eqnarray}
    &&\rho_\phi (t_{BB}) = \frac{1}{2} m_\phi^2 M_G^2 
      \left( \frac{ \phi_0 }{ M_G } \right)^2,\\
    &&\rho_{\varphi_I} (t_{BB})  \simeq 3 m_\phi^2 M_G^2,\\
    &&\rho_R (t_{BB} ) \simeq \frac{6}{5}m_\phi \Gamma_{\varphi_I} M_G^2.
\end{eqnarray}
Here we used the fact that the energy of the inflaton's oscillation
dominates the universe for $H_{EI} > H(t) > m_\phi$ and
\begin{eqnarray}
    H (t_{BB})^2 \simeq m_\phi^2
    \simeq H_{EI}^2 \left( \frac{ R(t_{EI}) }{ R(t_{BB}) } \right)^3
    = \frac{ V_I }{ 3 M_G^2 } 
    \left( \frac{ R(t_{EI}) }{ R(t_{BB}) } \right)^3.
\end{eqnarray}
The cosmic temperature at $t = t_{BB}$ is estimated as
\begin{eqnarray}
    T_{BB} \simeq
    \sqrt{ \frac{6}{\pi} }
    g_\ast (T_{BB})^{-1/4}
    m_\phi^{1/4} \Gamma_{\varphi_I}^{1/4} M_G^{1/2}.
\end{eqnarray}

The thermal inflation starts at $t = t_{STI}$ when the vacuum energy
$V_0$ of the flaton $\chi$ begins to dominate the energy of the
universe.  Since we are considering the case $m_\phi \ge H_{TI}$,
$t_{STI}$ should be $t_{STI} > t_{BB}$.  Then, for $t_{BB} < t <
t_{STI}$ the energy of the universe is almost dominated by the
inflaton energy as follows:
\begin{eqnarray}
    \rho (t) 
    &=& 
    \rho_{\varphi_I}(t) + \rho_\phi (t) + \rho_R (t),\nonumber \\
    &\simeq&
    \rho_{\varphi_I}(t) + \rho_\phi (t),\nonumber \\
    &\simeq&
    \left[ 3 m_\phi^2 M_G^2 ~+~ \frac{ 1 }{ 2 } m_\phi M_G^2 
        \left( \frac{ \phi_0 }{ M_G } \right)^2
    \right]
    \left( \frac{ R(t_{BB}) }{ R(t) } \right)^3,\nonumber \\
    &\simeq&
    3 m_\phi^2 M_G^2 \left( \frac{ R(t_{BB}) }{ R(t) } \right)^3
    = \rho_{\varphi_I}(t).
\end{eqnarray}
Therefore we obtain each energy density at $t = t_{STI}$ as
\begin{eqnarray}
    &&\rho_\chi (t_{STI}) = V_0 \\
    &&\rho_\phi (t_{STI}) \simeq \frac{1}{6} V_0 
          \left( \frac{\phi_0}{M_G} \right)^2 \\
    &&\rho_{\varphi_I} (t_{STI}) \simeq V_0 \\
    &&\rho_R (t_{STI} ) \simeq \frac{2 \sqrt{3} }{5} 
                        \Gamma_{\varphi_I} V_0^{1/2} M_G.
\end{eqnarray}
We find that the vacuum energy of the flaton when the thermal
inflation starts is comparable to the energy of the inflaton's
coherent oscillation.  The cosmic temperature at $t=t_{STI}$ is
\begin{eqnarray}
    \label{T_STI}
    T_{STI}
    \simeq 
    \left( \frac{ 12 \sqrt{3} }{ \pi^2 } \right)^{1/4}
    g_\ast (T_{STI})^{ -1/4}
    \Gamma_{\varphi_I}^{1/4} V_0^{1/8} M_G^{1/4}.
\end{eqnarray}

While the thermal inflation lasts, the Hubble parameter takes the
constant value given by
\begin{eqnarray}
    H(t) = H_{TI} = \frac{ V_0^{1/2} }{ \sqrt{3} M_G },
\end{eqnarray}
and the evolution of the energy densities are governed by 
the following equations 
\begin{eqnarray}
    && {\dot \rho}_{\varphi_I} + 3 H_{TI} \rho_{\varphi_I} 
    = - \Gamma_{\varphi_I} \rho_{\varphi_I},\\
    && {\dot \rho}_\phi + 3 H_{TI} \rho_\phi = 0,\\
    && {\dot \rho}_R + 4 H_{TI} \rho_R 
    =  \Gamma_{\varphi_I} \rho_{\varphi_I}.
\end{eqnarray}
Here we have neglected the effect of the modulus decay.
The solutions are
\begin{eqnarray}
    && \rho_{\varphi_I}(t) \simeq V_0 
       \left( \frac{ R(t_{STI}) }{ R(t) } \right)^3,\\
    && \rho_\phi(t) \simeq \frac{1}{6} V_0 
       \left( \frac{ \phi_0 }{ M_G } \right)^2
       \left( \frac{ R(t_{STI}) }{ R(t) } \right)^3,\\
    && \rho_R(t) \simeq  \sqrt{3}
       M_G \Gamma_{\varphi_I} V_0^{1/2}
       \left( \frac{ R(t_{STI}) }{ R(t) } \right)^{3}
       \left[ 1 - \frac{3}{5}
           \left( \frac{ R(t_{STI}) }{ R(t) } \right)
       \right].    
\end{eqnarray}

The thermal inflation ends when the the cosmic temperature becomes $T
= T_c$.  At this time ($t = t_{ETI}$) the energy densities are
given by
\begin{eqnarray}
    &&\rho_{\chi} (t_{ETI}) = V_0,\\
    &&\rho_{\phi} (t_{ETI}) \simeq \frac{1}{6} V_0 
       \left( \frac{ \phi_0 }{ M_G } \right)^2
       \frac{ \frac{ \pi^2 }{ 30 } g_\ast (T_c) T_c^4 }
            { \sqrt{3} \Gamma_{\varphi_I} V_0^{1/2} M_G},\\
    &&\rho_{\varphi_I} (t_{ETI}) \simeq V_0
       \frac{ \frac{ \pi^2 }{ 30 } g_\ast (T_c)T_c^4 }  
            { \sqrt{3} \Gamma_{\varphi_I} V_0^{1/2} M_G },\\
    &&\rho_R (t_{ETI}) =  \frac{ \pi^2 }{ 30 } g_\ast( T_c)T_c^4.        
\end{eqnarray}
Therefore, the ratio between the energy density of the 
big-bang modulus and the entropy density is estimated as
\begin{eqnarray}
    \frac{ \rho_\phi }{ s }
    \simeq
    \frac{ T_c V_0^{1/2} }{ 8 \sqrt{3} \Gamma_{\varphi_I} M_G }
    \left( \frac{ \phi_0 }{ M_G } \right)^2.
\end{eqnarray}

Below $T = T_c$ the flaton rolls down to its true minimum and
oscillates around it.  By the flaton decay (and by the $R$-axion
decay) the reheating process of the thermal inflation completes and
the universe is finally reheated to the temperature $T=T_R$.  Then the
huge entropy is produced by a factor $\Delta$ given by
Eqs. (\ref{Delta-Raopen})
and (\ref{Delta-mti}).  Therefore, the present abundance of the
big-bang modulus is given by
\begin{eqnarray}
    \RBB \simeq
    \frac{ T_c V_0^{1/2} }{ 8 \sqrt{3} \Gamma_{\varphi_I} M_G }
    \left( \frac{ \phi_0 }{ M_G } \right)^2
    \times \frac{ 1 }{ \Delta }.
\end{eqnarray}
In the original thermal inflation model, the entropy production factor
(\ref{Delta-Raopen}) leads to
\begin{eqnarray}
    \left( \frac{ \rho_\phi }{ s } \right)_{BB}
    &\simeq&
    4.8 \frac{ m_0^4 T_R }{ \Gamma_{\varphi_I} V_0^{1/2} M_G }
    \left( \frac{ m_\chi }{ 2 m_a } \right)
    \left( \frac{ T_c }{ m_0 } \right)^4
    \left( \frac{ \phi_0 }{ M_G } \right)^2.
\end{eqnarray}
On the other hand,  in the modified model,
Eq. (\ref{Delta-mti}) gives
\begin{eqnarray}
    \left( \frac{ \rho_\phi }{ s } \right)_{BB}
    &\simeq&
    4.8 \frac{ m_0^4 T_R }{ \Gamma_{\varphi_I} V_0^{1/2} M_G }
    \left( \frac{ T_c }{ m_0 } \right)^4
    \left( \frac{ \phi_0 }{ M_G } \right)^2.
\end{eqnarray}


\end{document}